\documentclass{jfm}
\usepackage{graphicx}
\usepackage{epstopdf, epsfig}
\usepackage{amsmath}
\usepackage{amssymb}
\usepackage{xcolor,colortbl}
\usepackage{multirow}
\usepackage{longtable}
\usepackage{array}

\makeatletter
\newcommand*{\rom}[1]{\expandafter\@slowromancap\romannumeral #1@}
\makeatother
\definecolor{Grey}{gray}{0.9}
\definecolor{LightRed}{rgb}{0.937,0.675,0.639}
\newcolumntype{x}[1]{>{\centering\let\newline\\\arraybackslash\hspace{0pt}}p{#1}}

\shorttitle{Freely rising spheroids at high {R}eynolds numbers}
\shortauthor{J. B. Will, V. Mathai, S. G. Huisman, D. Lohse, C. Sun, and D. Krug}

\title{Kinematics and dynamics of freely rising spheroids at high {R}eynolds numbers}
\author{Jelle B. Will\corresp{\email{j.b.will@utwente.nl}}\aff{1},
	Varghese Mathai\aff{2},
	Sander G. Huisman\aff{1},
	Detlef Lohse\aff{1,3},
	Chao Sun\aff{4}
    \and Dominik Krug\corresp{\email{d.j.krug@utwente.nl}}\aff{1}}

\affiliation{\aff{1}Physics of Fluids Group, Max Planck UT Center for Complex Fluid Dynamics, Faculty of Science and Technology, MESA+ Institute, and J.M. Burgers Centre for Fluid Dynamics, University of Twente, P.O. Box 217, 7500 AE Enschede, The Netherlands
	\aff{2}Department of Physics, University of Massachusetts, Amherst, MA 01003, USA
	\aff{3}Max Planck Institute for Dynamics and Self-Organization, 37077 G{\"o}ttingen, Germany
	\aff{4}Center for Combustion Energy, Key Laboratory for Thermal Science and Power Engineering of Ministry of Education, Department of Energy and Power Engineering, Tsinghua University, Beijing, China}

\begin{document}
	
	\maketitle
	
	\begin{abstract}
We experimentally investigate the effect of geometrical anisotropy for buoyant spheroidal particles rising in a still fluid. All other parameters, such as the Galileo number (the ratio of gravitational to viscous forces) $Ga \approx 6000$, the ratio of the particle to fluid density $\Gamma \approx 0.53$  and the dimensionless moment of inertia $\mathsfbi{I}^*= \mathsfbi{I}_p/\mathsfbi{I}_f$ (with $\mathsfbi{I}_p$ being the moment of inertia of the particle and $\mathsfbi{I}_f$ that of the fluid in an equivalent volume), are kept constant. The geometrical aspect ratio of the spheroids, $\chi$ , is varied systematically from $\chi$ = 0.2 (oblate) to 5 (prolate). Based on tracking all degrees of particle motion, we identify six regimes characterised by distinct rise dynamics. Firstly, for $0.83 \le \chi \le 1.20$, increased rotational dynamics are observed and the particle flips over semi-regularly in a ``tumbling''-like motion. Secondly, for oblate particles with $0.29 \le \chi \le 0.75$, planar regular ``zig-zag'' motion is observed, where the drag coefficient is independent of $\chi$. Thirdly, for the most extreme oblate geometries ($\chi \le 0.25$) a ``flutter''-like behaviour is found, characterised by precession of the oscillation plane and an increase in the drag coefficient. For prolate geometries, we observed two coexisting oscillation modes that contribute to complex trajectories: the first is related to oscillations of the pointing vector and the second corresponds to a motion perpendicular to the particle's symmetry axis. We identify a ``longitudinal'' regime ($1.33 \le \chi \le 2.5$), where both modes are active and a different one, the ``broadside''-regime ($3 \le \chi\le 4$), where only the second mode is present. Remarkably, for the most prolate particles ($\chi = 5$), we observe an entirely different ``helical'' rise with completely unique features.
	\end{abstract}
	
\section{Introduction}
A single particle rising or falling in a still fluid is an experiment of striking simplicity. Yet, the complexity arising from the interaction between the particle and fluid is astounding and has captivated a number of prominent scientists and engineers over the last centuries such as Leonardo da Vinci \citep[see the review by][]{marusic2020} and Isaac Newton \citep{newton1999}. Besides the fundamental scientific appeal, the problem is also relevant in numerous practical applications. This holds both in natural and industrial settings, where understanding and modelling of particle dynamics and wake structures can be of critical importance. One example is the use of particles in the chemical industry to mix a flow more efficiently enhancing reaction rates or heat transfer \citep{risso2018}. The problem is further of relevance to the sedimentation of granular naturally shaped particles in rivers and oceans \citep{Lowe1982,meiburg2010}, as well as to the precipitation of snow, hail, and rain in the atmosphere \citep{byron2015,gustavsson2016,gustavsson2017}.

Most practical applications involve some interaction with strong background turbulence or particle-particle interactions (either directly through collisions or indirectly via the particle wakes). Overviews on these aspects of dispersed flows are provided by \citet{toschi2009}, \citet{balachandar2010}, and \citet{mathai2020}. \citet{Voth:2017} reviewed results related to particle geometrical anisotropy for light, neutrally buoyant and heavy particles in turbulent flows, specifically for small particles in turbulence. However, besides all these more complicated scenarios, the study of isolated particles in quiescent fluid is still very relevant (especially for buoyant particles), since the motion is often unaffected by the presence of turbulence or adjacent bodies \citep{Magnaudet2000,risso2018}.

Directly motivated by practical problems, much of the previous work on the topic has dealt with particles of a higher density than that of the carrier fluid. In contrast, in the current work we will focus on light, rising, particles and on the influence of shape anisotropy on their rise behaviour. This expands the explored parameter space significantly and provides valuable insight on the role of the most important parameters governing the particle behaviour. These follow from the Newton-Euler equations applied to a submerged body, where the dimensionless control parameters are the density ratio $\Gamma \equiv {\rho_p}/{\rho_f} $, the dimensionless moment of inertia tensor $\mathsfbi{I}^* \equiv {\mathsfbi{I}_p} /{\mathsfbi{I}_f}$, the particle Galileo number $Ga\equiv {\sqrt{|1-\Gamma| g l^3}}/{\nu} $, and indirectly also the particle geometry. Here, $\rho_{f}$ and $\rho_{p}$ are the densities of the fluid and the particle, respectively, $g$ is the acceleration due to gravity, $l$ is a characteristic length-scale of the geometry, $\nu$ is the kinematic viscosity of the fluid and $\mathsfbi{I}_f$ and $\mathsfbi{I}_p$ are the rotational moment of inertia tensor of the displaced fluid and the particle. Note that the Galileo number is used rather than the particle Reynolds number, $Re \equiv Vl/\nu$, because the relevant velocity scale, $V$, is not known \textit{a priori} but is in fact an output parameter. In the definition of  $Ga$ the buoyancy velocity, $V_b = \sqrt{|1-\Gamma|gl}$, is used instead. The main focus of the current work will be the on the dependence on particle geometry.

The simplest case of a freely rising body is that of a spherical (isotropic) particle, which has been studied widely \citep[see e.g.][]{preukschat1962,Murrow:1965,Karamanev:1992,veldhuis2009,Horowitz:2010,Auguste2018}. For freely rising and falling spheres, the onset of path instabilities was characterized numerically by \cite{Jenny2004} and was found to occur between $Ga = 150$ and $Ga = 225$. It should be noted that the onset of path instabilities is only weakly dependent on $\Gamma$ and $\mathsfbi{I}^*$. However, the influence of these parameters is significant for the complex dynamics and rise-patterns at even higher $Ga$. This complexity arises from the laminar separation of the boundary layers past a blunt body, which leads to significant unsteadiness in the wake due to vortex shedding \citep{Achenbach:1974}. As a consequence, the variety of particle behaviours observed when varying $Ga$ or $\Gamma$ is rich, even for the simplest (i.e. isotropic) geometry \citep{veldhuis2007,Horowitz:2010}. 

Path oscillations of isotropic bodies are induced by horizontal asymmetries in the pressure field and are unaffected by particle orientation. Furthermore, it should be kept in mind that the rotation of the particle is coupled to translation, which can be modelled using the Kelvin-Kirchhoff equations \citep{Ern:2012,mathai2020}. Particle rotation can induce a Magnus lift force on the body causing additional pressure forcing which can even affect regime transitions \citep{Mathai2018}. Thus, in this manner, skin friction affects the horizontal oscillations indirectly. Therefore, the rotational moment of inertia of the body needs to be controlled when performing these types of experiments. This effect, combined with a potential dependence on $\Gamma$, could be one of the factors responsible for the large spread in drag coefficient data observed for spherical geometries at high $Re$ \citep[figure 2]{Horowitz:2010}, as as previously suggested by \citep{Mathai2018}.

In practical situations, the particle's shape is almost never perfectly spherical. Therefore, studying the effect of anisotropic geometries is of significant relevance \citep{Corey1949,Dupleich1949,alger1964,dietrich1982}. \citet{fernandes2007} showed that the onset of path instabilities is very similar for anisotropic bodies compared to spheres, albeit the critical Reynolds numbers are slightly altered. Beyond the transition point, however, the dynamics are fundamentally different since for non-spherical particles the centre of pressure, in general, does not coincide with the geometric centre of the particle. This creates an additional coupling between the fluctuating pressure in the wake and the rotational dynamics of the particle. Furthermore, the particle alignment relative to the incoming flow induces additional circulation around the body which results in a lift force contribution that is non-existent for spheres. The recent work by \cite{sheikh2019} underlines the importance of fluid inertia for the alignment of particles settling in turbulence. To date, related studies have focused on the transitional dynamics at low $Ga$. A classifications into regimes is reported for spheroids in the numerical work by  and \citet{Zhou:2017a}. Similarly, experimental investigations have been performed for both cylinders \citep{toupoint2019} and disks \citep{Fernandes:2005,fernandes2007,fernandes2008,Zhong2011,Zhong2013,Lee2013,Auguste2013}. For disks, a classification of the behaviour into regimes was first presented by \cite{Willmarth:1964} and later extended by others \citep[e.g.][]{Field1997,Zhong2011}.

It is the goal of this paper to elucidate how varying degrees of anisotropy affect the rise behaviour of light particles. Such an undertaking requires a precise definition and control of the ``anisotropy'' in the particle shape - and evidently also a restriction of the infinitely many possible particle shapes. To achieve both, we investigate spheroids, ellipsoids of revolution, ranging from oblate (disks) to prolate (needles). Naturally, these shapes are well defined mathematically. Compared to for instance disks or cylinders, an additional benefit is that spheroids allow for a gradual transition from the isotropic sphere towards both, prolate and oblate particles all the way to their extremes.

Among the specific aspects we want to address is the dependence of the drag coefficient on particle geometry. Both for spheres and anisotropic particles, it is known that the drag is dependent on $Re$. For low $Re$ the drag force is dominated by skin friction, this is called ``Stokes drag regime''. For $Re \gtrapprox 1000$, the dominant contribution to the particle drag is from vertical asymmetry in the pressure distribution on the body's surface resulting from flow separation and the formation of a wake. Here, the overall drag becomes independent of $Re$ (Newton's drag regime). This has extensively been documented in studies on natural, settling, particles \citep{alger1964,dietrich1982,Haider:1989,Ganser:1993,Holzer:2008}. In these papers the drag coefficient in Newton's regime is also shown to depend on parameters that characterize geometrical anisotropy. In Newton's regime, a potential dependence of the drag coefficient on $\Gamma$ and on $\mathsfbi{I}^*$ has not been investigated yet. This is one of the main aspects to which the present study aims to contribute. 

Another point to consider is under what circumstances a particle can tumble, i.e.``flip over''. Related studies have characterised this transition in terms of $Re$ and a nondimensionalised moment of inertia.  Thus far, these works have predominantly focused on the dynamics of flat plates and strips \citep{Smith1971,Lugt1983,tanabe1994,Belmonte:1998,Mahadevan:1999,Andersen:2005b,Andersen:2005a} or disks \citep{Willmarth:1964,Field1997,Zhong2011,Zhong2013,Lee2013}. \citet{Belmonte:1998} identified a critical Froude number (defined as the ratio of characteristic times scale for downward motion and pendular oscillations) of $Fr \approx 0.67$, which governs the transition from flutter-to-tumble for quasi 2-dimensional plat plates. Here, we will explore how the transition from flutter-to-tumble changes for non-slender geometries.
Recently, \citet{essmann2020} investigated the importance of added mass on the dynamics of ellipsoids with a set initial  rotational and translational kinetic energy in both inviscid and viscous fluids.

Finally, we would like to point out that most of the work on anisotropic bodies has thus far focused on either heavy or close to neutrally buoyant particles i.e. $\Gamma \geq 1$. The value of $\Gamma$ in the present work is significantly lower. Our results can therefore shed light on the importance of the effects of density ratio and moment of inertia for anisotropic bodies. A lower particle mass will result in larger amplitude translations and rotations for the same pressure distribution around the geometry, which could result in different regimes for the same geometry and $Re$. Thus, a stronger coupling between fluid forcing and particle motion is expected. No experimental data (or otherwise) is available, especially for Newton's drag regime. Therefore, the study of light particles is of fundamental and general interest in understanding the mechanisms resulting in specific regimes at any density ratio.

\section{Experimental method}\label{sec:Exp_method}
\subsection{Particle characteristics}
\begin{figure}
	\centerline{\includegraphics[width=1.0\textwidth]{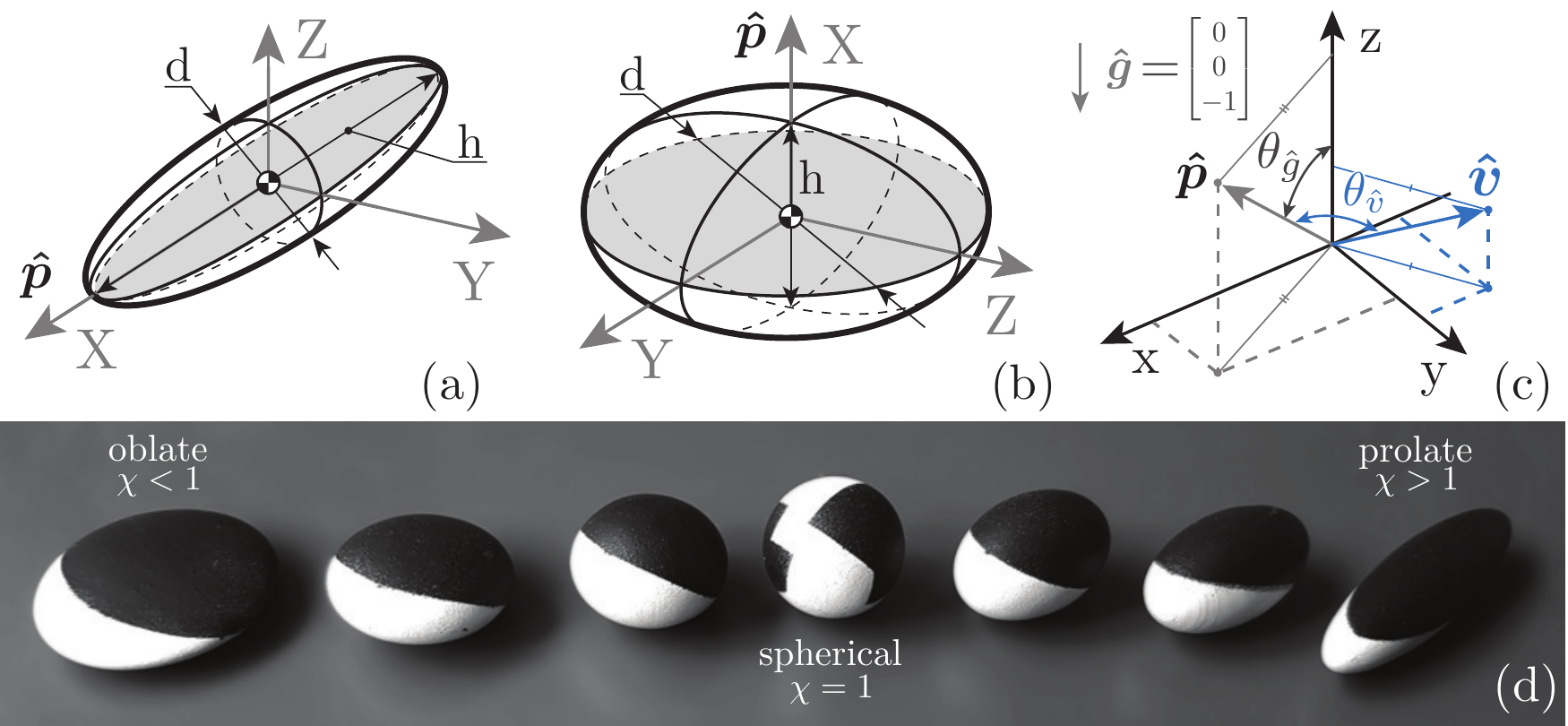}}
	\caption{Schematic of ({\it a\/}) prolate ($\chi >$1) and ({\it b\/}) oblate ($\chi <$ 1) particles along with the relevant length scales and the pointing vectors. In both cases the grey shaded areas indicate the maximum cross sectional area of the geometry. ({\it c\/}) The lab reference frame in which an arbitrary particle pointing vector and unit velocity vector are shown. The angles $\theta_{\hat{g}}$ and $\theta_{\hat{v}}$ represent the angle between the pointing vector and, respectively, the vertical $-\boldsymbol{\hat{g}}$ and the unit velocity vector $\boldsymbol{\hat{v}}$. ({\it d\/}) A picture of a selection of particles with varying aspect ratio. The particles have a pattern painted on their surface to aid with the orientation tracking.}
	\label{fig:painted}
\end{figure}

All particles used in this study are spheroids. Their shape is defined by $4(X^2/h^2 + Y^2/d^2 + Z^2/d^2) = 1$, where $h$ and $d$ specify the full lengths of the major and minor axes and capital letters indicate the particle coordinate system, which is aligned with the three perpendicular planes of symmetry (see figure \ref{fig:painted}\,({\it a,\,b\/})). Thus, the geometry can be defined by the aspect ratio $\chi \equiv h/d$. For $\chi < 1$, the geometry is called oblate (disk shaped), $\chi = 1$ corresponds to a sphere, and for $\chi > 1$ the spheroids are prolate (needle shaped). The particle orientation can be expressed in terms of a pointing vector $\boldsymbol{\hat{p}}$, which by definition aligns with the axis of rotational symmetry ( $\hat{}$ throughout this work will be used to denote unit vectors). Frequently, it is useful to consider the alignment of the particle with respect to the direction of gravity and with respect to the direction of motion of the body. These angles, which are schematically depicted in figure \ref{fig:painted}\,({\it c\/}), will be denoted by $\theta_{\hat{g}}$ and $\theta_{\hat{v}}$, respectively. The rotation around $\boldsymbol{\hat{p}}$ is denoted by the rotation angle $\psi$. 

Throughout all experiments presented here, only the aspect ratio ($\chi$) was varied in 23 steps from 0.2 to 5. All other particle characteristics, namely the volume, $V$, and the particle mass, $m_p$, were kept constant. Introducing the volume equivalent sphere diameter,  $D \equiv \sqrt[3]{6V/\pi}$, as the relevant length scale we obtain a buoyancy velocity $V_b = \sqrt{|1-\Gamma| gD}$ and the Galileo number 
\begin{equation}
Ga \equiv \dfrac{\sqrt{\left|1-\Gamma\right|gD^3}}{\nu}. \label{eq:defGalileo}
\end{equation}
The average value of $Ga$ is nominally constant across all particles and aspect ratios at approximately 6000 with a standard deviation of 122. The mean density ratio of the particles was kept as close to constant as possible with a mean $\langle \Gamma \rangle_{all} = 0.53$ and a standard deviation of 0.015. Here, $\langle \cdot \rangle_{all}$ is the average over all particles used in the experiments. The average volume equivalent diameter was $D = 19.90 \pm 0.3$ mm. Finally, the internal structure of the particles is designed to mimic a particle produced from a uniform, homogeneous, material with density $\rho_p$. This implies that the dimensionless moment of inertia tensor in the particle coordinate system in all cases equal is equal to $I^*_{ij} = \Gamma \delta_{ij}$, where $\delta_{ij}$ is the Kronecker delta. A full list of particle properties is provided in Appendix \ref{sec:App_properties}.

The rigid spheroidal particles were created using 3D printing on a RapidShape S30 SLA printer with a vertical resolution of 25 $\mu$m and a horizontal resolution of 21 $\mu$m. A non-porous resin was used with a density of approximately 1130 kg m$^{-3}$. The particles shells were printed in two parts and glued together forming a watertight seal. The particles were smoothened by hand to remove any edges at the adhesive joint and layers due to the printing process. A base coat of white paint was applied. Next, the particle was masked and black paint was applied producing the patterns (see figure \ref{fig:painted}\,({\it d\/})) which facilitates the orientation tracking. Two types of patterns were used: a complex one identical to that employed by \citet{Mathai:2016} for 0.83 $\le \chi \le $ 1.20 and a simpler one (as shown in figure \ref{fig:painted}\,({\it d\/})) for the others. The mass of the paint was found to be negligible. The primary dimensions of each particle and its mass were measured to obtain accurate values for the control parameters. No effect of surface roughness was found when testing different painting methods or unpainted particles. Surface roughness is often an important parameter when dealing with transitioning boundary layers, however in the present work we suspect that the effect of roughness is limited as the Reynolds numbers encountered are too low to trigger a transition to turbulent boundary layers, which, for a fixed sphere occur at substantially higher values of $Re \approx 3\cdot 10^5$ \citep{Achenbach:1972}.

\subsection{Setup and measurement procedure}\label{sec:setup_procedure}
\begin{figure}
	\centerline{\includegraphics[width=0.9\textwidth]{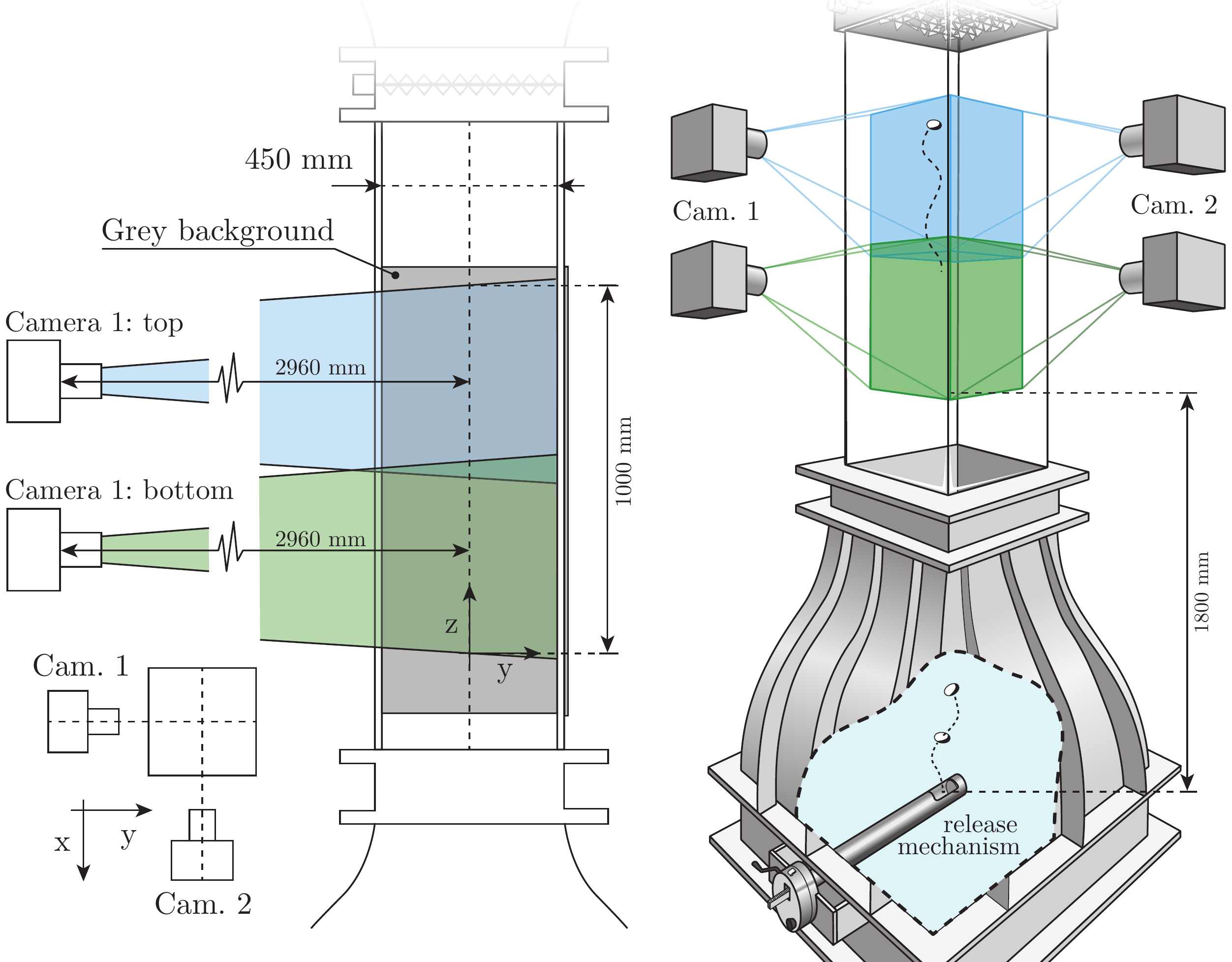}}
	\caption{Left: Schematic representation of the experimental setup showing  side and top view of the measurement section as well as the camera positions. Right: three-dimensional graphic showing the full setup including the release mechanism at the bottom of the TWT.}
	\label{fig:SetupTWT}
\end{figure}

The experiments were performed in the approximately 3 m high test section of the Twente Water Tunnel (TWT) depicted in figure \ref{fig:SetupTWT}. The temperature in the laboratory was kept constant at 20$^\circ$C and the water had ample time to equilibrate, therefore density and kinematic viscosity of water are assumed to be constant at 998 kg m$^{-3}$ and 1.00$\times$10$^{-6}$ m$^2$ s$^{-1}$, respectively. The particles were released using a release mechanism 1.8 meters (90$D$ for the particles used here) below the measurement domain in order to obtain a statistically steady state, as verified by means of the vertical acceleration statistics. The release mechanism consists of a water lock that allows the particle to be inserted without draining the tank. The particle is pushed to the release position inside a ``basket''. The end of the pipe in which the basket rests is open at the top, such that the particle can be released by  rotating the basket slowly. Between subsequent experiments the fluid was given 10 minutes to settle. No notable changes in particle behaviour were observed when extending the waiting period for up to an hour. Furthermore, we excluded any runs that were contaminated by bubbles on the surface of or around the particle. The weight of the particles was checked after every experiment in order to make sure no water had leaked into the particle shell.

During the experiments, we tracked the position and orientation of the particles as they rose through the quiescent fluid. The lab coordinate frame is defined as shown in figure \ref{fig:SetupTWT}. The origin is located at the base of the measurement volume, 1.8 m above the release. The $z$-direction is defined upwards in the vertical direction. The translational and rotational tracking was performed by image analysis of recordings made using two stationary Photron AX200 cameras with 1024$\times$1024 resolution at 256 grey levels and a recording rate of 250 fps. This recording rate was found to be sufficient to resolve the particle dynamics. The two cameras were positioned perpendicular to one another, as shown in the top view in figure \ref{fig:SetupTWT}, to track the 3 dimensional motion of the particles. The cameras were aligned with the axes of the global coordinate system along the $x$- and $y$-axes, perpendicular to the glass walls of the setup to minimize optical distortion. In order to obtain longer particle tracks a second set of cameras with the same specifications was positioned above the first. The two sets were arranged such that their fields of view slightly overlapped in the vertical direction. The ``overlap region'', where the particle could be seen by all four cameras simultaneously, was approximately 60 mm in height. All four cameras were positioned almost 3 meters away from the centre of the tunnel and outfitted with objectives with a 100 mm focal length. Each camera had approximately a 580$\times$580 mm$^2$ field of view in the central plane of the tunnel resulting in a spatial resolution of around 0.560 mm per pixel. The difference in magnification between the individual cameras was less than 0.0025 mm per pixel. The particle tracks obtained using this setup extend approximately 1 m in the vertical direction.

The tank was illuminated by eight continuous LED light sources. The lights were positioned such that the illumination is as homogeneous as possible while avoiding shadows in the particle images which would render the tracking inaccurate. As a backdrop two grey PVC plates were used. The camera apertures are adjusted such that the background corresponds to a grey value of approximately 128 out of the 256 to maximize the contrast with the black and white patterns on the particles. We further ensured that the depth of focus was sufficiently large to cover the full test section.

The tracking algorithm is based on the work of \citet{Mathai:2016} with only moderate modifications. Therefore, a complete description of the data processing method used to obtain the particle position and orientation is relegated to Appendix \ref{sec:App_tracking}.

\subsection{Data processing}\label{sec:data_processing}
\begin{figure}
	\centerline{\includegraphics[width=1\textwidth]{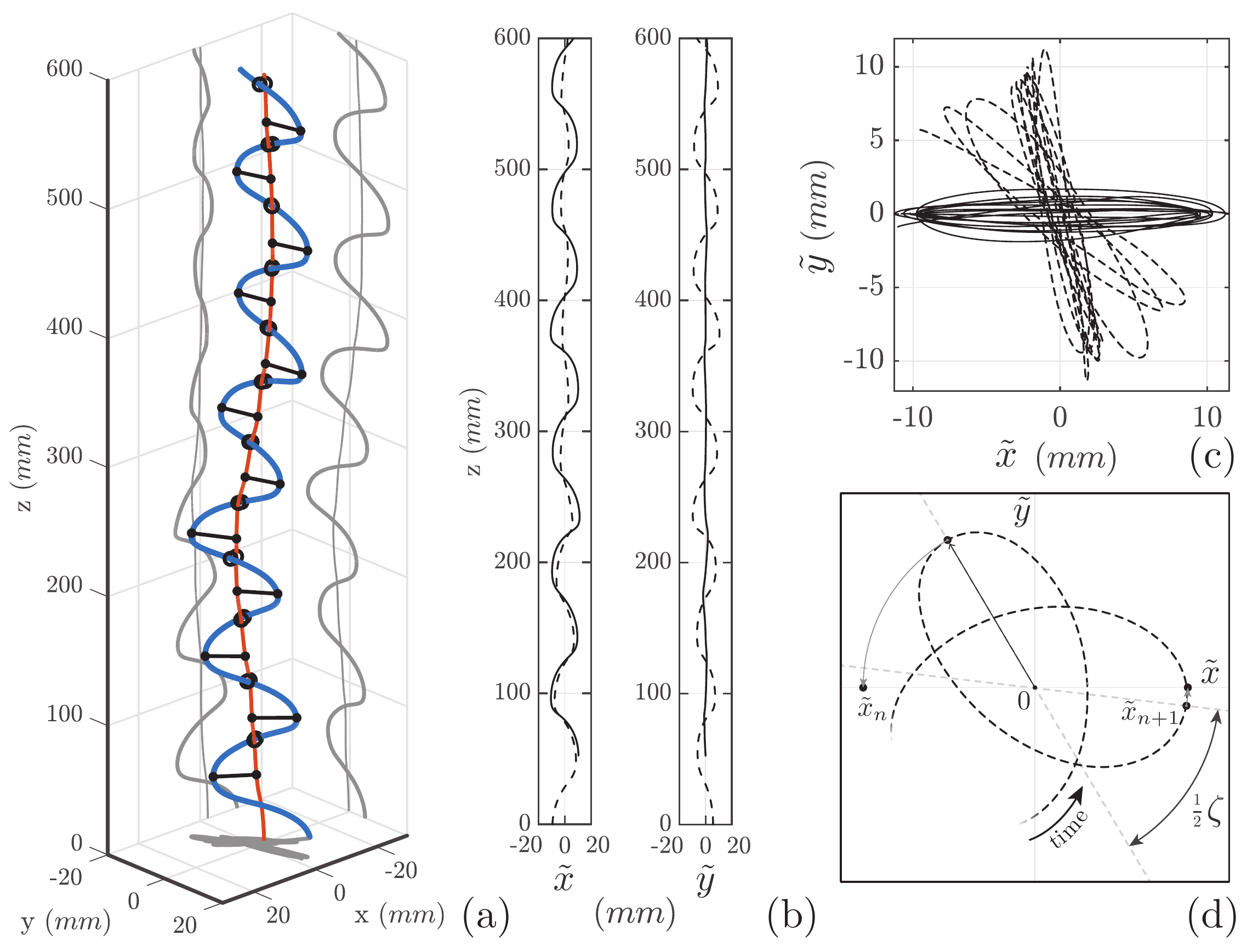}}
	\caption{({\it a,\,b,\,c\/}) Show the same particle trajectory, for $\chi$ = 0.25 ($Ga \approx 6263$ and $\Gamma \approx 0.53$), in various states of processing. ({\it a\/}) Shows the raw data in blue, and the period-averaged trajectory in red. Points of maximum and minimum amplitude are also shown (small black symbols). In ({\it b\/}) and ({\it c\/}) we show the drift corrected trajectory (dashed line), and the precession corrected path (solid line) from respectively the side and the top. The schematic in ({\it d\/}) illustrates the precession correction based on the points of maximum amplitude, the symbols match those in ({\it a\/}). The points $\tilde{x}_n$ for the precession corrected trajectory are identical to the maxima in amplitude of the position vector in the horizontal plane.}
	\label{fig:trajectories}
\end{figure}
The position of the centre of mass was extracted from the recorded frames. Consequently, the position data was smoothed in time using a convolution with a Gaussian kernel. Similarly, derivatives of the Gaussian kernel were used to obtain the first and second derivatives with respect to time \citep{Mordant:2004}. For the position, a kernel with standard deviation of 4 and window size of 13 frames was used. These values were increased to $(6,21)$ for the velocity and $(12,33)$ for the acceleration. These values were determined following the approach proposed by \cite{Mathai:2016} to filter out the high frequency noise but not the much lower frequency particle position, velocity and accelerations. This was possible thanks to temporal oversampling such that the filtering leaves the particle dynamics and its statistics unaffected. Additionally, the orientation was extracted in terms of three Euler angles $\phi$, $\theta$, and $\psi$. For the remainder of this work, we will express the orientation in terms of the pointing vector $\boldsymbol{\hat{p}}$, i.e. the direction of the particles axis of rotational symmetry (see figure \ref{fig:painted}), and a  rotation around $\boldsymbol{\hat{p}}$, which we call $\psi$. The orientation data was smoothed using the same parameters as the position data.

Further processing of the trajectories is required in order to extract properties such as the frequency and amplitude of oscillation. The simplest methods, such as fitting periodic functions to the data as employed in \cite{fernandes2007}, are not suited for the more chaotic trajectories encountered at higher $Ga$ (see figure \ref{fig:top_trajectories} and the videos in the supplementary materials). It was therefore necessary to develop an alternate approach in order to determine properties consistently across the full range of aspect ratios considered. This will be described  based on a sample trajectory as shown in \ref{fig:trajectories}\,({\it a,\,b,\,c\/}) in the following.

As a starting point, the smoothed trajectories (blue line in figure \ref{fig:trajectories}\,({\it a\/})) and its derivatives were used. First, the autocorrelation functions of the horizontal particle velocity components in the lab coordinate frame were computed. By determining the interval between subsequent peaks of this function, a first estimate for the particle oscillation period was obtained. This procedure, however, does not yield the correct frequency in cases where the trajectory is also precessing, as in illustrated in figure \ref{fig:trajectories}\,({\it d\/}).

To overcome this issue, we defined a period averaged trajectory (shown in red in figure \ref{fig:trajectories}\,({\it a\/})), where the window size of the moving average was based on the frequency obtained previously from the velocity autocorrelation. Subtracting the phase-averaged trajectory from the original one yields a drift-corrected trajectory (dashed line in figure \ref{fig:trajectories}\,({\it b,\,c\/}). We defined new coordinates $\tilde{x},\tilde{y}$, for which the origin coincides with the position of the phase-averaged trajectory.
Next, we searched for local maxima in the distance between the phase-averaged and the original trajectory (i.e. the norm of the position vector in the $\tilde{x},\tilde{y}$-frame). The magnitude of the distance at these points corresponds to the maximum amplitudes $a_n$, where $n$ is an index (see black lines in figure \ref{fig:trajectories}\,({\it a\/})). Based on the locations of the maxima, the trajectory was corrected for precession as shown by the solid lines in \ref{fig:trajectories}\,({\it b,\,c\/}). This was done by point-wise rotating segments of the trajectory around the origin (corresponding to the period-averaged trajectory), such that the points of maximum amplitude lie on the $\tilde{x}$-axis i.e. $\tilde{x}_n = a_n$, as schematically shown in figure \ref{fig:trajectories}\,({\it d\/}). In this subfigure, the designated half period is rotated by $\zeta$/2 to end up across from the previous point of maximum amplitude, here $\zeta$ is the precession angle over a complete particle oscillation cycle.

Finally, the precession corrected velocity data in the $\tilde{x}$-direction was used to obtain a new estimate of the oscillation frequency and the mean was taken over all runs. This whole process is then repeated using this new frequency as the smoothing timescale for the phase-averaging until the results converge (typically within 3--4 iterations). Statistics of the frequencies determined in this way form the basis for the discussion in the subsequent section \S \ref{sec:frequency} and Appendix \ref{sec:App_horizontal_motion}.

\subsection{Data set}\label{sec:data_set}
In total the data set consists of 269 runs. For each run the particle rises through the approximately 1 m high measurement domain. A minimum of 9 runs have been obtained per aspect ratio. For each aspect ratio a minimum of two particles were produced in order to rule out potential flaws in the production process. No aberrant behaviour was found that correlated to a single particle. Generally, the experiments show repeatable behaviour, however, in some cases there do appear to be multiple states that the particle motion can be in; transitioning from one to the other and then back akin to non-linear dynamical systems. This appears to be a property of the particle dynamics and not a flaw of the experiments.

In the supplementary materials we provide videos of the extracted motion and orientation of the particles as they rise through the quiescent fluid for 10 aspect ratios. These provide a better understanding of the dynamic behaviour of the spheroids and we will refer back to them throughout this work.


\section{Vertical motion: rise-velocity, Reynolds number and drag coefficient}\label{sec:Vertical_motion}
In this section, we will focus on the particle's vertical motion, most importantly their rise velocities.
We will show that the emergent behavior of the particles is strongly dependent on $\chi$. We
classify this dependence into six distinct regimes of the particle motion. Their definitions, characteristics, and crossovers will be further elaborated and supported as we study various properties of the rise patterns in more detail throughout this work.

\subsection{Reynolds number}\label{sec:ReynoldsNumber}
Following the work by \cite{fernandes2007} on rising disks, we define the particle Reynolds number
\begin{equation}
Re = \dfrac{v_z d_A}{\nu}. \label{eq:Reynolds_number}
\end{equation}
\begin{figure}
	\centerline{\includegraphics[width=1\textwidth]{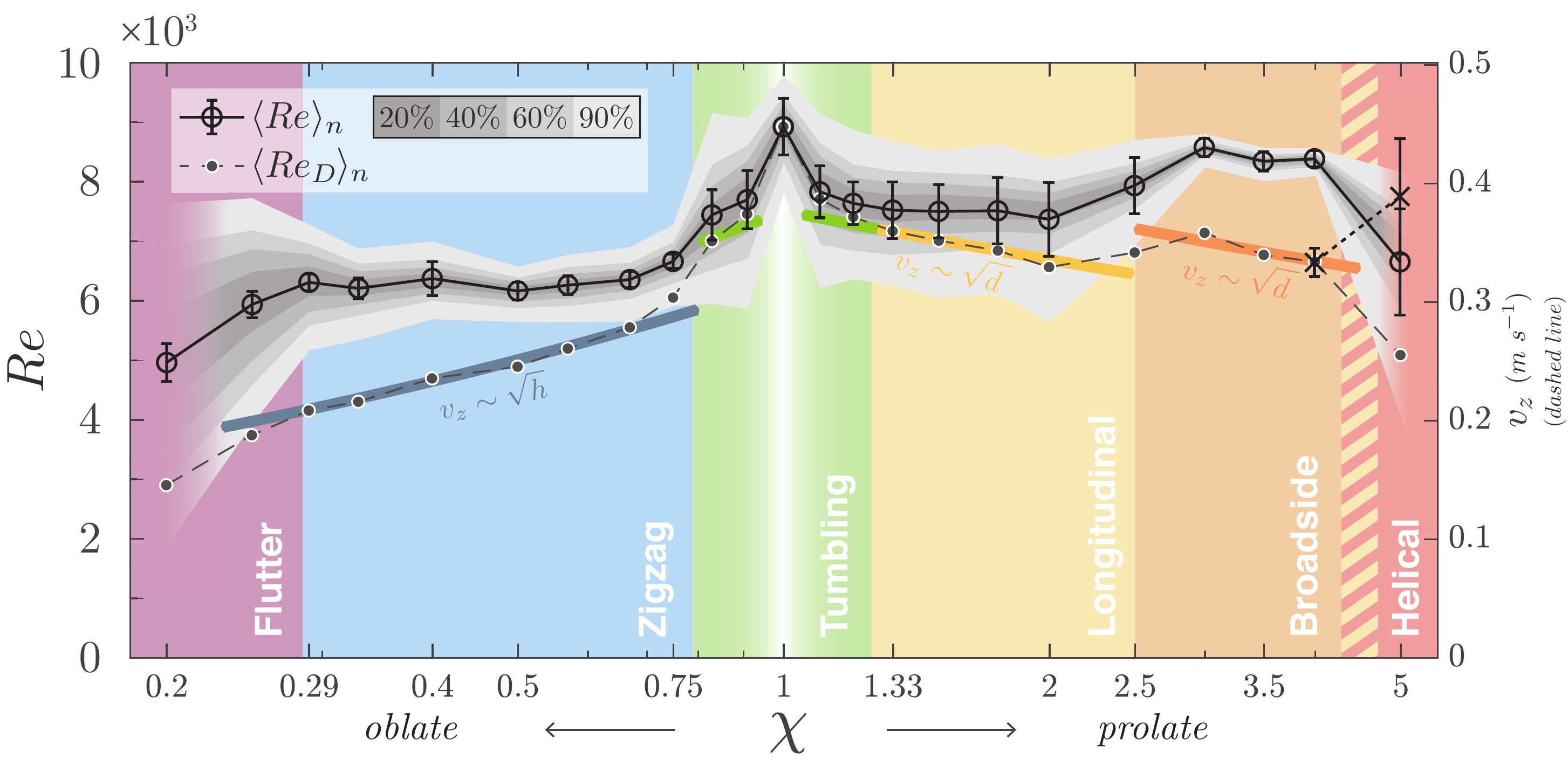}}
	\caption{Reynolds number as a function of aspect ratio for $Ga \approx 6000$. The grey symbols show $\langle Re_D \rangle_n$ and black symbols indicate the alternative definition $\langle Re \rangle_n$, using the length scale based on the maximum cross-flow area. Note that $Re_D$ is a scalar multiple of the mean rise-velocity, which is indicated on the right hand side axis. The error bars represent the standard deviation of the phase-averaged fluctuations in $Re$. The grey shaded areas show the fraction of instantaneous data points that fall within these respective regions. For $\chi = 4$ and $5$ two values are shown; these correspond to the two modes occurring simultaneously as described in \S \ref{sec:prolate_helix}, the data points marked with crosses mark the helical regime. The coloured regions indicate the different regimes that are defined based on the analysis of the particle kinematics.}
	\label{fig:ReynoldsNumber_chi}
\end{figure}
Here, $v_z$ is the vertical velocity and the length scale $d_A$ denotes the area equivalent disk diameter, which is based on the maximum cross-sectional area of the geometry ($A^+ (\chi)$) such that $d_A$ = $\sqrt{4A^+ /\pi}$ and consequently
\begin{equation}
    d_A = 
	\begin{cases}
	d \quad          & \text{for} \quad  \chi < 1\\
	D \quad          & \text{for} \quad  \chi = 1\\
	\sqrt{dh} \quad  & \text{for} \quad   \chi > 1.\\
	\end{cases}
\end{equation} 
Assuming a constant $\Gamma$ and a balance of buoyancy ($\sim d^2h$) and pressure drag ($\sim v_zd_A^2$) forces, $Re(\chi) = \text{const.}$ implies the scalings $v_z\sim\sqrt{h}$ for oblate and $v_z\sim\sqrt{d}$ for prolate geometries, respectively. Note that $A^+$ corresponds to the grey shaded areas shown in figures \ref{fig:painted}\,({\it a,\,b\/}). The choice of the length scale $d_A$ is motivated by the facts that for blunt bodies pressure drag dominates and that, according to potential flow theory, spheroids orient with the largest area perpendicular to the incoming flow \citep[see][article 124]{Lamb:1932}. 

In figure \ref{fig:ReynoldsNumber_chi}, we show $\langle Re \rangle_n$ (black symbols), where the ensemble average $\langle \cdot \rangle_n$ denotes averaging over all data points obtained for a specific aspect ratio, i.e. an average over time and across runs.
To complement the results for the average, we also consider two types of fluctuations in $Re$. The first kind relates to using the instantaneous value of $v_z$, as presented in equation (\ref{eq:Reynolds_number}). The corresponding results are indicative of the overall fluctuations in $Re$ and they are represented by the grey-shaded regions in figure \ref{fig:ReynoldsNumber_chi}. The different shadings relate to the quantiles of the distribution of data that lie within a certain region, e.g. for the 90\% region 5\% of the instantaneous data had a higher and 5\% a lower $Re$ value than the bounds of this region. This method of visualizing the result will be used  throughout this paper. Note that fluctuations in $Re$ are largely due to velocity fluctuations during an oscillation cycle. Additionally, we quantify how much the mean $Re$ varies over individual oscillations cycles of the particles. To this end, the error-bars in the figure show the standard deviation of $\langle Re \rangle_p$, where $\langle \cdot \rangle_p$ indicates a moving averaged $Re$ over one oscillation period.

The variation of $Re$ across different aspect ratios evident from figure \ref{fig:ReynoldsNumber_chi} unveils the signature of distinct regimes of particle motion, which will be introduced in the following. These regimes are indicated by colours in the background of the figure.

Particles close to isotropic (0.83 $\leq\chi\leq$ 1.2) are observed to be in the ``tumbling''-regime (green). Here, $\langle Re \rangle_n$ as a function of $\chi$ attains a global maximum for $\chi = 1$. On both, the oblate and the prolate sides, $\langle Re \rangle_n$ drops significantly once the particle becomes anisotropic. The characteristic feature of the tumbling regime is that the pointing vector can have any orientation and the particles flip over as will be shown in \S \ref{sec:alignment} and \S \ref{sec:tumble}. For even more oblate particles (lower $\chi$), we encounter the ``zigzag''-regime (blue) for 0.29 $\leq \chi \leq$ 0.75. In this regime, the motion of the particles is spiraling with varying degrees of eccentricity, ranging from nearly circular to almost planar orbits (see figure \ref{fig:Amp_and_Ecc}\,({\it b\/})) and Appendix \ref{sec:ecc}). The motion is very regular, as evidenced by the small error-bars denoting the variation in the period averaged Reynolds number. The most important property of the zigzag regime is the near constant value of $\langle Re \rangle_n$, indicating that $v_z \sim \sqrt{h}$. The most extreme oblate particles investigated here ($\chi \leq 0.25$) again display a different behaviour, which we refer to as the ``flutter''-regime (purple). Particle motion in this regime resembles inverted falling leaves \citep{pesavento2004}. Horizontal motions become dominant and the vertical velocity varies more strongly throughout an oscillation cycle. The latter is evident from the wide spread in the instantaneous values of $Re$ while the period-to-period variation, $\langle Re \rangle_p$, remains small. As a consequence, $\langle Re \rangle_n$ decreases strongly with decreasing $\chi$ within the flutter regime.

The rise patterns of prolate particles with aspect ratios between approximately $1.25 \leq \chi \leq 2.5$ are characterized by strong oscillations of the pointing vector (the ``long'' direction of these particles). However, unlike the tumbling prolate particles, spheroids in this regime do not ``flip over''. Therefore, we call this the ``longitudinal''-regime (yellow). Trajectories in this regime are more chaotic when compared to their oblate counterparts in the zigzag regime. This is related to the fact that with the symmetry axis oriented perpendicular to the velocity vector on average, there are two competing cross-flow length scales in the longitudinal regime as will be discussed in detail in \S \ref{sec:prolate_ellipsoids}. In this regime $\langle Re \rangle_n$ also appears constant, thus $v_z \sim \sqrt{d}$. At more extreme prolate aspect ratios ranging from $ 2.5 < \chi \leq 4.5$, we encounter the ``broadside''-regime (orange). Here, the oscillations of the pointing vector almost completely vanish. The dynamics appear consistent with the forcing by vortex shedding on an infinite fixed cylinder. The resulting horizontal translation of the particle is almost exclusively in the broadside direction of the particle, giving the regime its name. Remarkably, the value of $\langle Re \rangle_n$ in this regime is consistently higher than in the longitudinal regime, even though  $v_z \sim \sqrt{d}$ continues to hold. Finally, we observed another transition to what we call the ``helical''-regime (red) for the most extreme prolate aspect ratio. Here, we find two rise-patterns that coexist, the first being similar to the longitudinal mode indicated in the figure by the circular markers and the second, a near-perfect helix indicated by the crosses. The helical rise-pattern was also found to exist for the particle of $\chi = 4$ and is in this case dependent on the release conditions. Not surprisingly, the values of $Re$ differ significantly between the two modes. We will revisit this regime in \S \ref{sec:prolate_helix}. It should also be noted that the onset of the helical pattern appears to be discontinuous, while all other regime transitions (as a function of $\chi$) are gradual.

It is instructive to also consider an alternate definition of the particle Reynolds number based on $D$,
\begin{equation}
    \langle Re_D \rangle_n = \dfrac{\langle v_z \rangle_n D}{\nu}.\label{eq:Re_D}
\end{equation}
 The resulting Reynolds numbers are also included in figure \ref{fig:ReynoldsNumber_chi}. Note that $\langle 
 Re_D \rangle_n$ only varies due to changes in $\langle v_z \rangle_n$, the mean rise-velocity (also provided for these data points in the same figure). 
 Two interesting observations can be made by comparing results for $Re$ and $Re_D$. First, the rise velocity continuously decreases with increasing anisotropy everywhere but in the broadside regime where a secondary local maximum, i.e. a peak in the mean rise-velocity, is observed. Secondly, it becomes obvious, that the approximately constant values of $Re$ in the zigzag, longitudinal and broadside regimes critically rely on re-scaling with the appropriate cross-flow length-scale.
\begin{figure}
	\centerline{\includegraphics[width=0.8\textwidth]{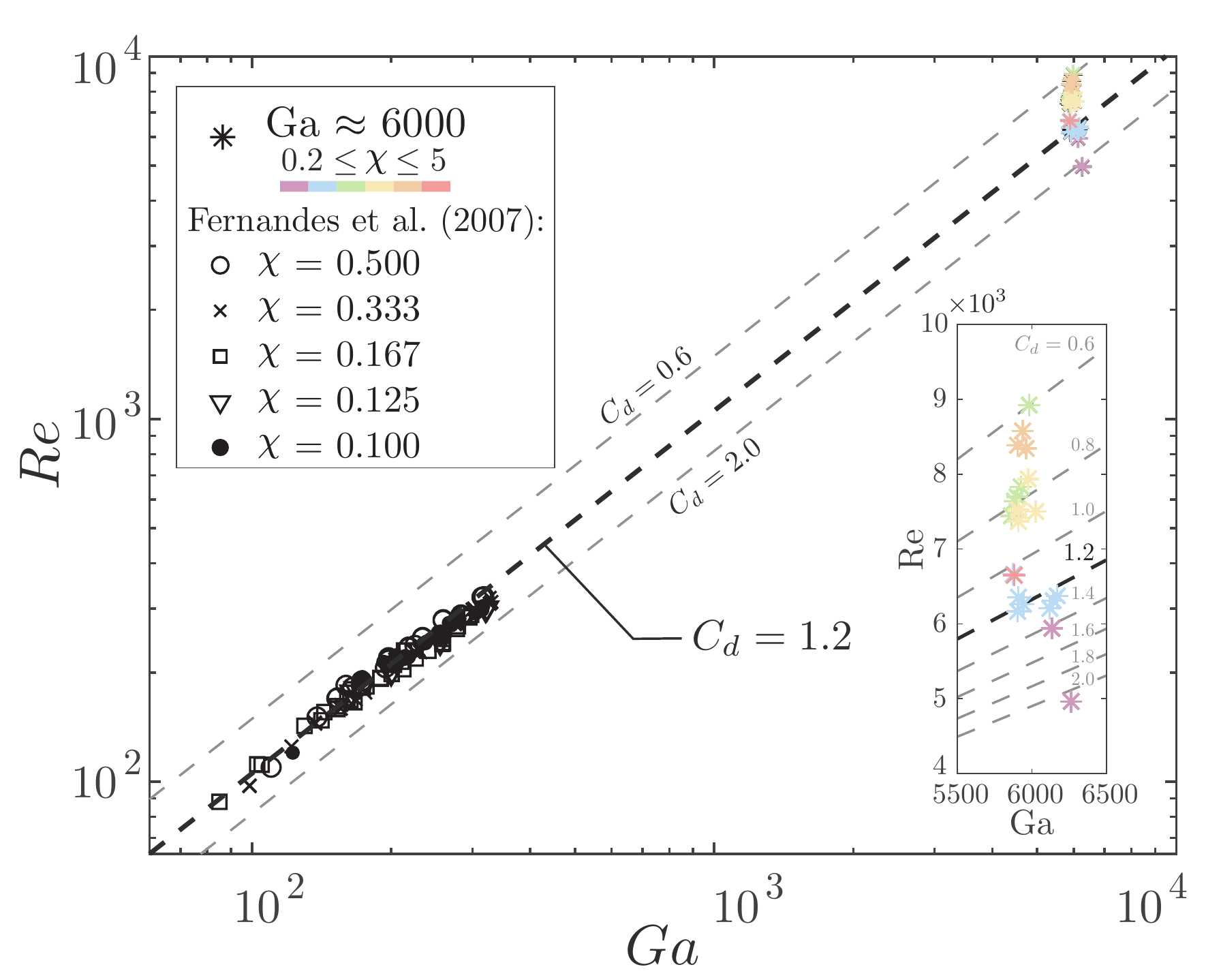}}
	\caption{ Particle Reynolds number versus Galileo number. The black markers show the data by \cite{fernandes2007}. The current results are indicated by the star shaped markers where the color indicates the regimes as defined in figure \ref{fig:ReynoldsNumber_chi}. The inset shows a more detailed overview of the current data set along with lines of constant drag coefficient. }
	\label{fig:Reynolds_Galileo}
\end{figure}

\subsection{Drag coefficient}
A key finding from figure \ref{fig:ReynoldsNumber_chi} is that neither $Re_D$ nor $Re$ are uniquely determined by $Ga$ but vary significantly with $\chi$. It is interesting to compare this result to the work by \citet{fernandes2007}, as is done in figure \ref{fig:Reynolds_Galileo}. This figure clearly shows that there is no discernible $\chi$ dependence for flat disks at lower $Re$ (80 $<Re<$ 330) for $0.1 \leq \chi \leq 0.5$, while the spread with $\chi$ is significant in the data at large $Re$. This indicates that the effect of aspect ratio variations is dependent on $Ga$ or $Re$. Nevertheless, also the difference in geometry (disks vs. spheroids) should be kept in mind when interpreting these results.

The ratio of $Ga$ to $Re$, i.e. the slope in figure \ref{fig:Reynolds_Galileo} is related to the commonly used drag coefficient $C_d$. To demonstrate this, we assume that the particles have reached their terminal velocity (as is the case for all results considered in this study). This implies that the sum of time-averaged forces applied to the body in vertical direction is equal to zero: $\langle \sum F_z \rangle_t = 0$. Here, the net driving force $\boldsymbol{F}_B = (\rho_f-\rho_p) V g \boldsymbol{\hat{z}}$ is balanced by the sum of all fluid forces. The fluid forces can be decomposed into two perpendicular components; the drag ($\boldsymbol{F}_d$) acting parallel to the direction of motion, and lift perpendicular to it. A convenient definition for rising or settling particles defines drag as the force balancing the net buoyancy force, such that the force balance reads:
\begin{equation}
\biggl< \sum F_z \biggr>_t = \underbrace{(1-\Gamma) \rho_f \dfrac{1}{6}\pi D^3 g}_{\boldsymbol{F}_B} - \underbrace{\dfrac{1}{8}\pi\rho_f \langle v_z \rangle^2_t C_d d_A^2}_{\boldsymbol{F}_d} = 0. \label{eq:Force_balance_z}
\end{equation}
Note that similar to (\ref{eq:Reynolds_number}) we use $d_A^2 \propto A^+(\chi)$ as the cross flow area in the definition of $C_d$ in (\ref{eq:Force_balance_z}). Additionally, it is worth mentioning that adhering to common practice we use   $\langle v_z \rangle^2_n$ in the definition of $C_d$, instead of $\langle v_z ^2 \rangle_t$. While the latter appears more appropriate conceptually, it cannot simply be deduced from the terminal velocity and is therefore not readily available in practice. From (\ref{eq:Force_balance_z}) it follows that $C_d$ relates $Re$ and $Ga$ according to
\begin{equation}
C_d = \dfrac{4(1-\Gamma)D^3g}{3 \langle v_z \rangle^2_t d_A^2} = \dfrac{4}{3} \dfrac{Ga^2}{\langle Re \rangle_n^2}. \label{eq:CdReGa}
\end{equation}
We include lines of constant $C_d$ in figure \ref{fig:Reynolds_Galileo}. \cite{fernandes2007} noted that for their range of $Ga$, $C_d$ was approximately constant at 1.2 for $0.1 \le \chi \le 0.5$. In  contrast, we find  that  the  particle  drag  coefficient varies from close to 0.59 for a sphere to around 2.08 for the most oblate disk. In the following sections we will investigate the behaviour of the drag coefficient in further detail, looking for physical mechanisms explaining these observations.

The inset of figure \ref{fig:Reynolds_Galileo} shows that $C_d \approx 1.2$ is also encountered for the present data in the zigzag-regime (blue). And indeed also other features  of the data in \cite{Fernandes:2005,fernandes2007} appear consistent with this regime. This suggests that for higher $Ga$, regime transitions occur for lower levels of anisotropy. This hypothesis will be further explored in \S \ref{sec:phase}. 

Specifically for the sphere, for which there is plenty of reference data in the literature, we  observe $C_d = 0.59$. This result lies in between the values  reported in \citet{Horowitz:2010} for what they call  the ``vertical'' ($C_d \approx 0.45$) and ``zigzag'' ($C_d \approx 0.75$) regimes. However, our sphere exhibits oscillations similar to the ``zigzag'' (Appendix E.1) and the drag coefficient is very similar to the results by \cite{preukschat1962} ($C_d \approx 0.54$) for a very similar density ratio ($\Gamma \approx 0.6$). Both these experimental results are contrary to the idea by \cite{Horowitz:2010} that low drag is associated with the ``vertical'' and high drag for the ``zigzag'' regime. Similar observations regarding the more subtle effects of $\Gamma$ and the importance of path oscillations on the drag of a sphere were made in the numerical work by \cite{Auguste2018}.

As a side note, we mention that the inset in figure \ref{fig:Reynolds_Galileo} shows the spread in $Ga$ for the particles used in the present experiments. Additionally, note that the variations in $Ga$ for different particles at the same $\chi$ are small (typically less than 2\%) (Appendix \ref{sec:App_properties}, table \ref{tab:particle_properties}), therefore the results are consolidated to one nominal $Ga$.

\begin{figure}
	\centerline{\includegraphics[width=1\textwidth]{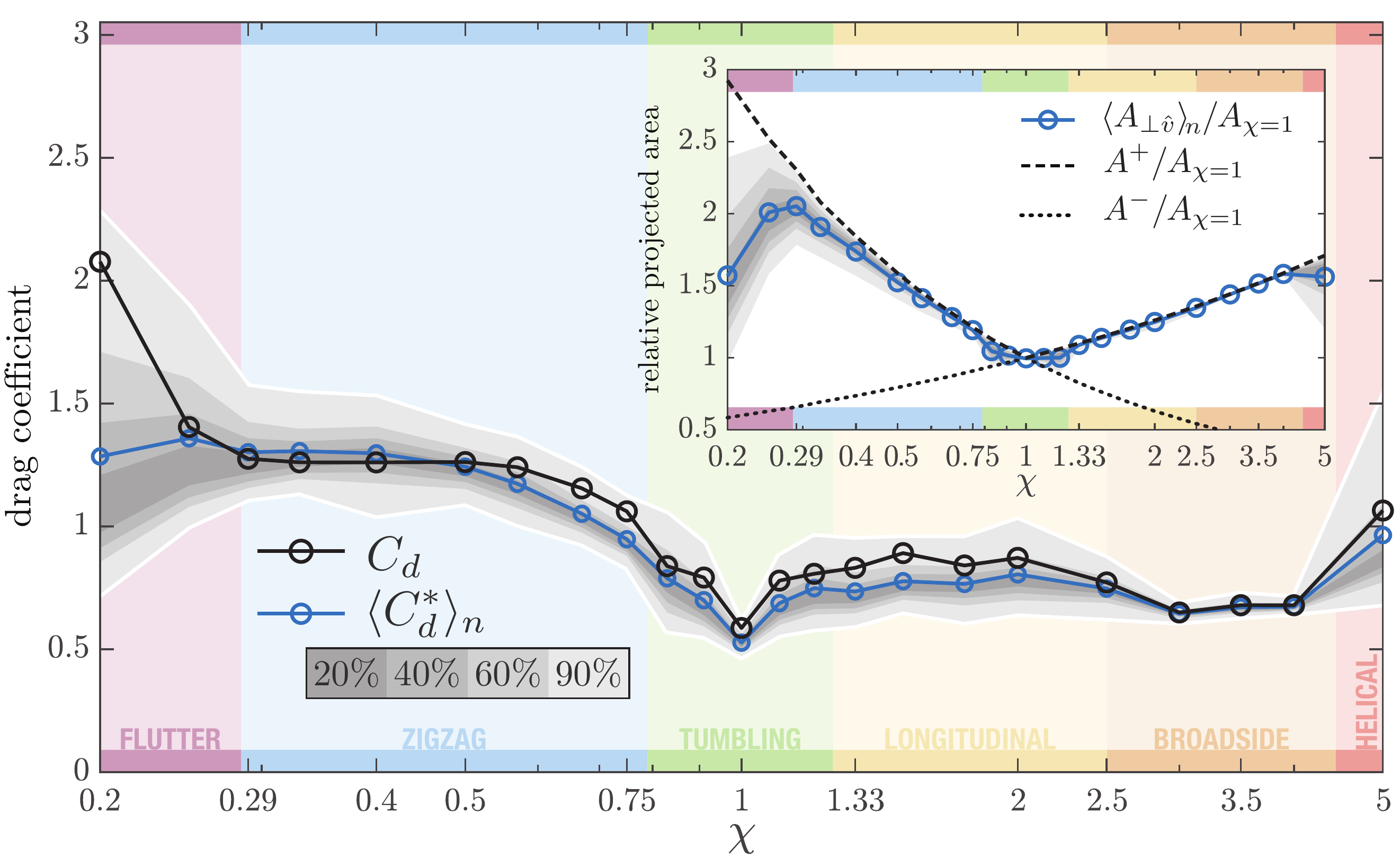}}
	\caption{Particle drag coefficient as a function of $\chi$ using two different definitions; $C_d$ (black circles) and $\langle C_d^* \rangle_n$ (blue circles) with the grey shaded area indicating the range of fluctuations in this quantity.	The inset (blue circles) shows the mean projected area perpendicular to the direction of the instantaneous particle velocity over $A_{\chi = 1} = 1/4\pi D^2$. The black lines indicate the minimum and maximum cross-flow area that is possible for a given aspect ratio. }
	\label{fig:drag_coefficient}
\end{figure}

In figure \ref{fig:drag_coefficient}, we plot $C_d$, as defined in (\ref{eq:CdReGa}), explicitly as a function of $\chi$. Note that according to (\ref{eq:CdReGa}) and with $Ga$ being constant, $C_d$ is directly related to $Re$. Hence, the regions of approximately constant $Re$ in figure \ref{fig:ReynoldsNumber_chi} show up as near constant $C_d$ in figure \ref{fig:drag_coefficient} and also other aspects of the discussion around figure \ref{fig:ReynoldsNumber_chi} apply. It is nevertheless useful to consider the drag coefficient separately. Firstly, because it is an important and widely used quantity in applications. Moreover, adopting the definition of $C_d$ has the added benefit of accounting for the slight variation in $Ga$ between different particles. But most importantly, studying the drag coefficient allows for direct comparison to related results in the literature and provides the opportunity to elucidate the cause for variations in the drag as a function of $\chi$.

\subsection{Instantaneous drag coefficient} 
The first factor we consider that will affect the behaviour of $C_d$ with $\chi$ is the relevant cross-sectional area, which we expect to be close to $A^+(\chi)$ as was already argued in the context of $Re$ in \S\ref{sec:ReynoldsNumber}. We check this assumption in the inset of figure \ref{fig:drag_coefficient} by showing the mean of $A_{\perp\hat{v}}$, which denotes the area that is perpendicular to $\boldsymbol{\hat{v}}$, the instantaneous direction of motion and its probability distribution. In this figure we also show the maximum ($A^+$) and minimum ($A^-$) cross sectional area as function of $\chi$. All results are normalized by the cross-section of the sphere $A_{\chi = 1}$. It becomes apparent that for almost all $\chi$ the $A_{\perp\hat{v}}$ follows $A^+$ very closely, with the notable exceptions of the flutter (purple), the helical (red), and the tumbling (green) regimes. For the latter however, the effect of this change is very small since the difference between $A^+$ and $A^-$ is small at $\chi \sim 1$. In general, these results confirm that even for very complex trajectories, $A^+$ is an appropriate choice for the cross-flow area in almost all cases.

One may further suspect that changes in the drag coefficient ($C_d$), which is defined using the vertical velocity only, are related to variations in the rise patterns. The drag might then be fairly uniform across $\chi$ when the velocity is considered \textit{along the path} instead. To test this hypothesis, we define an instantaneous drag coefficient $C_d^*$, which is based on the instantaneous force balance of the particle along its trajectory
\begin{equation}
	m_p\dfrac{\text{d}\boldsymbol{v}}{\text{d}t}\boldsymbol{\cdot}\boldsymbol{\hat{v}} = \boldsymbol{F}_B \boldsymbol{\cdot}\boldsymbol{\hat{v}} - \dfrac{1}{2}\rho_f\boldsymbol{v}^2A_{\perp \hat{v}}C_d^*. \label{eq:instantaneousDrag}
\end{equation}
Note that we retain the acceleration term, which only vanishes in the mean (when there is no horizontal drift) and we use the instantaneous cross sectional area $A_{\perp\hat{v}}$ for consistency.  Rewriting (\ref{eq:instantaneousDrag}) to obtain an explicit expression for $C_d ^*$ yields
\begin{equation}
	C_d^* = \dfrac{\pi D^3\left((1-\Gamma)\boldsymbol{g} - \Gamma \dfrac{\text{d}\boldsymbol{v}}{\text{d}t}\right)\boldsymbol{\cdot}\boldsymbol{\hat{v}}} {3\boldsymbol{v}^2A_{\perp\hat{v}}}. \label{eq:Cd_star}
\end{equation}
In this definition, $C_d^*$ accounts for \emph{all} fluid forces including added mass in the direction of instantaneous motion.
The ensemble-average $\langle C_d ^* \rangle_n$ is calculated for all aspect ratios and included in figure \ref{fig:drag_coefficient} (blue circles). We observe that in the ``flutter''-regime using $C_d ^*$ indeed eliminates the strong increase in the mean drag coefficient that was observed for $C_d$ (black symbols) and we find $\langle C_d ^* \rangle_n \approx$ 1.2, which agrees with the neighbouring zigzag regime. This suggests that, when correcting for the reduced cross-flow area and increased path oscillations, the behaviour is similar between these two regimes. Note, however,  that instantaneous values of $C_d^*$ fluctuate significantly in the ``flutter''-regime as indicated by the grey shaded regions. Furthermore, we find that the instantaneous drag coefficient, $C_d^*$, is approximately single-valued in the alignment $\boldsymbol{\hat{p}}(t)\boldsymbol{\cdot \hat{v}}(t)$ (not shown), suggesting that the balance in (\ref{eq:instantaneousDrag}) captures the essential dynamics. Besides in the purple regime, we do not observe a significant difference between $C_d^*$ and $C_d$. This implies that differences between the path velocity and $\langle v_z \rangle$ are effectively compensated by changes in the alignment and the changes in the relevant cross section associated with the latter. This observation indicates that using the maximum cross-flow area and the vertical velocity in fact are good approximations for spheroidal geometries, which is convenient as both quantities are easily obtained. We also conclude that for all but the most oblate particles variations in drag are not simply caused by the geometry of the trajectories but must have different origins, e.g. the structure of the wake.

On the basis of yet another definition of the drag coefficient (using $D$ as the length scale) a comparison of our results can be made to empirical drag models developed for heavy particles. This is presented in  Appendix \ref{sec:App_settling}, where it can be seen that significant differences arise between the two, highlighting the importance of the particle density ratio $\Gamma$ as a control parameter.

\subsection{Contribution of the added mass force on particle dynamics}\label{sec:AddedMass}
Using our data, we can also single out the effect the added mass force \citep{Maxey:1983,Magnaudet:1997,limacher2018} has on the particle motion. To this end, we consider the classical Kelvin-Kirchhoff equation for the translation of rigid body in a fluid
\begin{equation}
    \left( m_p \mathsfbi{I} + \mathsfbi{A} \right) \boldsymbol{\cdot} \dfrac{\text{d} \boldsymbol{v}'}{\text{d} t} + \boldsymbol{\Omega}' \times \left( m_p \mathsfbi{I} + \mathsfbi{A}  \right) \boldsymbol{\cdot v}' = \boldsymbol{F}'_\omega + \boldsymbol{F}'_B.\label{eq:KK}
\end{equation}
Here, $\mathsfbi{I}$ is the identity tensor and $\mathsfbi{A}$ is the added mass tensor. Furthermore, (\ref{eq:KK}) is defined in a coordinate frame (indicated with superscript $'$) with the origin fixed in space but whose axis are parallel to the particle coordinate frame and rotate with the particle rotation rate $\boldsymbol{\Omega}$. This choice ensures that the components of the added mass tensor $\mathsfbi{A}$, which can be obtained analytically from potential flow solutions as derived in the work by \citet{Lamb:1932} (see Appendix \ref{sec:App_added_mass_tensor} for details), remain constant in time. The transformation from the lab coordinate frame to the instantaneous particle orientation is given by the rotation matrix $\mathsfbi{R}(t)$. On the right hand side of (\ref{eq:KK}), the term $\boldsymbol{F}'_\omega$ represents all fluid forces on the particle (besides the added mass force) and can be decomposed into drag (acting along $\boldsymbol{\hat{v}}$) and lift (perpendicular to it): $\boldsymbol{F}'_\omega = \boldsymbol{F}'_d + \boldsymbol{F}'_l$. For completeness, results for the lift component are included in terms of a lift coefficient  $\langle C_l^* \rangle_n = \langle 2||\boldsymbol{F}_l||/\rho_f \boldsymbol{v}^2 A_{\perp\hat{v}} \rangle_n$ in table \ref{tab:rise_properties} in Appendix \ref{sec:App_properties}.

Typically, added mass contributions are lumped in with lift and drag, but here we aim to specifically consider their contributions to the particle dynamics. Remarkably, we find that the net contribution of the added mass terms cancels when taken along the path for all $\chi$, i.e.
\begin{equation}
    \bigg\langle \Big( \underbrace{\mathsfbi{A} \boldsymbol{\cdot} \dfrac{\text{d}\boldsymbol{v}'}{\text{d}t}}_{\boldsymbol{F}_{a1}'} + \underbrace{\boldsymbol{\Omega}' \times \left[ \mathsfbi{A} \boldsymbol{\cdot v}'\right]}_{\boldsymbol{F}_{a2}'} \Big) \boldsymbol{\cdot \hat{v}}' \bigg\rangle_n \approx 0.\label{eq:added_mass_path}
\end{equation}
 This implies that, on average, the added mass does not contribute to the motion along the path under steady state conditions. Noting that also $\left( \boldsymbol{\Omega}' \times \left[ m_p \boldsymbol{v}' \right] \right)\boldsymbol{\cdot \hat{v}}' \equiv 0$, we thus find that the drag coefficient along the path, $\langle C_d^* \rangle_n$ as defined in (\ref{eq:Cd_star}), is in fact largely unaffected by added mass and inertial effects. Thus $\langle C_d^* \rangle_n \approx \langle 2||\boldsymbol{F}_d||/\rho_f \boldsymbol{v}^2 A_{\perp\hat{v}} \rangle_n$.\\

\begin{figure}
	\centerline{\includegraphics[width=1\textwidth]{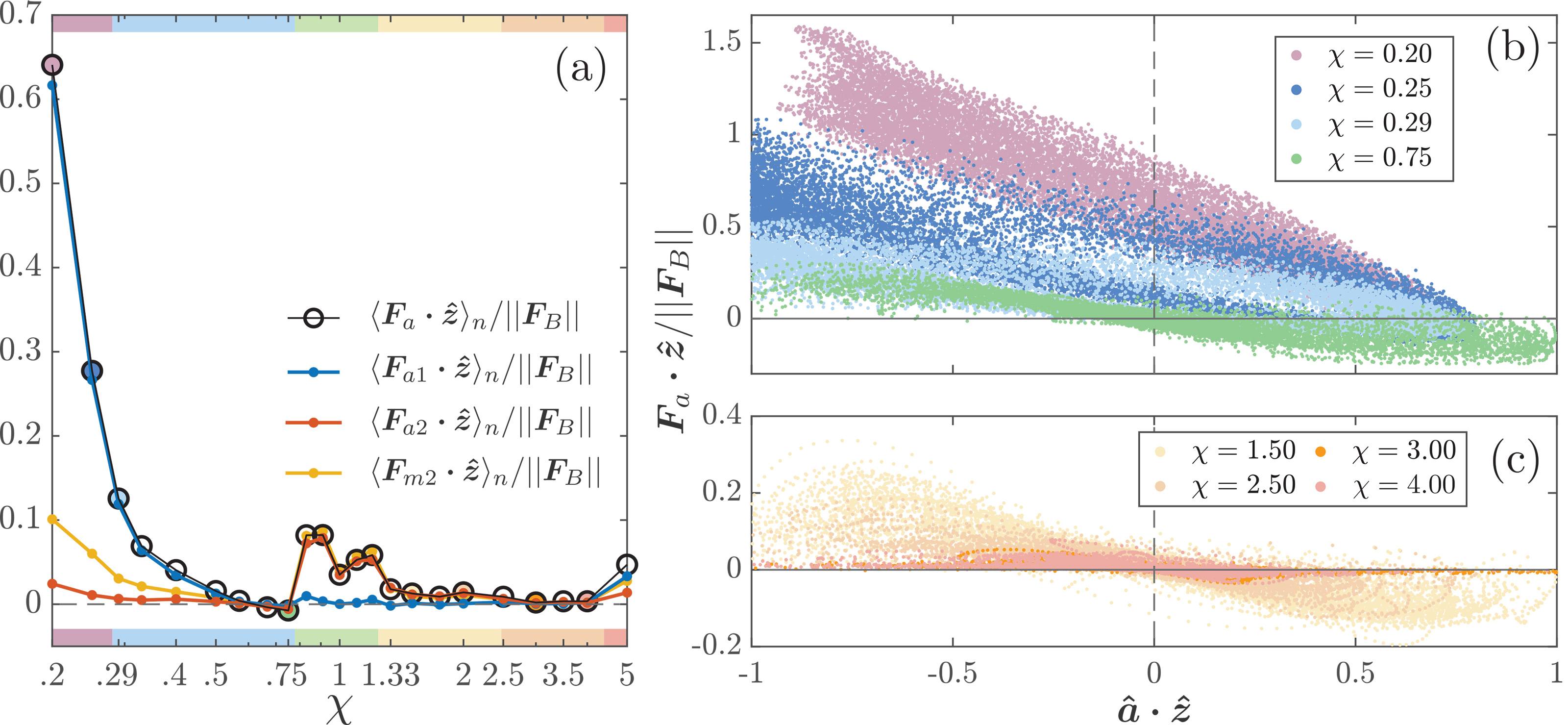}}
	\caption{({\it a\/}) Ensemble averaged added mass contributions as function of $\chi$. ({\it b,\,c\/}) Scatter plots showing the instantaneous added mass force ($\boldsymbol{F}_{a1} + \boldsymbol{F}_{a2}$) along the $z$-direction and alignment between the direction of particle acceleration and the $z$-direction. In ({\it b\/}) four oblate and in (c) four prolate aspect ratios are shown.   }
	\label{fig:added_mass}
\end{figure}

Finally, we evaluate if there is a net contribution of the added mass terms to the vertical motion of the particle. In doing so, we consider a decomposition of the added mass force $\boldsymbol{F}_a = \boldsymbol{F}_{a1}+ \boldsymbol{F}_{a2}$, with the components defined in (\ref{eq:added_mass_path}). In figure \ref{fig:added_mass}\,({\it a\/}) we show the mean contribution of these terms in the vertical direction ($\hat{\boldsymbol{z}}$) normalized by the magnitude of the net driving force $||\boldsymbol{F}_B||$ as a function of $\chi$. For $\boldsymbol{F}_a$, we observe that for almost all spheroids, besides the most oblate ones, the net contribution remains small, accounting for less than $10\%$ of the net gravitational force. For the flutter regime, however, the contribution is much greater, reaching a value as high as 0.65 for $\chi = 0.2$. This effect is almost entirely due to $\boldsymbol{F}_{a1}$ as can be seen from the blue line in figure \ref{fig:added_mass}\,({\it a\/}). Further, there is a small but noticeable  net vertical contribution of $\boldsymbol{F}_{a2}$ in the tumbling (green) regime, while this term remains insignificant for more anisotropic particles. For completeness, we also show $\boldsymbol{F}_{m2} \boldsymbol{ \cdot \hat{z}}/ ||\boldsymbol{F}_B||$, where $\boldsymbol{F}_{m2} = \mathsfbi{R}^T \boldsymbol{\cdot} \left( \boldsymbol{\Omega}' \times m_p \boldsymbol{v}' \right)$ in figure \ref{fig:added_mass}\,({\it a\/}). This quantity is seen to be almost identical to $\boldsymbol{F}_{a2}$ for $\chi \approx 1$. This result is consistent with the facts that the added mass tensor for these particles is almost a scalar multiple of the identity tensor and that the translational added mass is approximately $0.5\rho_f$, which is close to $\rho_p$ used in our experiments. However, towards the fluttering regime we notice that the contribution of $\boldsymbol{F}_{m2}$ outgrows the added mass effect. This implies that the particle rotation plays a noticeable role in the vertical force balance of tumbling and fluttering particles.

In figures \ref{fig:added_mass}\,({\it b,\,c\/}), we examine the combined vertical added mass contribution more closely to explain the asymmetry that causes a net vertical force. To this end, we present scatter plots of the normalized vertical component of the added mass force, $\boldsymbol{F}_a \boldsymbol{\cdot \hat{z}}/ ||\boldsymbol{F}_B||$, vs. $\boldsymbol{\hat{a} \cdot} \boldsymbol{\hat{z}}$, i.e. the alignment between the particle acceleration vector and the vertical.

Initially focusing on the results for $\chi = 0.75$ in figure \ref{fig:added_mass}\,({\it b\/}), we observe that negative and positive excursions of the added mass force are more or less symmetrical, i.e. when the particle accelerates downwards $\boldsymbol{\hat{a} \cdot} \boldsymbol{\hat{z}} < 0$ the added mass force is positive and when $\boldsymbol{\hat{a} \cdot} \boldsymbol{\hat{z}} > 0$  it is negative and of comparable magnitude. As expected, this symmetry results in a zero net contribution of the added mass to the vertical added mass force as is shown in \ref{fig:added_mass}\,({\it a\/}). The same applies to the results for prolate particles in figure \ref{fig:added_mass}\,({\it c\/}).

Now we consider the remaining oblate particles, with more extreme anisotropy, shown in figure \ref{fig:added_mass}\,({\it b\/}). For increasing anisotropy the symmetry in $\boldsymbol{F}_a$ completely vanishes. Remarkably, even when $\boldsymbol{\hat{a} \cdot} \boldsymbol{\hat{z}} < 0$  the direction of the added mass force is in the positive $z$-direction for a large portion of the data points. There are two effects at play that cause this asymmetry. First, an asymmetric distribution is produced if the  particle orientation (relative to $\boldsymbol{\hat{a} \cdot} \boldsymbol{\hat{z}}$) is different when the particle is accelerating vs. decelerating. Second, it is important to note that $\left(\mathsfbi{R}^{T}\boldsymbol{\cdot} (\mathsfbi{A} \boldsymbol{\cdot} \mathsfbi{R})\right) \boldsymbol{\cdot} \text{d}\boldsymbol{v}/\text{d}t$ is not parallel to $\text{d}\boldsymbol{v}/\text{d}t$. Combined with the fact that the particle acceleration in the horizontal direction becomes more and more dominant for increasing anisotropy, this can result in an upward added mass force even when the particle decelerates in the vertical direction. Both these effects can be observed in the videos in the supplementary materials showing the particle instantaneous accelerations and alignment along the path, for instance when comparing the video for $\chi = 0.57$ to that for $\chi=0.25$.

The correlation between $\boldsymbol{\hat{a} \cdot} \boldsymbol{\hat{z}}$ and $\boldsymbol{F}_{a} \boldsymbol{\cdot} \boldsymbol{\hat{z}}/||\boldsymbol{F}_{B} ||$ translates into the net force in the vertical direction shown in figure \ref{fig:added_mass}\,({\it a\/}). For the most oblate particle,  $\langle \boldsymbol{F}_{a} \boldsymbol{\cdot \hat{z}} \rangle_n$ amounts to more than half of the buoyancy force, but the effect quickly decreases with increasing $\chi$. Entirely consistent with the observations made in figure \ref{fig:added_mass}\,({\it b\/}); $\langle \boldsymbol{F}_{a} \boldsymbol{\cdot \hat{z}} \rangle_n$ is essentially zero for $\chi \gtrapprox 0.5$ or all non-tumbling cases. We observe that when added mass contributes significantly $\langle \boldsymbol{F}_{a} \boldsymbol{\cdot \hat{z}} \rangle_n >0$, which implies that the added-mass acts along the rise-direction, thus leading to an increased $\langle v_z \rangle_n$, compared to a hypothetical case without added mass.

\begin{figure}
	\centerline{\includegraphics[width=0.94\textwidth]{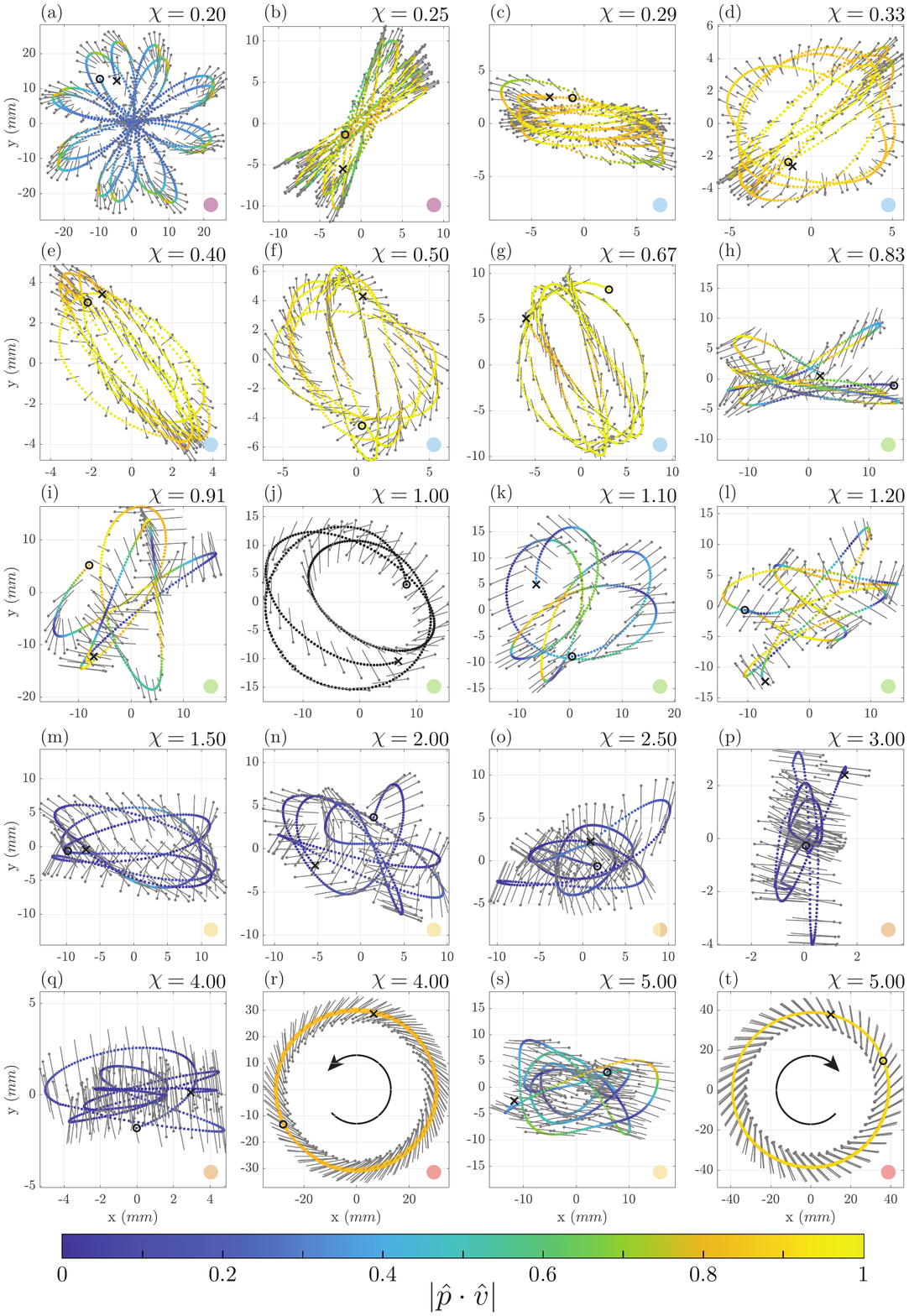}}
	\caption{Each subfigure shows a representative trajectory for a specific $\chi$ as seen from the top with the particle moving from the black circle to the cross. Particle position is indicated by the dots, the colour indicates the degree of  alignment between the pointing vector and the velocity vector. The solid grey lines are shown every 0.02 seconds (5 frames) and indicate the horizontal projection of the pointing vector, with the dot at the end indicating the upward direction. The coloured dot in the bottom right of each subfigure indicates the assigned regime.}
	\label{fig:top_trajectories}
\end{figure}

\section{Horizontal periodic motion of the particles}\label{sec:Periodic_motion}
In this section, we will focus on the motion in the horizontal plane, which is induced by unsteady motions in the wake of the rising particles. We plot horizontal projections of representative particle trajectories for 18 different aspect ratios in figure \ref{fig:top_trajectories}, in order to illustrate the richness of the patterns of motion. The data is smoothed and corrected for drift. 

In general, the patterns for the oblate particles (figure \ref{fig:top_trajectories}\,({\it a--i\/})) are regular and periodic, but even there the differences with varying $\chi$, e.g. in amplitude or eccentricity, are significant. For prolate particles (figure \ref{fig:top_trajectories}\,({\it k--t\/})), the trajectories are a lot more irregular compared to their oblate counterparts. The shape of the trajectories appears similar between the longitudinal (yellow) and the broadside (orange) regimes, but the amplitude of the excursions varies significantly between these two. What clearly stands out against the seemingly chaotic tracks observed otherwise in this regime is how stable and close to circular the helical modes for $\chi = 4$ and $\chi=5$ (figure \ref{fig:top_trajectories}\,({\it r,\,t\/})) are, this in addition to the near constant particle alignment. It should be noted, that the existence of two rather distinct rise patterns is not limited to the most prolate particles. Albeit less pronounced, another example for which two different rise patterns are encountered can be seen in figure \ref{fig:top_trajectories}\,({\it d\/}) for $\chi = 0.33$. Here, an almost planar oscillation switches to a close to circular one as the particle rises. However, in this case the transition is reversible. The distinct patterns therefore appear to be transient states of the particle dynamics in this instance. Such behaviour is similar to that observed for rising bubbles that transition from planar to helical motion\citep{mougin2006,shew2006}. We will revisit the interesting case of the most prolate spheroids in \S \ref{sec:prolate_helix}. First, we proceed to describe and classify the horizontal motions in terms of  frequency ($f$) in the next section. Results for other basic properties (the amplitude, $a$, and eccentricity, $\eta$, of the oscillations) are presented in appendix \ref{sec:App_horizontal_motion} to keep the main text succinct.

\subsection{Frequency analysis}\label{sec:frequency}
\begin{figure}
	\centerline{\includegraphics[width=1\textwidth]{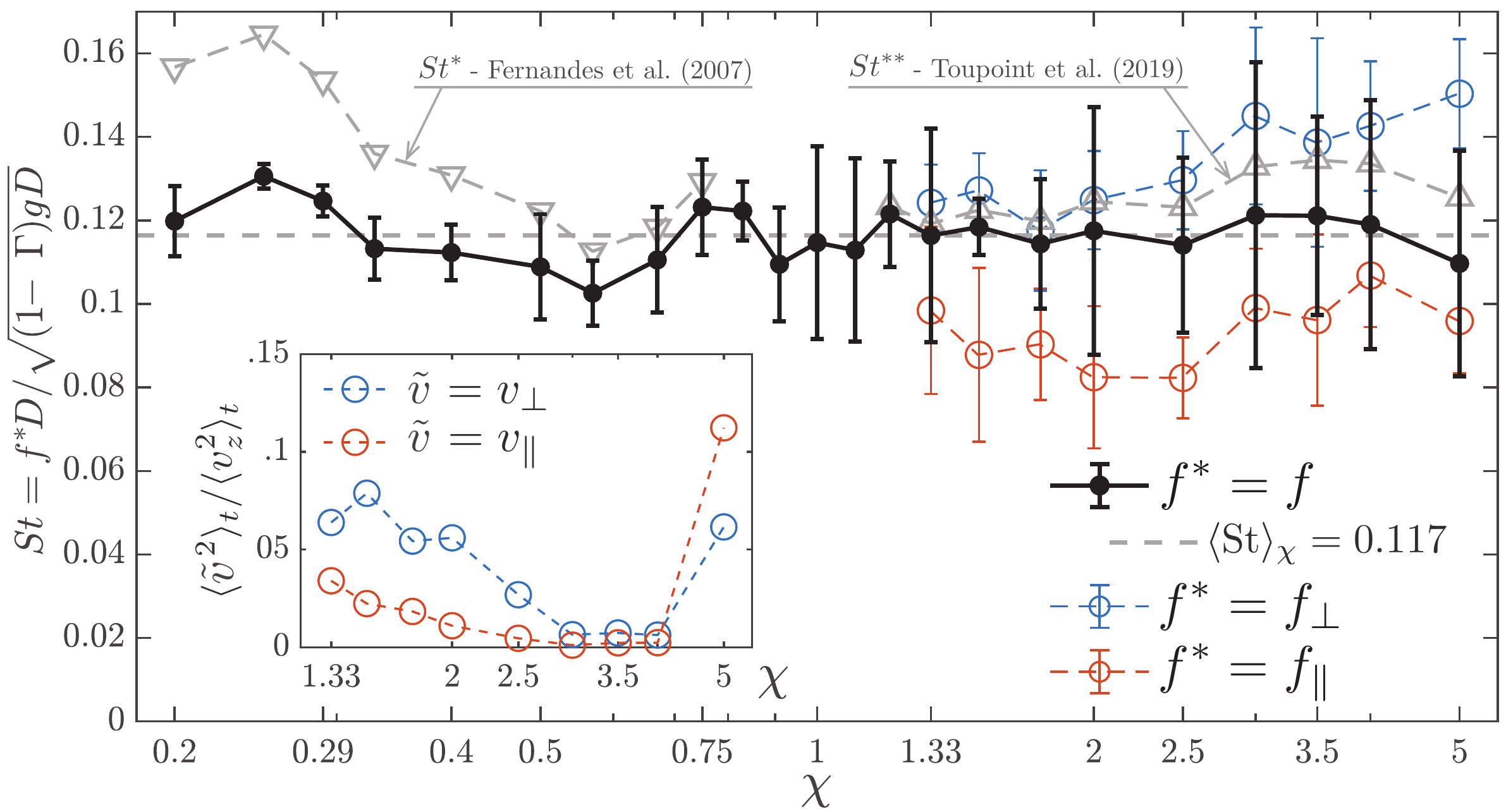}}
	\caption{ Oscillation frequency as a function of particle aspect ratio presented in terms of the Strouhal number using the volume equivalent sphere diameter and the buoyancy velocity scale. The black filled circles show the Strouhal number using the frequency of the precession corrected trajectory. For prolate particles,  $St$ is also shown using the frequencies $f_{\perp}$ (blue) and $f_{\parallel}$ (red) of the two different oscillation modes identified in \S\ref{sec:prolate_ellipsoids}. The grey lines at the top shows the current results using the $\chi$ dependence as proposed in the literature. The inset of the figure depicts (for prolate particles) a measure of the kinetic energy in the two respective oscillation modes normalized by the vertical kinetic energy. Showing that, except for $\chi$ = 5, the oscillation associated with $v_{\perp}$ is dominant.}
	\label{fig:frequency}
\end{figure}
We present results for the oscillation frequency of the particles (see figure \ref{fig:frequency}) in terms of the dimensionless Strouhal number given by
\begin{equation}
St \equiv \dfrac{f D}{V_b}.\label{eq:defStrouhal}
\end{equation}
This definition makes use of the buoyancy velocity scale $V_b$ and the volume equivalent sphere diameter $D$, which are both independent of $\chi$. The frequency $f$ is the one obtained using the approach detailed in \S\ref{sec:data_processing}.

Figure \ref{fig:frequency} shows that $St \approx$ 0.12 and this value is almost constant for the full range of aspect ratios investigated here. Data from \cite{Horowitz:2010} for rising spheres show $St \approx$ 0.1, which is within the scatter of our experiments, albeit slightly lower than the mean observed here. The mean over all $\chi$ is $St = 0.117$ , corresponding to an oscillation frequency of 1.77 Hz with a standard deviation of 0.1 Hz. This result is very surprising and indicates that the frequency is neither dependent on $\chi$ nor on $Re$ in the present case. The constant frequency for varying $\chi$ is different from related findings in literature. \citet{fernandes2007} found that for oblate disks the relevant length scale at low $Ga$ is $(dh)^{1/2}$ with a weak dependence on $Re$. This scaling is derived from work on tumbling cards by \citet{Mahadevan:1999} and later by \citet{JonesShelley2005}, giving an empirical dependence of the Strouhal number on $Re$ (or equivalently on $C_d$) and $\chi$ according to $St^* = St(2/C_d)^{1/2}\chi^{1/2} = fd/\sqrt{(\Gamma -1)gd}$. The latter is found to collapse all their data in the range $80<Re  < 330$ and $0.1<\chi<0.5$. In the work by \cite{toupoint2019} similar results are found for $St$ for heavy prolate cylinders (which behave similar to our longitudinal regime): $St^{**} =f\sqrt{hd}/\sqrt{(\Gamma -1)gd}$. Both definitions, $St^*$ (for oblate) and $St^{**}$ (for prolate particles), clearly indicate a dependence on $\chi$. For comparison, we show both definitions in figure \ref{fig:frequency} as grey lines with triangular markers. For both oblate and prolate particles, the use of $St^*$ and $St^{**}$ does not collapse the results obtained here. It should be mentioned however, that for prolate particles, both expressions lie within the spread in the observed frequencies (as evidenced by the error bars). Therefore no definitive conclusions can be drawn regarding the scaling even though $D$ appears to be the more appropriate choice based on the present results. However, the fact that the scaling observed in \citet{fernandes2007} does not apply here is not surprising, since the results by \citet{Mahadevan:1999,JonesShelley2005} are valid for heavy particles $\Gamma = \mathcal{O}(10^3)$ while the current results are for low $\Gamma$. Nevertheless, the results agree for $\Gamma \approx 1$ \citep{fernandes2007}.

In interpreting the apparently different frequency scalings, it is important to note that for both \citet{fernandes2007} and  \cite{toupoint2019}, the density ratios ($\Gamma \approx 1$ and $\Gamma \approx 1.16$, respectively) are significantly higher than the one considered here ($\Gamma \approx 0.53$). It therefore appears that the stronger coupling to the fluid motions are a likely explanation for the different behaviour observed in the present case. This is in line with conclusions drawn for spherical bodies by \cite{Karamanev:1992}, \cite{Horowitz:2010}, \cite{Mathai2018} and, \cite{Auguste2018} who found regime changes related to changes in the density ratio, and the rotational moment of inertia. Details, especially on how the apparent insensitivity to geometry variations at low density ratios emerges, remain nevertheless unclear at this point.

\subsection{Prolate spheroids; two interacting modes}\label{sec:prolate_ellipsoids}
\begin{figure}
	\centerline{\includegraphics[width=1\textwidth]{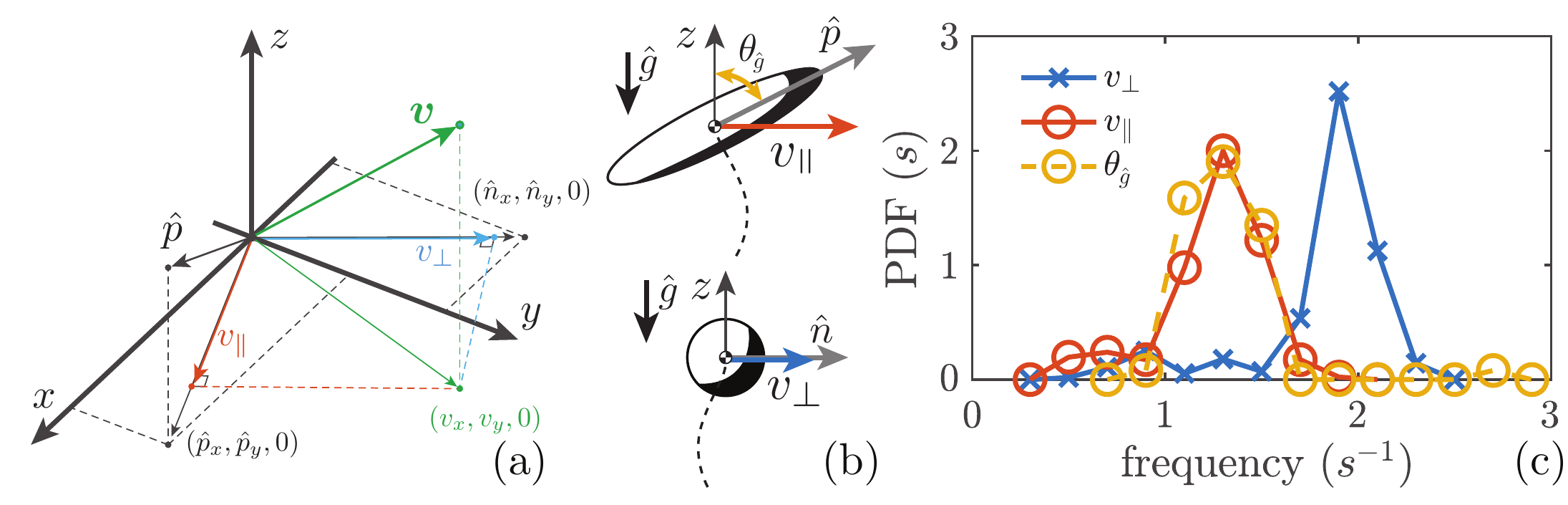}}
	\caption{({\it a\/}) Decomposition of the velocity vector in the horizontal plane into a component along ($v_\parallel$) and perpendicular ($v_\perp$) to the pointing vector. ({\it b\/}) Graphic showing the oscillation modes associated with the decomposition of the velocity; the top sketch shows the longitudinal and the bottom one the broadside oscillations for the same particle. ({\it c\/}) PDF of the frequencies obtained for aspect ratio $\chi = 2$. Two distinct frequency peaks are visible which are associated with the two modes of the geometry.}
	\label{fig:lagrangian_frame_explained}
\end{figure}
Next, we disentangle the complexity of the motion of the prolate particles, which leads to the rather irregular patterns in the trajectories displayed in figure \ref{fig:top_trajectories}\,({\it k--q,\,s\/}). The general tendency of particles to rise with their largest cross-section perpendicular to gravity exposes an ellipsoidal cross-section to the flow in the case of prolate spheroids. This is in contrast to oblate particles, where this area is circular. Hence, instead of only one length scale for oblate particles (the diameter), two unique length scales, i.e. the long ($h$) and the short axis ($d$) of the ellipsoidal cross-section, are relevant for the periodic motions of prolate spheroids. In this section we will show that this gives rise to two distinct oscillation modes that coexist and interact with each other producing the observed dynamics. Each of these modes will be shown to have a distinct frequency.

For the analysis, we consider projections of the velocity $\boldsymbol{v}$ and of the pointing vector $\boldsymbol{\hat{p}}$ onto the horizontal plane, as indicated in figure \ref{fig:lagrangian_frame_explained}\,({\it a\/}). We decompose the horizontal velocity into a component along the pointing vector ($v_{\parallel}$) and a component perpendicular to it ($v_{\perp}$). Without loss of generality we define the perpendicular direction as $\boldsymbol{\hat{n}}= \boldsymbol{\hat{z}}\times\boldsymbol{\hat{p}} / || \boldsymbol{\hat{z}} \times \boldsymbol{\hat{p}} ||$, so that the velocity $v_{\perp}$ attains both positive and negative values. This decomposition results in two periodic velocity signals that individually describe the motion of the particle in a plane spanned by $\boldsymbol{\hat{p}}$ and $\boldsymbol{\hat{z}}$ ($v_{\parallel}$) and perpendicular to it ($v_{\perp}$) as illustrated in figure \ref{fig:lagrangian_frame_explained}\,({\it a,\,b\/}). On this basis, it is now possible to discern the frequencies $f_\perp$ and $f_\parallel$  of these different motions by considering the primary peak of the autocorrelation of $v_{\perp}$ and $v_{\parallel}$, respectively. Note that no evidence for significant interactions of the two modes at frequencies corresponding to $f_{\perp} + f_{\parallel}$ was observed. Results at $\chi = 2$ are displayed in figure \ref{fig:lagrangian_frame_explained}\,({\it c\/})  in terms of a probability density function (PDF) of frequencies corresponding to individual periods.

It is clearly evident that the periodic motion for $f_\perp$ and $f_\parallel$ are distinctly different for this particle. Furthermore, we included the PDF of frequencies corresponding to oscillations of $\theta_{\hat{g}}$, i.e. the angle between the pointing vector and the $z$-axis (see figure \ref{fig:lagrangian_frame_explained}\,({\it b\/})) as yellow symbols in figure \ref{fig:lagrangian_frame_explained}\,({\it c\/}). We see that the frequency of particle alignment is identical to that of $v_{\parallel}$. This is similar to oblate bodies where these frequencies also match and are related to the same flow mechanism of vortex shedding inducing both horizontal translation and a periodic oscillation in the particle orientation. The oscillation in the $v_{\perp}$ direction, on the other hand, more resembles the pattern induced by vortex shedding behind an infinitely long cylinder. The latter might also induce a rotation around the $\boldsymbol{\hat{p}}$ ($\psi$), which remains inconsequential to the particle alignment, however.

The mean of $f_\perp$ and $f_\parallel$ is non-dimensionalised using equation (\ref{eq:defStrouhal}) and plotted in figure \ref{fig:frequency} for all prolate non-tumbling particles. 
This figure confirms that the frequencies corresponding to motions in different planes are indeed different at all $\chi$ considered. When rescaled according to (\ref{eq:defStrouhal}), both $f_\perp$ and $f_\parallel$ individually also exhibit a significant dependence on $\chi$, making it even more surprising that this normalization works well without the decomposition (the black datapoints).

For $v_{\perp}$ we see the Strouhal number increasing with $\chi$. This is consistent with the notion that $d$ is the relevant length-scale in this case, since $d$ decreases with $\chi$, while $D$, as employed in (\ref{eq:defStrouhal}), is constant. In the case of $v_{\parallel}$, we observe a general decrease of the Strouhal number for increasing anisotropy in the longitudinal regime, which is consistent with the increase of the length-scale $h$. This suggests that for the oscillation perpendicular to the pointing vector the length scale $d$ is appropriate and similarly $h$ is characteristic for the oscillation parallel to $\hat{p}$. This yields a more constant value of $St_{\perp}$ and $St_{\parallel}$ in the longitudinal regime, however the behaviour in the broadside regime (orange) is not consistent for $f_\parallel$, this could be explained by the fact that the oscillations in this direction are almost non-existent in this regime.

Apart form their frequencies, it is also relevant to consider how the two modes differ in energy. To this end, we show a measure for the kinetic energy in the respective oscillation modes in the inset of figure \ref{fig:frequency}. Here, the values are normalized by the mean kinetic energy in the vertical according to $\langle \tilde{v}^2 \rangle_t / \langle v_z^2 \rangle_t$, where $\tilde{v}$ stands for either $v_{\perp}$ or $v_{\parallel}$. In general, the energy in the oscillation along $\boldsymbol{\hat{n}}$ is larger than the one in the plane containing $\boldsymbol{\hat{p}}$, such that the vortex-shedding mechanism that induces oscillations normal to $\boldsymbol{\hat{p}}$ is the dominant one for most particles. It is also important to note that for both, $v_{\perp}$ or $v_{\parallel}$, the energy of the horizontal motions decreases to very low values towards the broadside regime. Additionally, the pointing vector oscillations are minute in this regime, yet measurable. Finally, the situation changes for $\chi = 5$, where the energy in the horizontal motion picks up significantly and $v_\parallel$ becomes the stronger mode, in \S\ref{sec:prolate_helix} this will be associated to the transition to the helical rising pattern.\\

A final point concerns the question whether the oscillations associated with $v_\perp$ are connected to rotations around the pointing vector, $\psi$. In the numerical work by \cite{namkoong2008} rising and falling cylinders were studied at particle Reynolds numbers ranging from 66 -- 185. They found that translations of the particle in the $\boldsymbol{\hat{n}}$-direction are indeed related to body rotations around its symmetry axis inducing a lifting force on the body, i.e. a Magnus lift force. In our experiments, we do not observe this direct connection as the rotation $\textrm{d}\psi/\textrm{d}t$ remains weak and is non-periodic as well as only weakly correlated to $v_\perp$. A similar observation was made by \cite{toupoint2019}, but was hampered by their limited accuracy in resolving rotations around $\boldsymbol{\hat{p}}$. Here, this accuracy is $\pm 0.5^\circ$, which puts this finding on stronger footing. We therefore conclude that strong path oscillations in the perpendicular direction are not always induced by a rotation induced Magnus lift force and thus are not necessarily coupled to the particle's rotational dynamics.

\section{Statistics of the particle orientation}\label{sec:orientations}
In this section we will consider particle orientations and alignment. For anisotropic bodies, the interaction between the particle's orientation and its direction of translation is a crucial aspect that determines the particle kinematics and dynamics. We will see that this effect is strongly related to the onset of regime changes. 

\subsection{Preferential orientation and alignment}\label{sec:alignment}
\begin{figure}
	\centerline{\includegraphics[width=1\textwidth]{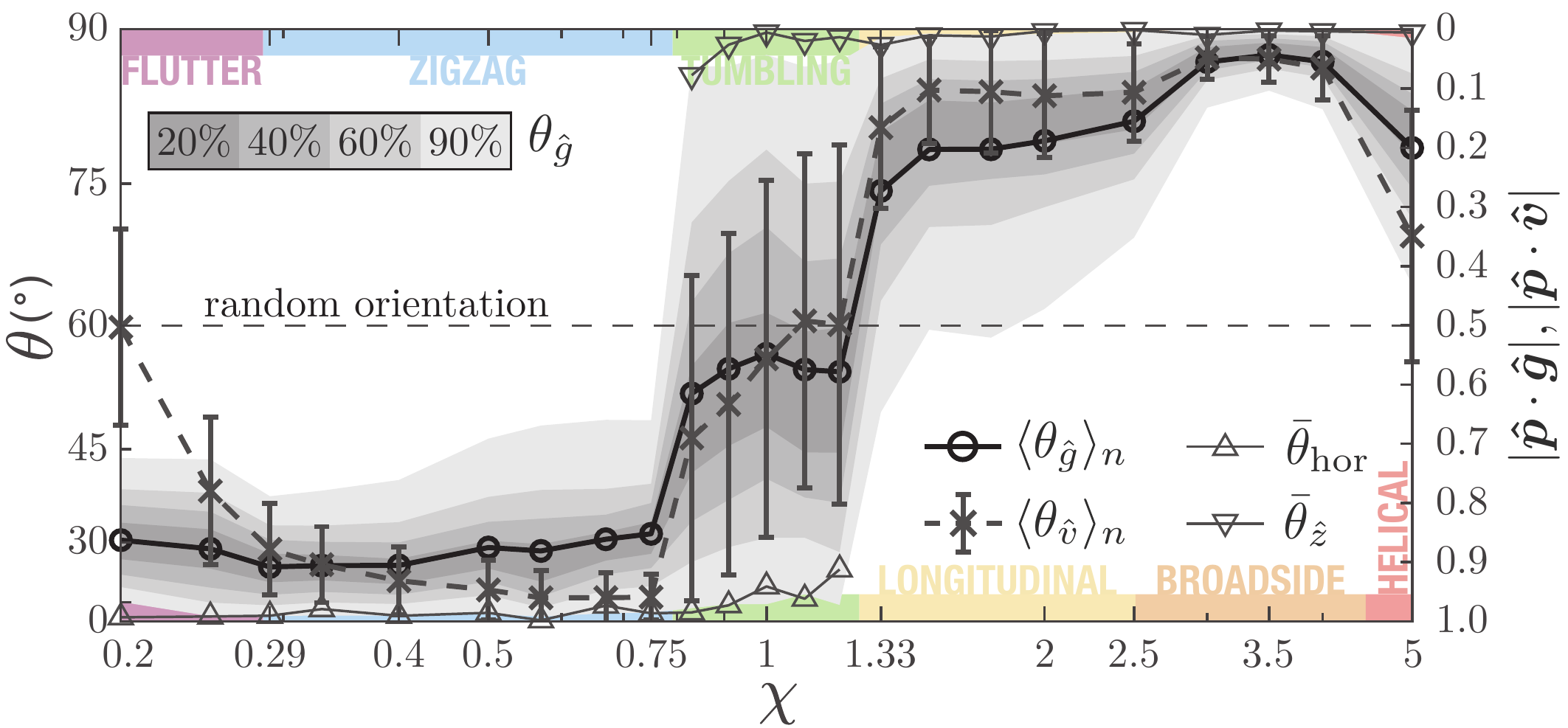}}
	\caption{ Statistics of particle pointing vector alignment, on the left $y$-axis in degrees and on the right in terms of the dot products. The black and grey markers show the ensemble-averaged absolute alignments of the pointing vectors with respectively the direction of gravity and that of the instantaneous particle motion. The shaded regions show the distribution of $\theta_{\hat{g}}$. Additionally, the statistics of the alignment of the reference orientation are shown indicated by the triangular mark. To observe the convergence of these quantities towards $0^\circ$ and $90^\circ$, respectively, would require more statistics for the tumbling regime.}
	\label{fig:alignment_theta}
\end{figure}

To capture the mean orientation of the particles around which the pointing vector oscillates, we consider the angle with respect to the vertical direction as defined in figure \ref{fig:painted}\,({\it c\/}) using the definitions
\begin{equation}
\bar{\theta}_{\hat{g}} =\left\{
\begin{matrix}
\quad \bar{ \theta }_ { \textrm{hor}} = \sin^{-1}(||\{ \langle\hat{p}_x\rangle_n, \langle \hat{p}_y \rangle_n\}||), & \qquad (\chi < 1) \\
\quad \bar{ \theta }_ { \hat{z}} = \cos^{-1}(| \langle \hat{p}_z\rangle_n|), & \qquad (\chi > 1)
\end{matrix}
\right.\ \label{eq:Strouhal_Decomp}
\end{equation}
for oblate and prolate particles, respectively (with overlap in the tumbling regime). Our results in figure \ref{fig:alignment_theta} confirm that the expected mean orientation of $\boldsymbol{\hat{p}}$ is pointing along $\hat{z}$, $\bar{\theta}_{\textrm{hor}} = 0$ for oblate, and perpendicular to it ($\bar{\theta}_{\hat{z}} = 90^\circ$) for prolate spheroids.

It is important to note, however, that the mean orientation might never actually be encountered during the rise of the particles. This applies in particular for oblate objects, where the pointing vector tends to orbit around the vertical direction without ever perfectly aligning with it. It is therefore insightful to consider additional orientation statistics including the mean alignment of the pointing vector with the particle motion ($\boldsymbol{\hat{v}}$), but also the mean offset from the reference angle $\bar{ \theta }_{\hat{g}}$. To this end, we introduce the angles
\begin{align}
\theta_{\hat{g}} &= \cos^{-1}{(| \boldsymbol{\hat{p}}(t) \boldsymbol{\cdot} \boldsymbol{\hat{g}} |)}, \\
 \theta_{\hat{v}} &= \cos^{-1}{(| \boldsymbol{\hat{p}}(t) \boldsymbol{\cdot} \boldsymbol{\hat{v}}(t) |)},
\end{align}
as a measure of the instantaneous alignments. For convenience, we also use $| \boldsymbol{\hat{p}}(t) \boldsymbol{\cdot} \boldsymbol{\hat{g}} | $ and $| \boldsymbol{\hat{p}}(t) \boldsymbol{\cdot} \boldsymbol{\hat{v}}(t) |$ without conversion to an angle and therefore these quantities are also included in figure \ref{fig:alignment_theta}. The absolute values are taken because of symmetry in the particle geometry, which renders the positive direction of $\boldsymbol{\hat{p}}$ arbitrary. The angles $\theta_{\hat{g}}$ and $\theta_{\hat{v}}$ are schematically shown in figure \ref{fig:painted}\,({\it c\/}). Slight differences arise in the interpretation of the statistics of $\theta_{\hat{g}}$ for $\chi<1$ and $\chi>1$ since  $\boldsymbol{\hat{p}} \boldsymbol{\cdot} \boldsymbol{\hat{g}}$ always changes sign along the trajectory for prolate particles. In essence, however, $\theta_{\hat{g}}$ remains a measure of the actual mean particle orientation encountered in both cases.

In figure \ref{fig:alignment_theta}, the ensemble-averaged alignments $\langle | \boldsymbol{\hat{p}}(t) \boldsymbol{\cdot} \boldsymbol{\hat{g}} |\rangle_n$ and  $\langle | \boldsymbol{\hat{p}}(t) \boldsymbol{\cdot} \boldsymbol{\hat{v}}(t) |\rangle_n$ are shown as black circles and grey crosses respectively. For $|\boldsymbol{\hat{p}} \boldsymbol{\cdot} \boldsymbol{\hat{g}}|$, we also show the probability density distribution contours indicating the range of angles covered during an oscillation cycle. The error bars for $|\boldsymbol{\hat{p} }\boldsymbol{\cdot} \boldsymbol{\hat{v}}|$ provide similar information.

The synopsis of the results in figure \ref{fig:alignment_theta} substantiates the previously defined regimes. Firstly, the tumbling (green) regime is evident from the fact that $|\boldsymbol{\hat{p}} \boldsymbol{\cdot} \boldsymbol{\hat{g}}|$ spans the full range from 0 to 1, which is the case for  $ 0.83 \le \chi \le 1.20$. It is surprising to observe that the statistical distribution of $|\boldsymbol{\hat{p}} \boldsymbol{\cdot} \boldsymbol{\hat{g}}|$ is almost the same throughout this regime even though the dynamics are very different for $\chi \neq 1$. In particular, the non-spherical tumbling particles show a much more dynamic rotational behaviour. A stronger dependence on $\chi$ is seen in the statistics of $\langle | \boldsymbol{\hat{p}} \boldsymbol{\cdot} \boldsymbol{\hat{v}} |\rangle_n$ in the tumbling regime, where a more gradual transition between oblate and prolate is evident.

For the special case of the sphere one expects random orientation for the pointing vector direction. This corresponds to $\langle |\boldsymbol{\hat{p}} \boldsymbol{\cdot} \boldsymbol{\hat{g}}|\rangle_n = 0.5$ or $\theta_{\hat{g}} = 60^\circ$. Our result slightly deviates from this value, namely we observe $\langle |\boldsymbol{\hat{p}} \boldsymbol{\cdot} \boldsymbol{\hat{g}}|\rangle_n = 0.55$. The difference is however within one standard deviation of the statistical convergence of the data,  as determined from considering random pointing vector orientations.

The mean alignment of $\boldsymbol{\hat{p}}$ with gravity is very strong and consistent throughout both non-tumbling oblate regimes  ($\chi \le 0.75$). In the zigzag (blue) regime, we observe a gradual decrease in $\langle\theta_{\hat{g}}\rangle_n$ with decreasing $\chi$. At the same time, also the distribution of $\theta_{\hat{g}}$ becomes narrower, indicating lower fluctuation levels, which is most likely associated with an increase in (added) rotational inertia. For the same range of aspect ratios, $\langle \theta_{\hat{v}} \rangle_n$ is seen to  steadily increase with decreasing $\chi$. The main reason for this trend is the increased phase delay between the oscillations of $\boldsymbol{\hat{p}}$ and $\boldsymbol{\hat{v}}$ as we will discuss in more detail in \S \ref{sec:phase}.\\
In the flutter (purple) regime, $\langle\theta_{\hat{g}}\rangle_n$ rises only slightly compared  to the zigzag regime. However, $\langle\theta_{\hat{v}}\rangle_n$,  increases considerably below $\chi =0.29$ which clearly distinguishes the two regimes. The angle $\langle\theta_{\hat{v}}\rangle_n$ reaches a mean value as high $60^\circ$ at $\chi = 0.20$. This implies that  the particle is predominantly moving `sideways' in the direction of its smallest cross section (consistent with the statistics of $A_{\perp v}$ reported in the inset of figure \ref{fig:drag_coefficient}). Note also that the particles never get close to $\theta_{\hat{v}} \approx  0$ in the flutter regime.

On the prolate side, the different regimes are distinctly noticeable in both the behaviour of $\langle\theta_{\hat{g}}\rangle_n$ and of $\langle\theta_{\hat{v}}\rangle_n$. In the longitudinal (yellow) regime, we observe large oscillations in the orientation of $\boldsymbol{\hat{p}}$ relative to $\boldsymbol{\hat{g}}$ as well as to $\boldsymbol{\hat{v}}$. The amplitude of this pointing vector oscillation decreases with increasing $\chi$, as can be seen from the angle $\langle\theta_{\hat{g}}\rangle_n$ which increases and the spread in the data which decreases, drawing closer to the reference orientation of $90^\circ$. When transitioning to the broadside (orange) regime, the averages $\langle\theta_{\hat{g}}\rangle_n$ and $\langle\theta_{\hat{v}}\rangle_n$ increase, while the fluctuations decrease considerably, such that the maximum deviation of the pointing vector from the horizontal becomes less than $10^\circ$.

At $\chi = 5$, data is shown for the rise patterns corresponding to the longitudinal (yellow) state. In this state the inclination $\langle \boldsymbol{\hat{p}} \boldsymbol{\cdot} \boldsymbol{\hat{g}} \rangle_n$ is identical to that in the yellow regime, however the value of $\langle\boldsymbol{ \hat{p}} \boldsymbol{\cdot} \boldsymbol{\hat{v}} \rangle_n$ is significantly higher (i.e. $\langle \theta_{\hat{v}} \rangle_n$ is lower). This implies that the particle is more aligned with the  $\boldsymbol{\hat{v}}$-direction which will be shown to play an important role in triggering the transition to the helical mode in section \S \ref{sec:prolate_helix}.
The particle alignment is also visualized in the videos in the supplementary materials, showing the coupling between orientation and the particle motion.

\subsection{Particle reorientation}\label{sec:phase}
\begin{figure}
	\centerline{\includegraphics[width=1\textwidth]{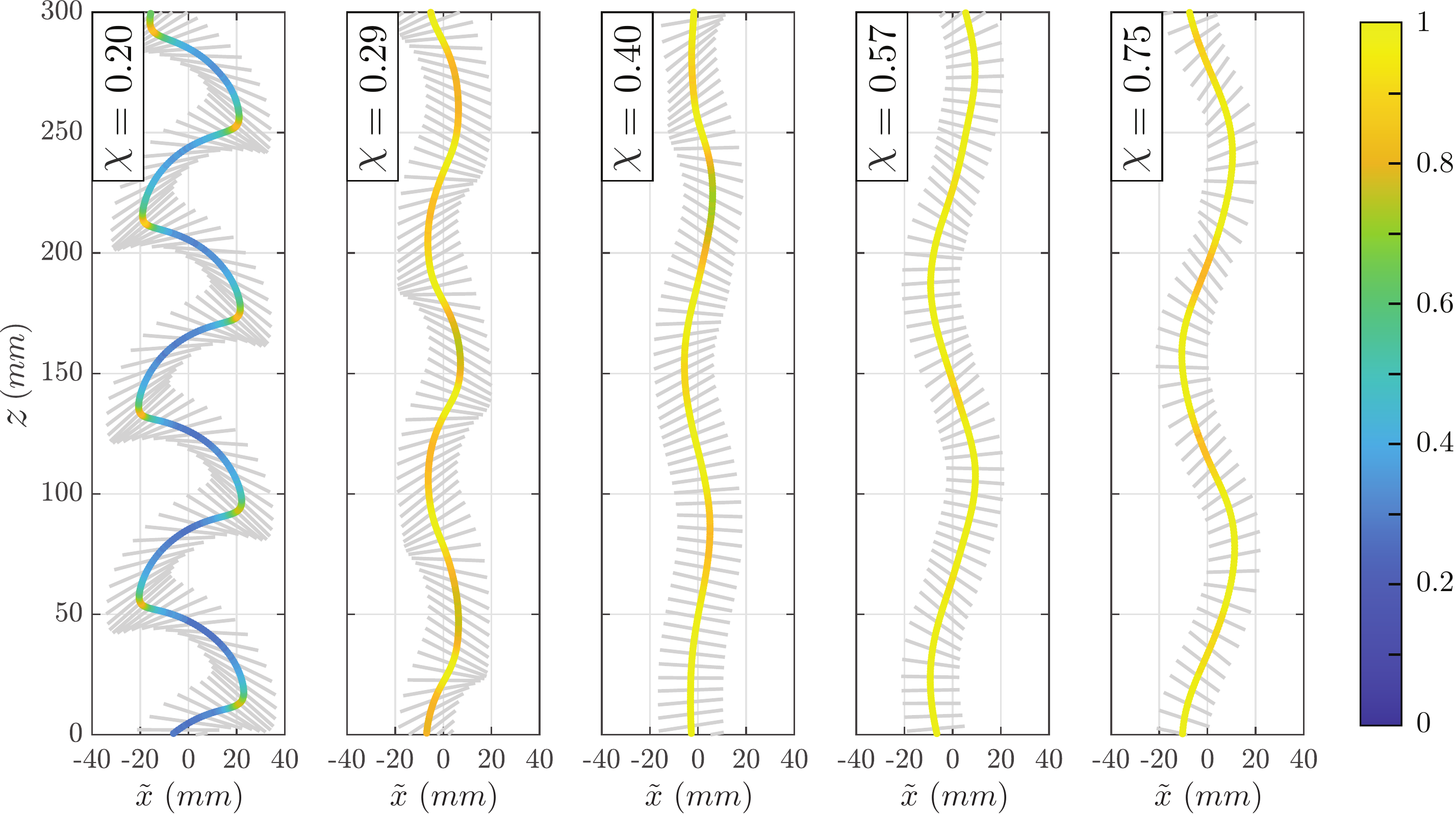}}
	\caption{ Representative precession corrected trajectories. Trajectories are color-coded with the instantaneous alignment $|\boldsymbol{\hat{p}} \boldsymbol{\cdot} \boldsymbol{\hat{v}}|$. The grey lines are shown every 5 frames (0.02 seconds) and indicate the major axis of the projected geometry, which is always perpendicular $\boldsymbol{ \hat{p} }$.} 
	\label{fig:trajectories_alignment_side}
\end{figure}
Within the Newton drag regime, where pressure induced drag and lift are due to a deflection of the flow, it is immediately obvious that an inclination of the particle's symmetry axis will lead to a lift force. The statistics of particle alignment provide some insight into this facet but lack details on the temporal sequence. The latter is important since it is the temporal dependence that leads, for example, to the net contribution of the added mass force observed in \S \ref{sec:AddedMass}. To complete the analysis, we will therefore examine the phase difference $\Delta \phi$ between the periodic motion of the $\boldsymbol{\hat{p}}$ and the $\boldsymbol{\hat{v}}$-directions in this section, which is defined such that a positive value of $\Delta\phi$ corresponds to a lagging pointing vector.

To illustrate this, we show examples of trajectories and alignments for five different oblate particles in figure \ref{fig:trajectories_alignment_side}. Let us first consider $\chi = 0.75$. In this case the rotation of the particle and the direction of motion are in phase. This means that when the particle's horizontal velocity is maximal the pointing vector is also at a maximum deflection from its reference orientation. And when the velocity is minimal the pointing vector direction is also close to the vertical. Note that this does not necessarily mean that $|\boldsymbol{\hat{p}} \boldsymbol{\cdot} \boldsymbol{\hat{v}}|$ remain close to 1 at all times. This can be seen close to the zero-crossings, where the horizontal particle position $\tilde{x} = 0$, here the particle orientation ``overshoots'' resulting in a slight dip in $|\boldsymbol{\hat{p}} \boldsymbol{\cdot} \boldsymbol{\hat{v}}|$. This overshooting is related to the tumbling motion, which is described in \S \ref{sec:tumble}. With increasing anisotropy, the orientation starts lagging behind the direction of motion more and more (see figure \ref{fig:trajectories_alignment_side}: $\chi \le 0.29$ ), which corresponds to a positive phase lag $\Delta \phi$. This causes stronger fluctuations and a general decrease in the alignment between particle orientation and path velocity as $\chi$ decreases in figure \ref{fig:trajectories_alignment_side}.

To extract quantitative information about the phase delay, we have to employ slightly different approaches for oblate and prolate particles to account for the different pointing vector alignments. In the oblate case, we consider the cross-correlation of the periodic signals  $\boldsymbol{\hat{p} }_{\tilde{x}}$  and $\boldsymbol{\hat{v}}_{\tilde{x}}$, i.e. the horizontal components of the pointing and of the velocity vectors in the precession corrected frame of reference. For prolate particles, where the pointing vector is oscillating around the horizontal plane, we use a cross-correlation between $\hat{v}_\parallel$ (as defined in figure \ref{fig:lagrangian_frame_explained}\,({\it b\/})) and  $-\boldsymbol{\hat{p}}_z$. Note that here we focus exclusively on the mode associated with $\hat{v}_\parallel$, since only this one occurs at the same frequency as the pointing vector oscillations (see \S \ref{sec:prolate_ellipsoids} and in particular figure \ref{fig:lagrangian_frame_explained}\,({\it c\/})). 

\begin{figure}
	\centerline{\includegraphics[width=1\textwidth]{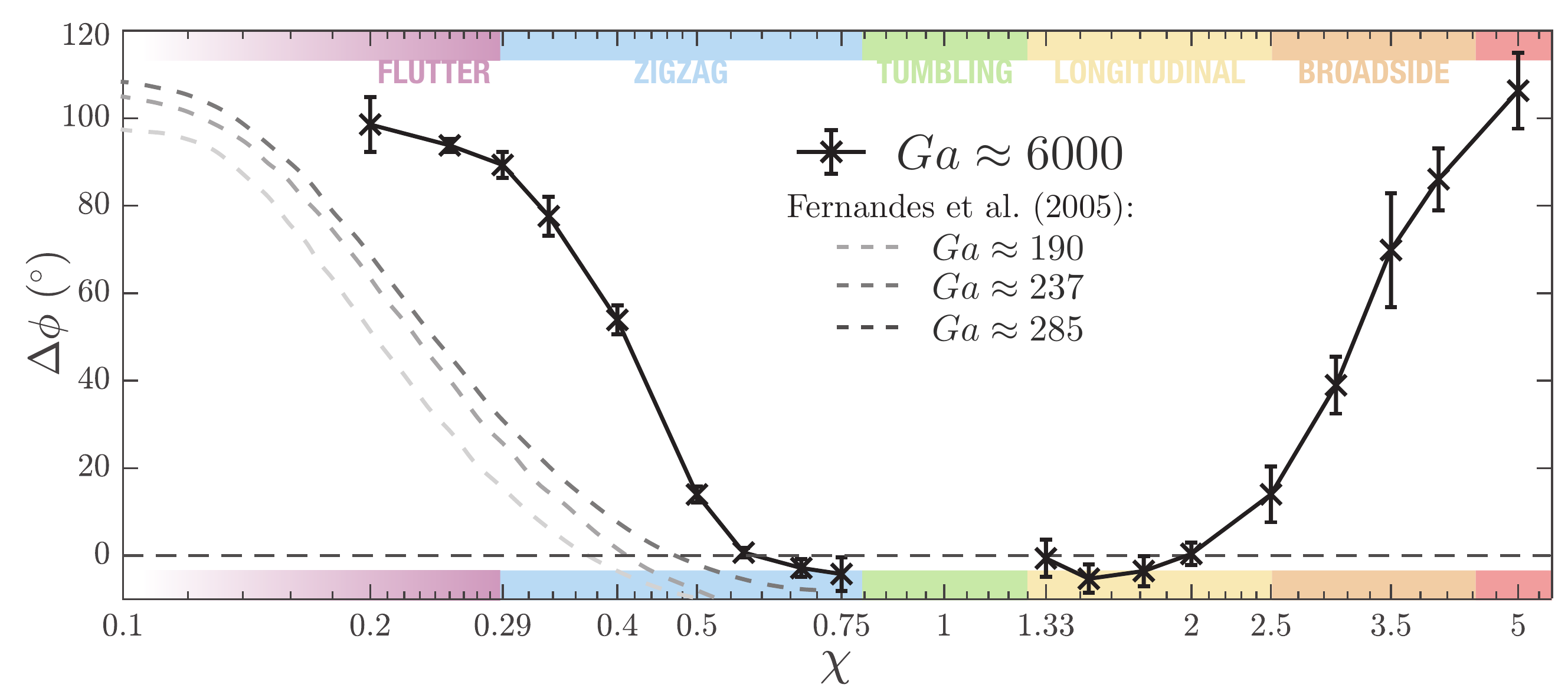}}
	\caption{ Phase angle between velocity and pointing vector oscillations at different aspect ratios. The results for the current data set ($Ga \approx 6000$) are represented by the black symbols, while the grey dashed lines are fits from the work by \cite{Fernandes:2005} for disks at lower $Ga$.  }
	\label{fig:phase_lag}
\end{figure}

Based on the peak of the cross-correlations defined in this way, we determine the phase lag. The corresponding results are shown in figure \ref{fig:phase_lag} in terms of the phase angle $\Delta\phi$. In this figure, we observe an interesting similarity between oblate and prolate geometries. Close to spherical, the phase lag is essentially zero for both cases. Towards more extreme anisotropy however, $\boldsymbol{\hat{p}}$ increasingly lags behind $\boldsymbol{\hat{v}}$. For oblate particles we observe that the regime change from zigzag (blue) to flutter (purple) coincides with a levelling-off of the phase lag at around $\Delta\phi \approx 100^\circ $. The flattening out of the $\Delta\phi$-curve also coincides with the onset of a strong precession at $\chi$ = 0.2. On the prolate side, the regime transition from longitudinal (yellow) to broadside (orange) oscillations is marked by the onset of a phase lag. The transition to the helical regime occurs at a value of $\Delta\phi$ similar to the one at which oblate particles level off. Yet, in the prolate case, $\Delta\phi$ appears to keep increasing towards more extreme anisotropy, which might be one of the factors that promote the transition to the helical pattern as described in \S \ref{sec:prolate_helix}.

It is further noteworthy that for both prolate and oblate geometries the intermediate phase lags, where the behaviour transitions from 0$^\circ$ lag to 100$^\circ$, are associated with minima in the amplitude of the particle oscillations (see Appendix \ref{sec:App_horizontal_motion}). For prolate particles oscillations of the pointing vector also almost vanish in this regime. It remains unclear, however, to what extent these trends are connected.

The observed trend in the phase lag with $\chi$ for oblate particles is in line with results by \cite{Fernandes:2005} for oblate disks at low Galileo number. Their fitted results are also shown in figure \ref{fig:phase_lag}. We observe that for increasing $Ga$ the curve for $\Delta\phi$ shifts towards smaller particle anisotropy. Intuitively this makes sense: for larger $Ga$ the pressure forcing on the particle surface becomes larger in magnitude. Neglecting changes in the rotational inertia, thus, the time-scale of the particle response becomes shorter with increasing $Ga$, causing the observed $Ga$ trend in figure \ref{fig:phase_lag}.

In \S \ref{sec:ReynoldsNumber} it was noted that the drag coefficient in our ``zigzag'' (blue) regime collapsed to the data from \cite{fernandes2007}. It appears from figure \ref{fig:phase_lag} that the experiments performed in the cited paper all would fall in the blue regime, thus potentially explaining the collapse of $C_d$ for their experiments, even though their particle anisotropy is much larger, going as low as $\chi = 0.1$. 

\section{Tumbling and helical rise -- a closer look}
In this last section, we will have a closer look at the two most outstanding regimes: tumbling particles with shapes close to spherical (\S \ref{sec:tumble}) and the helical pattern observed for the most prolate spheroids (\S \ref{sec:prolate_helix}).
\subsection{Tumbling regime: Froude number dependence}\label{sec:tumble}
\begin{figure}
	\centerline{\includegraphics[width=0.6\textwidth]{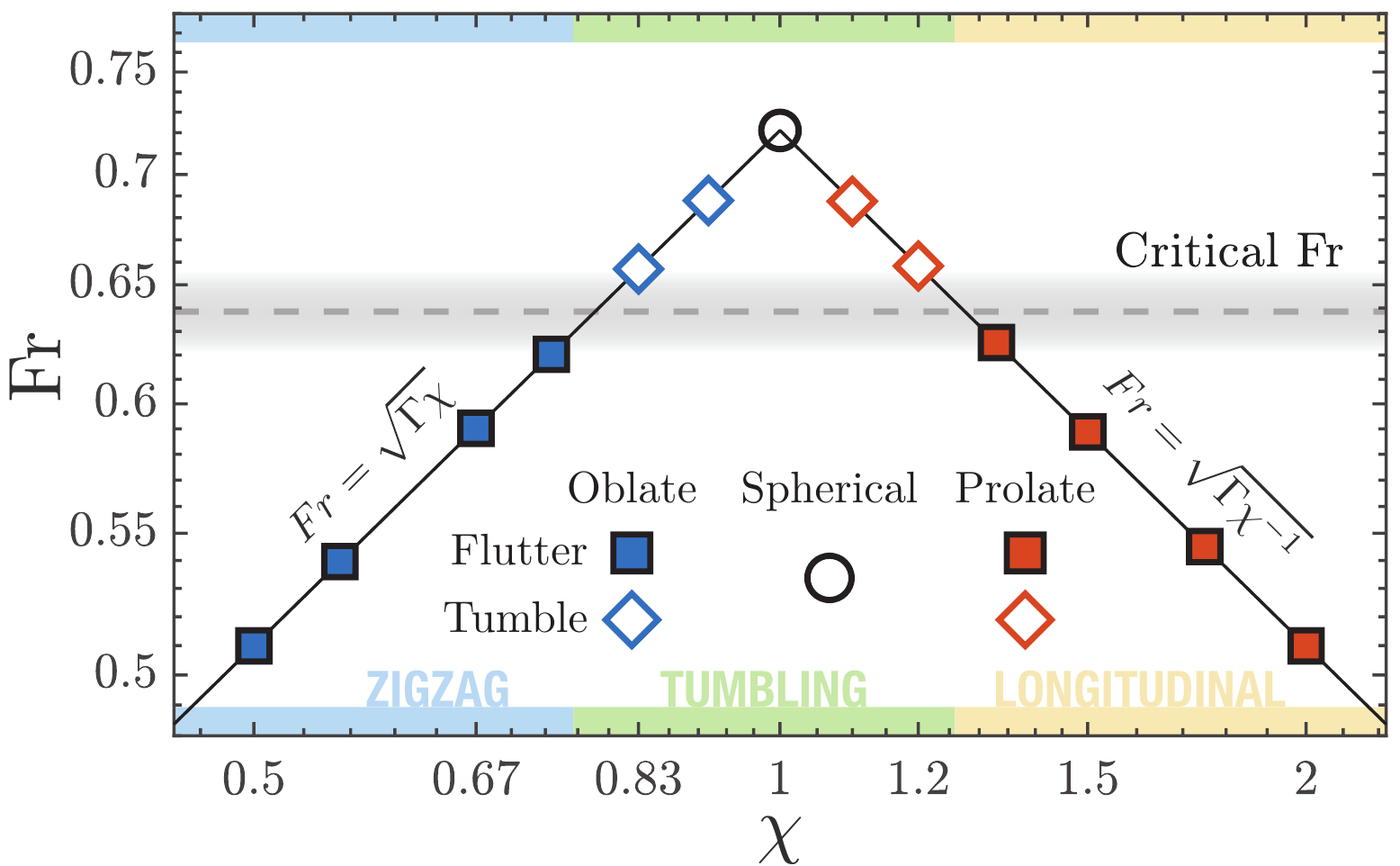}}
	\caption{Particle Froude number as calculated using the scaling approach by \cite{Belmonte:1998} for particles close to the transition to tumbling. The symbols indicate the observed behaviour and particle type  and the dashed horizontal line shows the estimated critical $Fr$ value above which tumbling occurs. The $\chi$ dependencies of $Fr$ indicated by the black lines follow from  \cite{Belmonte:1998} (see text for details). Note also the log-scaling on the $\chi$-axis.}
	\label{fig:Froude_Chi}
\end{figure}
In the current data, there is a gradual transition from a fluttering to a tumbling state. This regime change has been extensively studied mostly for thin bodies, where the particle length scale in the flow direction is much smaller than the one perpendicular to it. It is common \citep[see e.g.][]{Dupleich1949,Willmarth:1964,Smith1971,Lugt1983} to characterize the phase-space and transition between regimes in terms of a dimensionless moment of inertia $\tilde{I} = I/I_{ref}$ where $I_{ref} $ is the moment of inertia of the cross-section of a reference geometry, often chosen to be a cylinder. This approach works for particles that are quasi two-dimensional, i.e. whose cross section does not vary along the axis around which they tumble. The flutter-to-tumble transition received additional attention more recently in experiments by \cite{Belmonte:1998}, \cite{Zhong2011} and \cite{Wang2013}, in numerical work \citep{Andersen:2005a,Lau:2018} or through a combination of both in \cite{Andersen:2005b}. Nevertheless, in all these works, the influence of the geometry, especially of a finite and varying thickness in the flow direction, has not been considered systematically yet. Therefore, the current experiments can provide additional insight by applying this analysis to particles with a significant thickness in the direction of the flow.

The approach using a dimensionless moment of inertia is not well suited for our particles, since there is no universal logical choice for $I_{ref} $ due to the variations of the cross section along the axis of rotation. Therefore, we follow the approach by \citet{Belmonte:1998}, and define a dimensionless Froude number 
\begin{equation}
Fr \equiv \dfrac{\tau_p}{\tau_r} = \sqrt{\dfrac{\Gamma d^*}{L}},\label{eq:Froude}
\end{equation}
as the ratio of two intrinsic particle time-scales $\tau_p$, which is the time-scale for a buoyant pendulum, and $\tau_r$, the settling time-scale. These are defined by
\begin{equation}
\tau_p = \sqrt{\dfrac{ L \Gamma}{(1-\Gamma)g}}\ \qquad\text{and} \qquad\tau_r = \dfrac{L}{v_0}. \label{eq:tau_r}
\end{equation}
Here, $v_0$ is a velocity scale based on the balance between particle drag force and buoyancy given by $v_0 \sim \sqrt{(1-\Gamma) g d^*}$ as derived from (\ref{eq:CdReGa}) and $d^*$ is the particle length scale in the direction parallel to the flow, i.e. $d^*$ = $h$ for $\chi <$ 1 and $d^*$ = $d$ for $\chi \geq$ 1. In (\ref{eq:Froude}, \ref{eq:tau_r}) we additionally introduce the length scale $L$, which is defined as $L = d$ for $\chi \leq$ 1 and $L = h$ for $\chi >$ 1 and characterises the largest length scale perpendicular to the mean flow direction. With these definitions (\ref{eq:Froude}) simplifies to  $Fr = \sqrt{\Gamma \chi}$ for oblate particles and $Fr = \sqrt{\Gamma \chi^{-1}}$ for prolate ones.

The Froude number as function of aspect ratio is shown in figure \ref{fig:Froude_Chi}\,({\it a\/}) for the tumbling regime as well as for the two neighbouring ones. In this figure, we mark particles that tumble with open and those that do not with filled symbols. We find that at least sporadic tumbling is observed for all particles with $Fr \gtrapprox 0.64$, whereas particles with lower values of $Fr$ never tumble. Hence, the critical $Fr$ appears to be the same regardless of whether the shape is  oblate or prolate. Moreover, the critical value of $Fr$ is also similar to that obtained by \citet{Belmonte:1998}, where the flutter-to-tumble transition for quasi-2D strips was observed to occur at $Fr = 0.67 \pm0.05$. This is somewhat surprising and may be coincidental  since the mechanism of flipping is actually different for strips and spheroids, as we will discuss later in this section.

\begin{figure}
	\centerline{\includegraphics[width=1\textwidth]{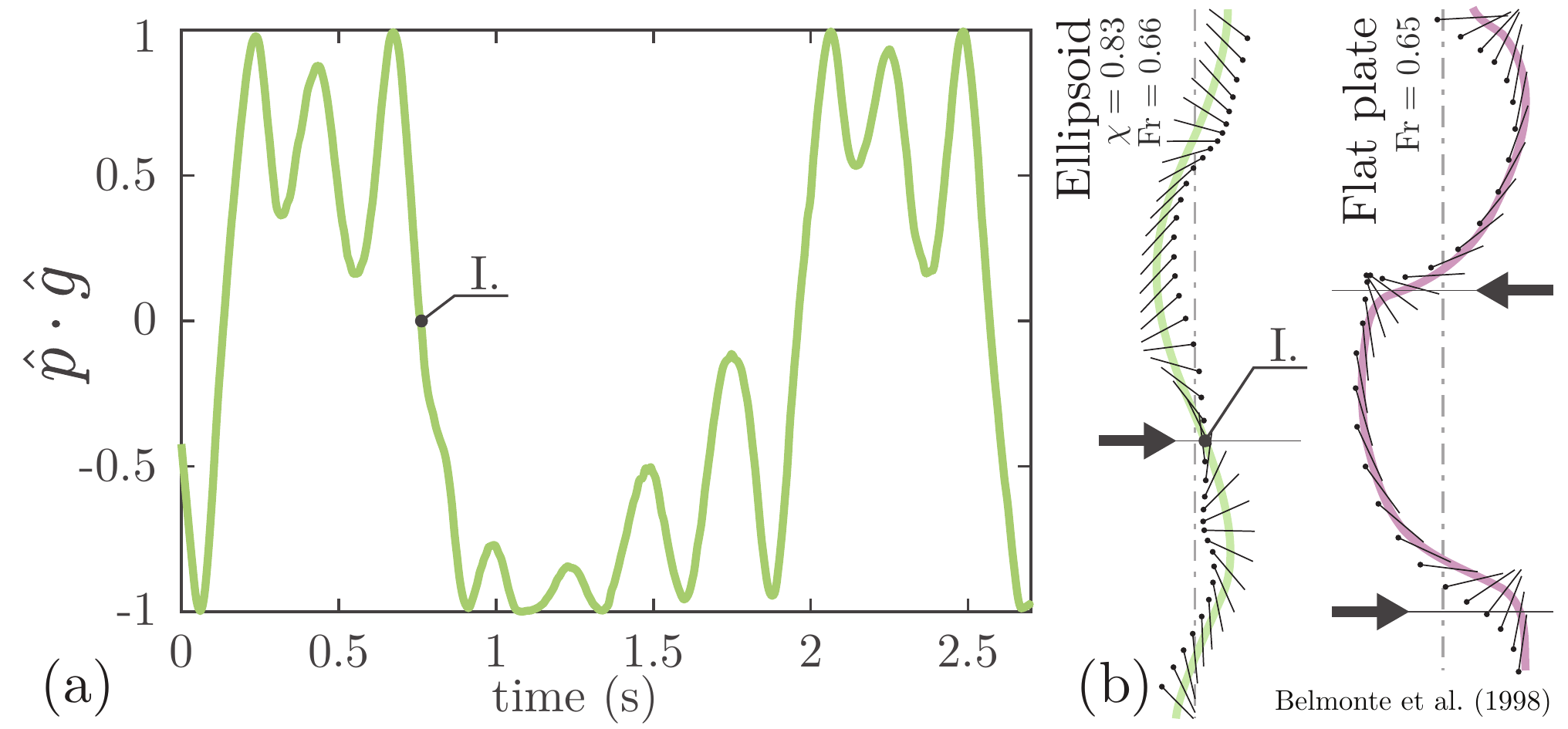}}
	\caption{({\it a\/}) Particle alignment $\boldsymbol{\hat{p}}  \boldsymbol{\cdot} \boldsymbol{\hat{g}}$ over time for a single experiment for $\chi$  = 0.83. ({\it b\/})  Schematic representation of the tumbling mechanisms for spheroids and flat plates. On the left a section of one of the trajectories for $\chi$ = 0.83 is shown and on the right the rise pattern of a strip  \citep[taken from][]{Belmonte:1998} is displayed very close to the tumbling transition (but not actually tumbling). The orientation of the particles largest cross section is indicated by the black lines. The points of the trajectory at which the particle (would) flip are indicated by the large arrows and the label  \rom{1} marks the same instances in ({\it a\/}) and ({\it b\/}).}
	\label{fig:tumbling behavior}
\end{figure}

To analyse the transition to tumbling in more detail, we consider the temporal evolution of the particle inclination in terms of $\boldsymbol{\hat{p}}  \boldsymbol{\cdot} \boldsymbol{\hat{g}}$ for which an example is shown in figure \ref{fig:tumbling behavior}\,({\it a\/}). From this it becomes clear that the tumbling of the particle, indicated by the change in sign of $\boldsymbol{\hat{p}}  \boldsymbol{\cdot} \boldsymbol{\hat{g}}$, does not occur for every single oscillation, which is unlike the behaviour of flat quasi-2D strips observed in \citet{Belmonte:1998} at similar $Re$ ($3\cdot 10^3 < Re < 4 \cdot 10^4$). A similar intermittent fluttering and tumbling behaviour was encountered in the ``apparently chaotic'' regime reported by \citet{Andersen:2005b,Andersen:2005a}. For the present data, a gradual increase in amplitude in the particle inclination, $\boldsymbol{\hat{p}} \boldsymbol{\cdot} \boldsymbol{\hat{g}}$, leading up to flipping events can be observed in figure 15\,({\it a\/}). It appears that the particle has to build up enough rotational momentum before it can eventually flip over. Only once the amplitude is large enough, after a couple of oscillations, a flip happens and $\boldsymbol{\hat{p}} \boldsymbol{\cdot} \boldsymbol{\hat{g}}$ changes sign. The direction of the flip (clockwise or counterclockwise) appeared to be random and both appeared equally probable.

However, differences between the tumbling motion of 2D strips and 3D spheroids are not restricted to the frequency of the tumble. Also the way in which the spheroids tumble differs significantly from the mechanism that is observed for thin bodies, which also renders the  tumbling trajectories fundamentally different. For thin plates, the tumbling motion is associated with oblique trajectories that exhibit a  very strong mean drift \citep[]{Belmonte:1998,pesavento2004,Andersen:2005a}, whereas we find that for spheroids the trajectory remains vertical. The two distinct mechanisms of tumbling are exemplified by the two particles trajectories in figure \ref{fig:tumbling behavior}\,({\it b\/}); the  left  one corresponds to a spheroid in the present work and the one on the right shows the trajectory of a thin plate close to tumbling as reported in \cite{Belmonte:1998}. The key difference is the point of the trajectory at which the flip happens. For spheroids, the flip occurs when the path oscillation is at a minimum, whereas for flat plates the tumbling occurs at the maximum path amplitude. The difference is related to the phase difference $\Delta\phi$. For a flat plate $\Delta\phi$ is close to $90^\circ$, thus the pointing vector is horizontal at points of maximum amplitude, whereas $\Delta\phi = 0$ in our case results in a flip at the centre. The difference is most likely related to the mechanism that causes the particle to rotate. For flat plates, the particle rotates primarily due to a torque induced by pressure forces. However, such torque is minimal for spheroids close to spherical ($\chi \sim 1$). In this case, the main mechanism is therefore the asymmetry in the skin friction caused by vortex shedding. Additionally, rotational added mass plays an important role for flat plates while its effect is negligible for particles close to spherical.

As mentioned, this results in markedly distinct trajectories from those of flat plates. The oscillation frequencies and amplitudes observed for tumbling spheroids are very close to those of the perfect sphere. However, the trajectories differ between tumbling spheroids and the sphere, as can be seen in \ref{fig:top_trajectories}\,({\it h--l\/}). The former show more sudden changes in direction which can be related to the strong rotations and changing alignment for these particles. The rise-velocity and drag are also clearly affected by this.

\subsection{The odd one out: The helical regime}\label{sec:prolate_helix}
\begin{figure}
	\centerline{\includegraphics[width=1\textwidth]{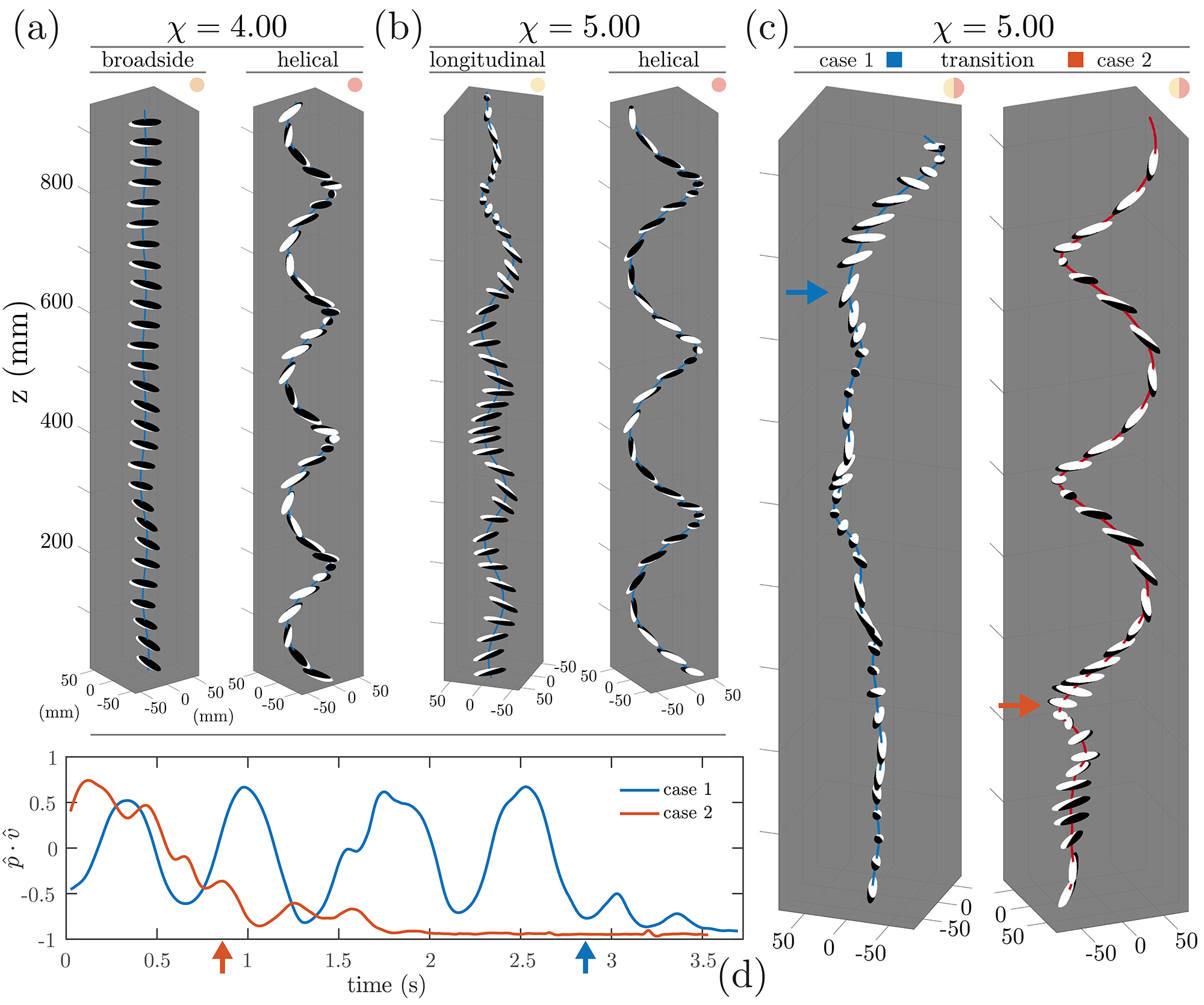}}
	\caption{Particle trajectories obtained from the experimental measurements with superimposed renders of the particles at their instantaneous orientation plotted every 0.1 seconds. ({\it a\/}) The two coexisting rise-patterns for $\chi$ = 4 are shown. These correspond to the broadside (orange) regime on the left where the particle rises almost vertically with minimal oscillations of the pointing vector and the helical (red) regime on the right. ({\it b\/}) The two regimes for $\chi$ = 5: On the left is the longitudinal (yellow) regime showing large amplitude oscillations in the pointing vector oscillations, on the right its helical (red) regime state. ({\it c\/}) Two cases for $\chi$ = 5 where a transition from the longitudinal regime to the helical one is observed (marked by arrows). ({\it d\/}) For the two transition cases in ({\it c\/}) we show the alignment $\boldsymbol{\hat{p}} \boldsymbol{\cdot} \boldsymbol{\hat{v}}$ as a function of time, clearly illustrating the transition as well as the distinct behaviour in the two regimes. The coloured arrows correspond to the marked points in ({\it c\/}).}
	\label{fig:spiral_transition}
\end{figure}
All cases discussed so far have in common that the pointing vector performs an oscillatory motion around the reference orientation for which the largest area of the geometry is perpendicular to gravity. Remarkably, however, we found that for $\chi \geq 4$ also a distinctly different rise pattern is possible. In what we refer to as the ``helical'' rise pattern,  it was found that the particles enter an orbit in which both angles $\theta_{\hat{g}}$ and $\theta_{\hat{v}}$ remain approximately constant in time. In the following we provide details on the helical orbit and on how the particles at $\chi = 4$ and  $\chi = 5$ transition to this pattern. All properties of this regime are tabulated along with the other results from this paper in table \ref{tab:rise_properties} of Appendix \ref{sec:App_properties}. 

The difference between the non-helical (figure \ref{fig:top_trajectories}\,({\it q,\,s\/})) and helical patterns (figure \ref{fig:top_trajectories}\,({\it r,\,t\/})) is stunningly clear from the particle trajectories alone. With $\langle \eta \rangle_n = 0.99$ for both $\chi = 4$ and $\chi = 5$,  the eccentricity (as defined in \S \ref{sec:ecc}) of the helical trajectories is very close that of a perfect circle for which $\eta = 1$.  Furthermore, the helical path is almost perfectly periodic, which is in complete contrast to all other observed prolate modes that generally display very chaotic and complex motion.

The 3D representation in figure \ref{fig:spiral_transition}\,({\it a--c\/}) shows all the different rise modes possible for the particles with $\chi \geq 4$ in the form of a render. For $\chi = 4$ (figure \ref{fig:spiral_transition}\,({\it a\/})) the broadside (orange) and helical (red) regimes coexist. In this case, the regime appears to be determined exclusively by the particles initial release since in all experiments performed we have never observed a regime transition during the rise. In case the particle is released with minimal disturbance and $\boldsymbol{\hat{p} } \boldsymbol{\cdot} \boldsymbol{\hat{g}}$ is close to 0, the particle tends to go into the broadside regime. However, if the release is less controlled or the particle is inclined at release, it was found to gain significant velocity along the direction of the pointing vector and enter the helical pattern. 

The dependence on the release conditions is similar at  $\chi = 5$. However, when not in the helical mode the particle at this aspect ratio displays a pattern that is reminiscent of the longitudinal regime rather than the broadside one with large oscillations of the pointing vector inclination (see figure\ref{fig:spiral_transition}\,({\it b\/})). Contrary to the case at $\chi = 4$,  we identified transitions from the longitudinal to the helical regime in two experiments at $\chi = 5$ (but never the reverse transition). The transitional trajectories are shown in  figure \ref{fig:spiral_transition}\,({\it c\/}) with the approximate point of transition indicated by the arrows.  

The gradual transition is also visible from  the corresponding plot of $\boldsymbol{\hat{p}} \boldsymbol{\cdot} \boldsymbol{\hat{v}}$ in figure \ref{fig:spiral_transition}\,({\it d\/}). Here, we initially see the large changes in particle inclination characterizing the longitudinal regime.  As time advances these fluctuations slowly decrease and the particle inclination subsequently remains largely constant at a value close to $|\boldsymbol{\hat{p}} \boldsymbol{\cdot} \boldsymbol{\hat{v}}| = 1$. This reduction in angular fluctuations is accompanied by a gradual growth of the particle horizontal velocity $v_{\parallel}$ (not shown), which may help to  stabilize the particle at a non-zero inclination.

It should also be remembered that the fact that $\boldsymbol{\hat{p}}$ comes close to aligning with $\boldsymbol{\hat{v}}$ is related to the increase of the phase lag $\Delta \phi$ with increasing $\chi$. Unlike for oblate particles, $\Delta \phi$ was seen to increase beyond $\Delta \phi \approx 100^\circ$ for prolate spheroids (see \S\ref{sec:phase}). Based on the above observations, we suspect that the longitudinal regime is not stable at $\chi = 5$ and that all particles may eventually  transition to the helical regime at this aspect ratio. Due to the small oscillations of the pointing vector alignment in the helical regime it is deemed unlikely that the particle will transition back to the longitudinal one without outside forcing.

\begin{figure}
	\centerline{\includegraphics[width=1\textwidth]{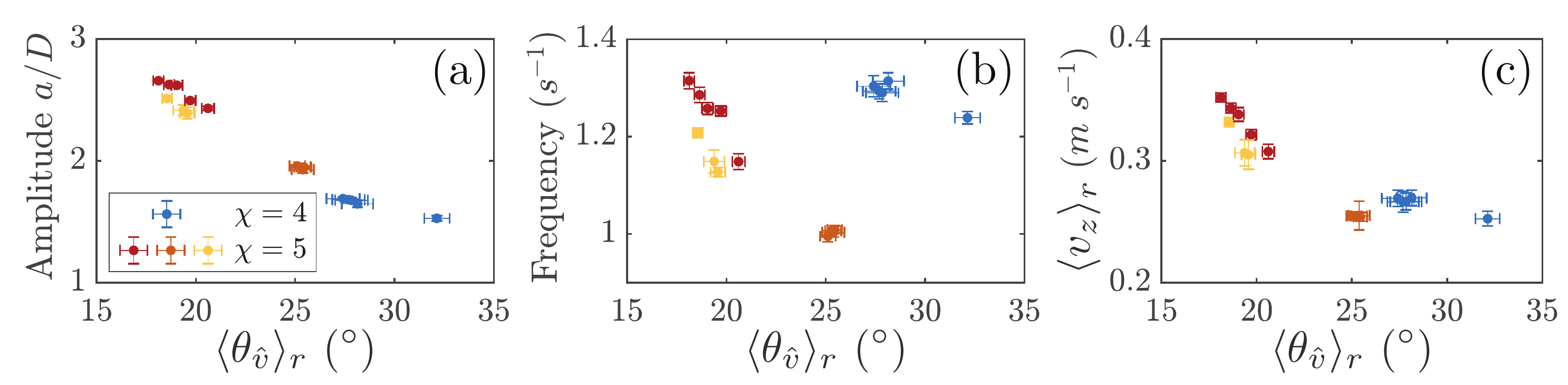}}
	\caption{ Properties of the helical trajectories for $\chi = 4$ and  $\chi = 5$ as a function of the angle  $\langle \theta_{\hat{v}}\rangle_r$:  (a) Amplitude $\langle a/D \rangle_r$, (b) oscillation frequency $\langle f\rangle_r$, and (c)  rise velocity $\langle v_z \rangle_r$. All averages are per run and error bars correspond to $\pm$ one standard deviation. Three different particles were used at $\chi=5$ and data corresponding to the same particle are marked by  different shades of red.}
	\label{fig:spiral_properties}
\end{figure}

To characterize the helical regime in more detail, we plot the amplitude ($\langle a/D \rangle_r$), the oscillation frequency ($\langle f\rangle_r$), and the rise velocity (($\langle v_z \rangle_r$) as a function of $\langle \theta_{\hat{v}}\rangle_r$ in figure \ref{fig:spiral_properties} for both $\chi = 4$ and $\chi = 5$. Note that the averaging here, $\langle \cdot \rangle_r$, is performed over individual runs to highlight run-to-run variations. Also included in the plot are the standard deviations around this mean for all quantities in the form of error bars. Furthermore, for some runs a rotation around $\boldsymbol{\hat{p}}$ was observed, while in other cases it was absent. Somewhat surprisingly, such rotations appeared to have no significant effect on the particle motion.

The first thing to note about $\langle a/D \rangle_t$ in figure \ref{fig:spiral_properties}\,({\it a\/})  is that the amplitude of all helical trajectories is much larger than for any other regime where $\langle a/D \rangle_n < 1$ consistently (see figure \ref{fig:Amp_and_Ecc}\,({\it a\/})). Moreover, the standard deviations for $\langle a/D \rangle_r$ but also for $\langle \theta_{\hat{v}}\rangle_r$ are minute (average of the per run standard deviation: 0.0210 $a/D$ for $\chi$ = 4 and 0.0195 $a/D$ for $\chi$ = 5), demonstrating how consistent the trajectories are. Another interesting observation is that there is quite some spread in the data between runs, and that the amplitude appears to be inversely correlated to to the angle $\langle \theta_{\hat{v}}\rangle_r$. This trend is consistent for both aspect ratios considered.

Also the frequency (see figure \ref{fig:spiral_properties}\,({\it b\/})) varies very little over the course of a single run. However, we do note a spread in the data between subsequent runs. The frequencies of the helical patterns are substantially lower than for all other regimes for which the average frequency is 1.77 Hz. Also $\langle f\rangle_r$ is seen to decrease with increasing $\langle \theta_{\hat{v}}\rangle_r$. However,  there is an offset between the data at $\chi = 4$ and $\chi = 5$.

Finally, we show the mean rise-velocity in figure \ref{fig:spiral_properties}\,({\it c\/}). Similar to the amplitude and frequency, we observe nearly constant velocities within individual runs with small standard deviations.  Although, again there is substantial variation between the runs that is inversely correlated with $\langle \theta_{\hat{v}}\rangle_r$. The mean rise-velocity of the particles for $\chi = 4$ is lower in the helical regime (0.265 m s$^{-1}$) compared to the broadside velocity (0.334 m s$^{-1}$). For $\chi = 5$, on the other hand, the mean vertical velocity is higher in the helical regime (0.297 m s$^{-1}$) than the longitudinal rise-velocity (0.255 m s$^{-1}$).

The observation that the particle inclination is so constant suggests that for this regime there is no periodic vortex shedding which would cause variations in rise-velocity or particle orientation. We would therefore expect a continuous wake behind the particle with minimal discrete vortex shedding events similar to that for slender ``hydrodynamic'' objects and it will be interesting to study this in the future.

To rule out that the helical pattern was caused by some imbalance  of the particle caused by inaccuracies in the  manufacturing, we tested three different particles at  $\chi = 5$. All three of them entered similar helical patterns upon release. 
Results corresponding to each individual particle are marked in figure \ref{fig:spiral_properties}, as indicated by the different shades of red, and reveal that the parameters of the helical orbit do vary somewhat per particle. However, also  for $\chi = 4$, where only a single particle was used in this regime, there are still significant differences between the runs. It appears therefore that the inclination angle is $\theta_{\hat{v}}$ is very sensitive to the particle geometry and mass distribution, but potentially also to the release conditions. The sensitivity to particle properties is interesting since it opens up possibilities to design for a specific desired  behaviour. It should be noted that in the non-helical rise pattern the dynamics and kinematics of the different particles at $\chi = 5$ were indistinguishable from one-another. 

\section{Conclusion and discussion}\label{sec:conclusion}
The main outcome of this study is a classification of the rich behaviour of freely rising spheroidal particles across a wide range of particle aspect ratios, at high Galileo number $Ga \approx 6000$. 

The spheroidal geometry allows for a gradual transition from spherical to (slightly) anisotropic particles. Our data reveals, that even for small levels of anisotropy in the shape have a profound impact on the particle dynamics. This is reflected in significantly reduced particle rise velocities and substantially increased rotational dynamics once $\chi$ deviates from 1. The latter can cause the particles to flip over, resulting in a tumbling motion for $0.83 \leq \chi \leq 1.20$. Interestingly, the bounds of the tumbling regime here are governed by the same time-scales ratio, captured by the Froude number $Fr$, as the ``flutter-to-tumble'' transition for flat plates. However, the precise nature of the tumbling motion differs in several aspects between spheroids and flat plates. For spheroids (\emph{i}) the tumbling does not result in oblique trajectories, (\emph{ii}) the torque inducing rotation is primarily due to skin friction and not pressure induced and (\emph{iii}) the tumble occurs at a different time during the oscillation cycle.

For non-tumbling oblate spheroids ($\chi \le 0.75$), period-to-period variations in the oscillation cycle are lower. In particular, the erratic tumbling motion gives way to regular oscillations of the pointing and the velocity vectors with a phase shift $\Delta \phi(\chi)$. This more regular motion is due to the fact that there is only a single transverse length scale, namely the diameter $d$ of the circular cross-section exposed to the flow. We found that for the ``zigzag'' regime (blue), ranging from $0.29 \lessapprox \chi \lessapprox 0.75$, the average cross-flow area is almost exactly the maximum cross-sectional area of the particle. This results in a constant drag coefficient within this regime of $C_d \approx 1.2$ ($v_z \sim \sqrt{h}$), identical to that found for disks at $Ga = 100-300$ by \citet{fernandes2007}, but 
unlike these authors we do not observe a constant velocity along the path.

At the most extreme oblate aspect ratios considered here  ($0.2 \leq \chi \lessapprox 0.29$), the direction of particle motion is no longer primarily perpendicular to the largest cross-section of the geometry. Instead, the particles perform periodic leave-like oscillations that lead to significant horizontal movement but also  significant fluctuations in the vertical velocity. 
As a consequence, $C_d$ is seen to increase in this ``flutter'' regime. Remarkably, the added mass force is found to significantly contribute to the vertical particle motion,even when terminal velocity has been reached. This effect amounts to more than $0.6||\boldsymbol{F}_B||$ for $\chi = 0.2$. In contrast, we found that along the path the added mass force does not result in a net drag contribution for any $\chi$ here.

We further observe a trend from planar  (at $\chi = 0.25$) to strongly precessing oscillations ($\chi = 0.2$)  within the flutter regime, which is very similar to findings by \cite{Zhong2011} for thin flat disks (from  ``planar-zigzag'' to ``transitional'' in their wording). Our transition occurs around the same $Re$ and for a similar dimensionless moment of inertia as in their data when plotted in the phase-space according to \cite{Field1997}. This analogy invites some speculation regarding a potential ``spiral'' regime for even ore oblate particles \citep[see][figure 5]{Zhong2011}. The existence of such a regime would give rise to a rather pleasing symmetry with helical trajectories for both, extremely prolate and oblate spheroids.

Generally, the trajectories for prolate particles appear more chaotic compared to their oblate counterparts. We were able to relate this to the fact that the mean cross-flow area of the geometry is no longer circular (as is the case for oblate particles) but ellipsoidal. This results in two transverse length scales that grow increasingly more distinct with increasing anisotropy. Our analysis revealed that these two length scales give rise to two different modes of oscillation, one in the plane containing the pointing vector and one perpendicular to it. Those oscillations occur at distinctively different frequencies and their interaction gives rise to the very complex and seemingly chaotic trajectories.

For $ 1.2 \lessapprox \chi \lessapprox 2.5$, we define the ``longitudinal'' regime which is characterised by large amplitude pointing vector oscillations. These oscillations disappear almost completely for $2.5 \lessapprox \chi \lessapprox 4.5$ , we refer to this range as the ``broadside'' regime. Both of these regime have a near constant drag coefficient of $C_d = 0.86$ and  $C_d =0.67$ respectively, i.e. $v_z \sim \sqrt{d}$. The fact that 
the drag actually decreases for increasing anisotropy is a remarkable result that is unique for the ``broadside'' regime. This gives rise to a  secondary local maximum in the particle rise velocity at $\chi \neq 1$.

Arguably, the most surprising observation was made for the most extreme prolate aspect ratio, $\chi = 5$. For this case we found that two completely different rise patterns coexists: one is an oscillation similar to the ``longitudinal'' regime but with a larger phase difference  $\Delta \phi \approx 110$, the other an almost perfectly helical trajectory. The helical rise-pattern is triggered by an alignment between the velocity direction and the pointing vector and this transition occurs naturally at $\chi =5$.  The same helical pattern can also be triggered for $\chi = 4$,  but it has to be forced via a large inclination at the particle release in this case.  The helical pattern truly stands out among all other trajectories observed since its the only case where the particle inclination is fixed and not oscillating around the orientation of maximum cross-flow area.

It is tempting to draw a parallel between the present observations and the helical trajectories reported for bubbles \citep{lindt1972,clift1978}. \citet{mougin2006} showed numerically that  deformed bubbles transition from planar type oscillations to a helical path and this was confirmed by \citet{shew2006} in a simplified model based on the Kelvin-Kirchhoff equations. Obviously there are differences compared to the present work. This concerns not only the boundary conditions at the particle/bubble surface, but also the alignment. For uncontaminated rising bubbles the major axis is aligned with the radial direction, i.e. normal to the path \citep[ch.$\sim$7]{veldhuis2007thesis}, whereas in our case the pointing vector is directed tangentially to the spiraling motion.

On the oblate side, it is possible to get insight into the $Ga$-variation of the regime boundaries by comparing the present results to those of \citet{Fernandes:2005,fernandes2007}. Across a wide range of different properties (Reynolds number, amplitude, eccentricity, phase difference), it appears consistently that transitions appear at lower levels of anisotropy (i.e. larger $\chi$ for oblate particles), the higher $Ga$. This is also in agreement with the trend over the (small) range of $Ga$ considered in \citet{Fernandes:2005} for the phase lag.

We find that results for $C_d$, can be divided into three approximately constant values: $C_d\approx 1.22$ for the ``zigzag'' regime, $C_d \approx 0.83$ for the ``longitudinal'' and ``tumbling'' regimes and $C_d \approx 0.67$ for the ``broadside'' regime. These values seem related to the relevant particle dimension setting the size of the vortex shedding system. For the ``zigzag'' regime, this is the major axis $d$ (oblate), which results in a large vortex system and thus the highest drag coefficient. For the ``broadside'' motions, the vortex shedding is associated with only the minor axis $d$ (prolate) resulting in the lowest drag. Finally, for the intermediate ``tumbling'' and ``longitudinal'' regimes, we observe oscillations associated with both particle dimensions $d$ and $h$ ( major and minor axes). Correspondingly, the resulting drag is intermediate between the values obtained for the other cases.

Another important aspect of the current work is that it sheds light on the effect of particle (rotational) inertia. In the present case, $\Gamma \approx 0.53$ and thus also the relative (rotational) inertia is smaller. This implies that the particle reacts more swiftly and stronger to the fluid forcing. We observe these effects indirectly by comparing our results to those of settling particles. From this we find that the drag coefficient of particles with similar geometry is much lower for heavy particles compared to the current data set (with the notable exception of the ``broadside'' regime, keep in mind that the settling data is compiled across a wide range of geometries. Therefore this does not necessarily imply that prolate particles rise faster than they settle).

Furthermore, we observe that the particle frequency of oscillation is largely independent of $\chi$. This is contrary to findings by \citet{fernandes2007} who identified a dependence on the cross flow length scale $d$ at lower $Ga$. This difference is most likely related to the lower density ratio ($\Gamma$) and rotational moment of inertia ($I^*$) in our case. Further research will be required to shed light on how the oscillation frequency depends on these parameters.

As a final remark, we would like to point out that  no significant rotations around the particle symmetry axis were observed for any aspect ratio in this work.

\section*{Acknowledgments}
 This work was supported by Natural Science Foundation of China under grant nos 11988102, 91852202, and the Netherlands Organisation for Scientific Research (NWO) under VIDI Grant No. 13477, STW, FOM, ERC, and MCEC.

\section*{Declaration of Interests}
The authors report no conflict of interest.

\appendix
\section{Tabulated: particle and rise properties}\label{sec:App_properties}
In this appendix we provide the information presented in this work in tabulated form for the readers convenience. In this work results have been presented in terms of $\chi$, results for a specific $\chi$ are composed of multiple experiments with multiple particles. Therefore, the input parameters ($\chi$,$\Gamma$ and $Ga$) have some spread in them and also a certain uncertainty is associated with the results. To clarify this we document in table \ref{tab:particle_properties} the statistics of the measured properties per $\chi$. The averaging $\langle \cdot \langle_\chi$ is introduced, this is the average over all particles with a specific $\chi$. Similarly $\sigma_\chi$ is the standard deviation. At the bottom of the table we provide the statistics over all used particles (denoted as $\langle \cdot \rangle_{all}$).

In addition to this, we also provide in table \ref{tab:rise_properties} the compiled results that are presented throughout this paper in one convenient table with all value as a function of $\chi$. The results for the helical rise-pattern from \S\ref{sec:prolate_helix} are marked in red in the table for $\chi$ = 4 and 5.

\begin{table}
\begin{center}
\begin{tabular}{x{1.4cm}|x{1.4cm}x{1.5cm}x{1.4cm}x{1.4cm}x{1.4cm}x{1.4cm}x{1.4cm}}
$\chi$ & $\langle \chi' \rangle_\chi$ & $\langle D \rangle _\chi [mm]$ & $\langle d \rangle _\chi [mm]$ & $\langle h \rangle _\chi [mm]$ & $\langle \Gamma \rangle_\chi$ & $\langle Ga \rangle_\chi$ & $\sigma_\chi(Ga)$ \\
\hline \hline
0.200 & 0.210 & 20.229 & 34.025 & 7.150      & 0.513         & 6263          & 46 \\
\rowcolor{Grey}
0.250 & 0.256 & 19.965 & 31.442 & 8.050       & 0.514         & 6135          & 76 \\
0.286 & 0.293 & 20.011 & 30.113 & 8.838       & 0.516         & 6143          & 34 \\
\rowcolor{Grey}
0.333 & 0.344 & 20.132 & 28.733 & 9.883       & 0.529         & 6120          & 24 \\
0.400 & 0.414 & 19.993 & 26.833 & 11.100       & 0.511         & 6171          & 93 \\
\rowcolor{Grey}
0.500 & 0.502 & 19.832 & 24.956 & 12.525       & 0.541         & 5904          & 43 \\
0.571 & 0.568 & 19.767 & 23.875 & 13.550       & 0.531         & 5943          & 6  \\
\rowcolor{Grey}
0.667 & 0.667 & 19.947 & 22.825 & 15.233       & 0.548         & 5909          & 70 \\
0.750 & 0.753 & 19.807 & 21.767 & 16.400       & 0.544         & 5875          & 23 \\
\rowcolor{Grey}
0.833 & 0.840 & 19.895 & 21.083 & 17.717       & 0.552         & 5864          & 41 \\
0.909 & 0.913 & 19.815 & 20.425 & 18.650       & 0.540         & 5901          & 75 \\
\rowcolor{Grey}
1.000 & 1.000 & 19.972 & 19.972 & 19.972       & 0.539         & 5982          & 97 \\
1.100 & 1.092 & 19.872 & 19.300 & 21.067       & 0.540         & 5928          & 44 \\
\rowcolor{Grey}
1.200 & 1.195 & 19.886 & 18.738 & 22.400       & 0.548         & 5884          & 36 \\
1.333 & 1.326 & 19.885 & 18.100 & 24.000     & 0.544         & 5910          & 5  \\
\rowcolor{Grey}
1.500 & 1.503 & 19.796 & 17.281 & 25.975       & 0.519         & 6027          & 34 \\
1.750 & 1.741 & 19.868 & 16.517 & 28.750       & 0.543         & 5911          & 39 \\
\rowcolor{Grey}
2.000 & 2.003 & 19.823 & 15.725 & 31.500       & 0.540         & 5909          & 60 \\
2.500 & 2.474 & 19.883 & 14.700 & 36.375       & 0.534         & 5976          & 1  \\
\rowcolor{Grey}
3.000 & 2.987 & 19.911 & 13.825 & 41.300       & 0.541         & 5937          & 6  \\
3.500 & 3.495 & 19.816 & 13.058 & 45.633       & 0.531         & 5960          & 116\\
\rowcolor{Grey}
4.000 & 3.997 & 19.887 & 12.531 & 50.088       & 0.545         & 5904          & 51 \\
5.000 & 5.014 & 19.782 & 11.558 & 57.950       & 0.542         & 5878          & 49 \\
\hline
 \multicolumn{2}{c}{$\langle \cdot \rangle_{all} $:}   & 19.903 & & & 0.534     &   5979    &       \\
 \multicolumn{2}{c}{$ \sigma_{all}(\cdot) $:}           & 0.111 & & & 0.015     &   122     &     
\end{tabular}
\caption{Average particle properties per aspect ratio group $\langle \cdot \rangle_\chi$, we provide the measured particle aspect ratio $\chi'$, density ratio, and the Galileo number along with the standard deviation per aspect ratio grouping $\sigma_\chi(Ga)$. The bottom two rows show the average and the standard deviation over all particles, i.e. not grouped by aspect ratio.} \label{tab:particle_properties}
\end{center}
\end{table}

\begin{table}
\begin{center}
\begin{tabular}{x{0.8cm}|x{0.9cm}x{0.7cm}x{0.9cm}x{0.9cm}x{0.9cm}x{0.6cm}x{1.1cm}x{1.0cm}x{1.1cm}x{1.0cm}}
$\langle \chi' \rangle_\chi$ & $\langle Re \rangle_n$ & $C_d$ & $\langle C_d^* \rangle_n$ & $\langle C_l^* \rangle_n$ & $f$[Hz] & $St$  & $\langle \theta_{\hat{g}} \rangle_n [^\circ]$ & $\Delta \phi  [^\circ]$ & $\langle a \rangle_n /D$ & $\langle \eta \rangle_n$ \\
\hline \hline
0.210                    & 4963 & 2.08 & 1.27 & 3.02 & 1.83                    & .120 & 30.49                                         & 98.30                   & 0.959 & 0.095                    \\
\rowcolor{Grey}
0.256                    & 5939 & 1.40 & 1.34 &1.55 & 1.99                   & .131 & 28.79                                         & 93.50                   & 0.494 & 0.093                    \\
0.293                    & 6309 & 1.28 & 1.29 & 0.99 & 1.93                   & .125 & 24.91                                         & 88.21                   & 0.327 & 0.170                    \\
\rowcolor{Grey}
0.344                    & 6208 & 1.26 & 1.30 & 0.81 & 1.69                   & .113 & 25.22                                         & 78.10                   & 0.283 & 0.194                    \\
0.414                    & 6373 & 1.26 & 1.30 & 0.70 & 1.75                    & .112 & 25.26                                         & 53.64                   & 0.250 & 0.247                    \\
\rowcolor{Grey}
0.502                    & 6165 & 1.26 & 1.24 & 0.73 & 1.66                   & .109 & 29.05                                         & 13.97                   & 0.300 & 0.467                    \\
0.568                    & 6262 & 1.24 & 1.17 & 0.65& 1.59                   & .103 & 28.36                                         & 0.46                    & 0.464 & 0.261                    \\
\rowcolor{Grey}
0.667                    & 6356 & 1.16 & 1.05 & 0.72 & 1.66                   & .111 & 30.66                                         & -2.47                   & 0.495 & 0.424                    \\
0.753                    & 6662 & 1.06 & 0.95 & 0.72& 1.86                   & .123 & 31.67                                         & -3.84                   & 0.529 & 0.441                    \\
\rowcolor{Grey}
0.840                    & 7442 & 0.84 & 0.79 & 1.03 & 1.83                   & .122 & 52.09                                         & --                       & 0.666 & 0.227                    \\
0.913                    & 7697 & 0.79 & 0.70 & 0.86 & 1.66                   & .110 & 55.05                                         & --                       & 0.776 & 0.291                    \\
\rowcolor{Grey}
1.000                    & 8924 & 0.59 & 0.53 & 0.37 & 1.70                   & .115 & 56.81                                         & --                       & 0.684 & 0.540                    \\
1.092                    & 7831 & 0.78 & 0.69 & 0.74 & 1.71                   & .113 & 54.97                                         & --                       & 0.752 & 0.290                    \\
\rowcolor{Grey}
1.195                    & 7637 & 0.81 & 0.75 & 0.86 & 1.83                   & .122 & 54.71                                         & --                       & 0.622 & 0.282                    \\
1.326                    & 7523 & 0.83 & 0.74 & 0.53 & 1.74                   & .116 & 74.17                                         & -0.64                   & 0.742 & 0.389                    \\
\rowcolor{Grey}
1.503                    & 7501 & 0.89 & 0.78 & 0.45 & 1.85                   & .118 & 78.31                                         & -6.32                   & 0.655 & 0.287                    \\
1.741                    & 7512 & 0.84 & 0.77 & 0.39 & 1.72                   & .114 & 78.32                                         & -3.92                   & 0.553 & 0.318                    \\
\rowcolor{Grey}
2.003                    & 7369 & 0.87 & 0.80 & 0.42 & 1.79                   & .118 & 79.15                                         & 0.12                    & 0.629 & 0.228                    \\
2.474                    & 7935 & 0.77 & 0.75 & 0.27& 1.75                   & .114 & 81.07                                         & 14.15                   & 0.434 & 0.281                    \\
\rowcolor{Grey}
2.987                    & 8576 & 0.65 & 0.64 & 0.11 & 1.84                   & .121 & 86.83                                         & 40.03                   & 0.181 & 0.430                    \\
3.495                    & 8343 & 0.68 & 0.67 & 0.12 & 1.83                   & .121 & 87.51                                         & 61.28                   & 0.135 & 0.276                    \\
\rowcolor{Grey}
3.997                    & 8385 & 0.68 & 0.67 & 0.12 & 1.81                   & .119 & 86.86                                         & 83.69                   & 0.161 & 0.418                    \\
5.014                    & 6653 & 1.07 & 0.96 & 0.67 & 1.63                   & .110 & 78.45                                         & 106.61                  & 0.694 & 0.511                    \\
\rowcolor{Grey}
\hline
\rowcolor{LightRed}
3.997                    & 6645 & 1.07 & 0.70 & 1.20 & 1.29                      & .087 & 69.02                                         & --                       & 1.632 & 0.990                    \\
\rowcolor{LightRed}
5.014                   & 7752 & 0.80 & 0.54  & 1.09 & 1.14                   & .075 & 67.29                                         & --                       & 2.274 & 0.992
\end{tabular}
\caption{A tabulated version of the results presented throughout this work for different properties as a function of the aspect ratio. The two red highlighted lines at the bottom show the results for the helical rise rise-pattern for the particles of aspect ratios $\chi =$ 4 and 5.} \label{tab:rise_properties}
\end{center}
\end{table}

\begin{table}
\begin{center}
\begin{tabular}{x{1.2cm}||x{0.9cm}x{0.9cm}x{0.9cm}x{0.9cm}x{0.9cm}x{0.9cm}x{0.9cm}x{0.9cm}x{0.9cm}}
$\chi$        & 1.33 & 1.50 & 1.75 & 2.00 & 2.50 & 3.00 & 3.50 & 4.00 & 5.00 \\
\hline
$f_\perp\ [\text{Hz}]$     & 1.86 & 1.99 & 1.77 & 1.90 & 1.99 & 2.20 & 2.09 & 2.17 & 2.23 \\
$f_\parallel\ [\text{Hz}]$ & 1.48 & 1.37 & 1.36 & 1.26 & 1.26 & 1.50 & 1.45 & 1.62 & 1.42
\end{tabular}
\caption{Tabulated results for the decomposed frequency oscillations of all non-tumbling prolate particles in the non-helical rising state.} \label{tab:frequency_decomposition}
\end{center}
\end{table}

\section{Particle translational and rotational tracking}\label{sec:App_tracking}
In this appendix we provide a complete overview of the methods used to track the particle motion and orientation.

\subsection{Translational tracking}\label{sec:trajectories}
Firstly, we obtained the position of the particle's center of mass in the lab coordinate frame. spheroids, i.e. ellipsoids of revolution, are orthotropic and spherically isotropic, which implies that they have three perpendicular symmetry planes and, further, a point symmetry about their geometric centres. Therefore, tracking the centre of mass is identical to tracking the centre of an arbitrary two-dimensional projection of the particle shape. Here we neglect the effect of perspective since the cameras are far enough away to almost get a perfect isometric projection of the geometry.

Background subtraction is applied to the recordings and the images are processed using thresholds to obtain the white and black parts of the geometry. The union of the two in dilated and eroded to determine the particle's geometric center in the image with sub-pixel accuracy. The black and white part are stored separately as well as the particle outline which will be used later for determining the orientation.

Next, we use our calibration data to determine the particle position in real space. Because the particles do not always rise exactly in the center of the tunnel, a correction based on the two camera angles was applied to correct for the parallax of the camera angles. The mean value of the two cameras was used to determine the vertical position of the particle. The maximum error was $\pm$0.5 mm or around a single pixel value. This is also approximately the noise observed in the unprocessed signals.

In order to verify if the terminal velocity has been reached the time-averaged vertical acceleration was considered. We note that this quantity does not necessarily average out for individual runs because a non-integer number of periods may be observed in the recording range. However, the mean across all runs was found to be very close to zero for all aspect ratios as expected after the start-up transient. Thus, we are confident that the particles have reached their terminal rise velocity and hence a statistically stationary state.

\subsection{Rotation tracking}
Secondly, in addition to tracking particle translation, the instantaneous orientation of the body was also obtained. The method used for this purpose is similar to the one pioneered by \cite{Zimmermann:2011} and subsequently improved by \cite{Mathai:2016} for rotational tracking of spheres. Both these techniques rely matching the video recording of the particle with a (computer generated) representation of the particle for which the orientation is known. Some modifications to the method used by \cite{Mathai:2016} are necessary since the latter relies on the projection being a circle and the pattern having a concise mathematical description both of which are inapplicable to the current set of particles. To extend the capabilities of the method to particles of arbitrary shape and pattern, the code developed here employs the \textit{stereolithography} (\textit{.stl}) file used for 3D-printing to also render the particles. These rendered images were then be compared to the recordings and the particle orientation follows from minimizing the difference between the two.

\subsubsection{Processing of the video recordings}\label{sec:App_video_images}
For the orientation tracking, we started from the video recordings. As a first step, the particle image was interpolated onto a $N\times N$ frame, such that the centre of the particle projection (obtained previously from the translational tracking) coincides with the centre of the frame. The size $N$ of the frame was chosen according to the maximum dimension of the particle in pixels in the video recordings. This ensures that the complete particle image lies within the $N\times N$ frame and the rest of the recording can be discarded. The particle position is sub-pixel accurate and a linear interpolation was used to determine the new grey-scale value of each pixel. Then, a threshold was applied to identify pixels as either background, or white and black portions of the particle and their greyvalues were set to 128, 255 or 0, respectively. The comparison between the rendered and the actual particle image required a fixed size of the particle image. Therefore, as a final step in the pre-processing, the particle image was enlarged (via a bicubic interpolation) such that it fits exactly within the $N\times N$ frame with a margin on the closest side of 2 pixels.

These processing steps were applied to the images from both cameras. We will designate the resulting particle images by the array $\mathbb{V}_i$. where the index $i$ is the camera direction, either along the $x$- or the $y$-axis.

\subsubsection{Generating the computer rendered images}
To generate the rendered image the 3D model of the particle was rotated to the desired orientation. The particle orientation with respect to the reference orientation, shown in figure \ref{fig:painted}\,({\it a,\,b\/}), was defined using three Euler-angles, $\phi$, $\theta$ and $\psi$. These angles relate to sequential rotations around the $X$, the $Z'$, and the $X''$ axes, respectively. Here, the upper case font denotes the particle coordinate system and the $'$ indicates the rotated frame of reference due to the previous rotations. We note that the particle coordinate system $X,Y, Z$, as shown in figure \ref{fig:painted}\,({\it a,\,b\/}), is chosen such that for both oblate and prolate spheroids the final rotation with angle $\psi$ is around the axis of symmetry of the particle.

Apart from the particle orientation in space, also the angle at which the camera views the particle will affect the pattern observed in the recording. The orientation of the 3D model was corrected accordingly based on the position information obtained in the translational tracking.

Finally, the rotated model was rasterised onto a 2D image with  a $N \times N$ resolution taking into account the same margins we imposed for the video image. Thus, the rendered images, $\mathbb{R}_i$, were obtained for both viewing directions.

\subsubsection{Minimization algorithm}
The determination of the particle orientation was achieved by varying the orientation such that the difference between the rendered and the actual particle images becomes minimal. In order to quantify the mismatch between the two, we define the error as
\begin{equation}
E_i = \sum_{n = 1}^{N} \sum_{m = 1}^{N} |\mathbb{R}_i(n,m)^2 - \mathbb{V}_i(n,m)^2|. \label{eq:error_images}
\end{equation}
Note that the difference of the squares of $\mathbb{R}_i$ and $\mathbb{V}_i$ was used to put more emphasis on the difference between the pattern on the particle itself compared to the background. Doing so was found to improve the orientation detection slightly. This error value was calculated for both of the camera angles, thus we obtain the metric, $E$, used for the optimization algorithm 
\begin{equation}
E = c_xE_x + c_yE_y, \label{eq:error_total}
\end{equation}
here $c_i$ is a scaling factor based on the number of pixels that the particle contour contains in the non-enlarged video frame  $\mathbb{V}_i$. The minimization problem to be solved is then to adjust the input Euler angles, $\phi$, $\theta$ and $\psi$, such that error $E$ becomes minimal. Identical to \cite{Mathai:2016}, this was achieved iteratively using the Nelder-Mead, gradient descent, algorithm. For the first frame of a particle track, a range of pre-rendered images covering a wide range across all angles was used to determine an initial guess. Subsequently, the angle of the previous frame was used as the starting point for the minimization routine.

\subsubsection{Validation and accuracy of the orientation tracking}
\begin{figure}
	\centerline{\includegraphics[width=1\textwidth]{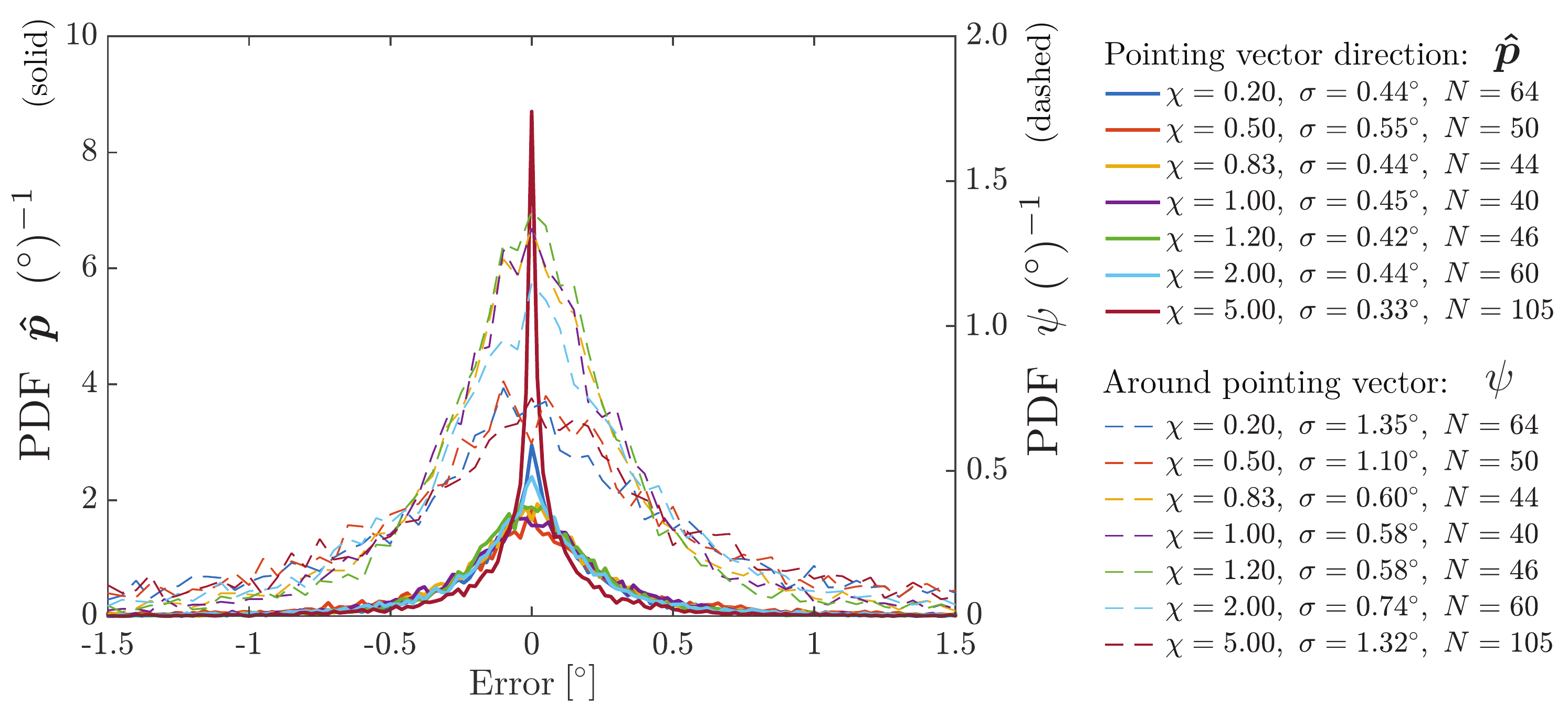}}
	\caption{Probability density functions showing the accuracy of the orientation tracking algorithm. The error plotted on the horizontal axis is the error between the rendered orientation of the particle and the detected orientation by the algorithm. The solid lines show the accuracy in the detection of the orientation of the pointing vector(left $y$-axis). The dashed line the tracking of the rotation around this vector (right $y$-axis). Here, $\sigma$ gives the standard deviation of the distribution. }
	\label{fig:rot_accuracy}
\end{figure}
The algorithm for the orientation tracking of anisotropic particles was validated using computer generated images at known orientations. The 3D models of the particles at various aspect ratios were rendered and rasterised at the appropriate resolution ($N\times N$ minus two times the margin of 2 pixels) to simulate video recordings. These artificial particle images were then processed in the same manner as outlined in \S\ref{sec:App_video_images} to obtain the particle orientations. Statistics of the resulting errors, as measured by the differences between the actual and the extracted orientation angles, are shown in figure \ref{fig:rot_accuracy}. In this figure the distributions of the errors  for a representative set of  7 out of a total of 23 aspect ratios tested are shown. Errors in detecting the direction of the pointing vector ($\hat{\boldsymbol{p}}$) and those related to the rotation around the pointing vector ($\psi$) are treated separately. This is done because the latter only relies on information regarding the painted pattern while the former additionally relies on the body outline. Therefore, $\hat{\boldsymbol{p}}$ can be more accurately determined, this becomes increasingly more relevant as $\chi$ moves further away from unity for which orientation tracking gets more reliable as the projected shape is more and more affected by the orientation of the particle. Furthermore, we see that the accuracy increases for increasing $N$ which is the number of pixels of the largest dimension of the particle. Finally we see that the angle $\psi$ is more accurate for particles with the complex painted pattern ($0.83 \leq \chi \leq 1.20$, the pattern is shown for the sphere in figure \ref{fig:painted}\,({\it d\/})) as is expected, but the others still perform well. For both types of patterns and all aspect ratios the accuracy will also depend on the particle orientation. Overall the accuracy of the alignment of $\hat{\boldsymbol{p}}$ appears to be accurate to within $1^\circ$ and of the rotation $\psi$ within $2^\circ$ from the actual orientation.

The test with artificial images presents a best case scenario. Unavoidably, the error levels in the actual experiment will be larger, e.g. due to inaccuracies in the patterns, noise in the images, or errors in the calibration. We can estimate the actual orientation errors in our experiments based on the results in the region of overlap between the two camera levels. In this region, the particle is recorded by all four cameras which yields two independent measurements. The raw, unsmoothed, data is compared for all measurements in this region and an error is defined as the mean absolute difference between the two measurements. The error in the pointing vector direction is 2.35$^\circ$ and 2.06$^\circ$ in the angle $\psi$, the rotation around the pointing vector. These errors are partly due to the noise in the raw signal. When smoothing the data  the mean errors reduce to 1.63$^\circ$ and 1.58$^\circ$ respectively, which is hence a more reliable estimate of the error in the orientation tracking. As an additional check, the mean pointing vector for all non-tumbling, particles was considered. The mean orientation of this vector should be exactly vertical for oblate particles and horizontal for prolate particles (except for the helical trajectory). It was found that for all cameras the detected mean angle deviated less than 0.5$^\circ$ from the vertical, thus giving an approximate error of the misalignment of the cameras with respect to the lab coordinate frame.

\section{Added mass tensor}\label{sec:App_added_mass_tensor}
Here we will briefly show how the added mass tensor in the particle coordinate system is calculated from the particle dimensions and the fluid density. These equations are taken from the works of \cite{Lamb:1932} and \cite{Brennen:1982}.  The directions of this tensor are identical to the particle coordinate system defined for oblate and prolate particles in figure \ref{fig:painted}\,({\it a,\,b\/}). First we start by defining the eccentricity $\epsilon$ according to:
\begin{equation}
    \epsilon = \left(1-\dfrac{h^2}{d^2}\right)^{0.5}, \qquad \text{for}\ \chi < 1,
\end{equation}
\begin{equation}
    \epsilon = \left(1-\dfrac{d^2}{h^2}\right)^{0.5}, \qquad \text{for}\ \chi > 1,
\end{equation}
for oblate and prolate spheroids respectively. Using these definitions we calculate the coefficients $\alpha$, $\beta$ and $\gamma$, first for oblate:
\begin{equation}
\left. \begin{array}{ll}  
\alpha = \dfrac{\left(1-\epsilon^2 \right)^{0.5}}{\epsilon^3}\sin^{-1}(\epsilon)- \dfrac{\left(1-\epsilon^2\right)}{\epsilon^2}\\
\beta = \dfrac{2}{\epsilon^2}\left( 1-\left( 1-\epsilon^2 \right) \dfrac{\sin^{-1}(\epsilon)}{\epsilon}\right)\\
\gamma = \alpha
 \end{array}\right\} \qquad \text{for}\ \chi < 1
  \label{eq:App_oblate_AM}
\end{equation}
and secondly for prolate:
\begin{equation}
\left. \begin{array}{ll}  
\alpha = \dfrac{1}{\epsilon^2} - \dfrac{1-\epsilon^2}{2\epsilon^3}\ln \left(\dfrac{1+\epsilon}{1-\epsilon}\right) \\
\beta = \dfrac{1-\epsilon^2}{\epsilon^3}\left( \ln\left( \dfrac{1+\epsilon}{1-\epsilon} \right) -2 \epsilon \right)\\
\gamma = \alpha
 \end{array}\right\} \qquad \text{for}\ \chi > 1.
  \label{eq:App_prolate_AM}
\end{equation}
For both of these oblate and prolate the coefficient $\beta$ is for accelerations along the $Y$-direction of the particle frame, which is in the direction of the particle pointing vector. The coefficients $\alpha$ and $\gamma$ deal with accelerations perpendicular to it and are therefore identical due to the symmetry in the particle geometry.

Using these coefficients we calculate the added mass tensor:
\begin{equation}
    \mathsfbi{A} = \dfrac{1}{6} \pi d^2 h \rho_f \begin{bmatrix}
        \alpha/(2-\alpha)& 0 & 0    \\
        0 & \beta/(2-\beta) & 0     \\
        0 & 0 & \gamma/(2-\gamma)
     \end{bmatrix}.
\end{equation}
The values of the coefficients are checked using the tables provided in the work of \citet{Brennen:1982}. For a sphere, $\chi = 1$, the coefficients are all equal $\alpha = \beta = \gamma = 2/3$, giving us the well known added mass equal to that of half the displaced fluid. Both (\ref{eq:App_oblate_AM}) and (\ref{eq:App_prolate_AM}) converge to this result for $\epsilon$ approaches $0^+$:
\begin{equation}
    \lim_{\epsilon \to 0^+} \alpha(\epsilon) = \lim_{\epsilon \to 0^+} \beta(\epsilon) = \lim_{\epsilon \to 0^+} \gamma(\epsilon) = \dfrac{2}{3}.
\end{equation}

\section{A comparison to settling particles}\label{sec:App_settling}
\begin{figure}
	\centerline{\includegraphics[width=0.9\textwidth]{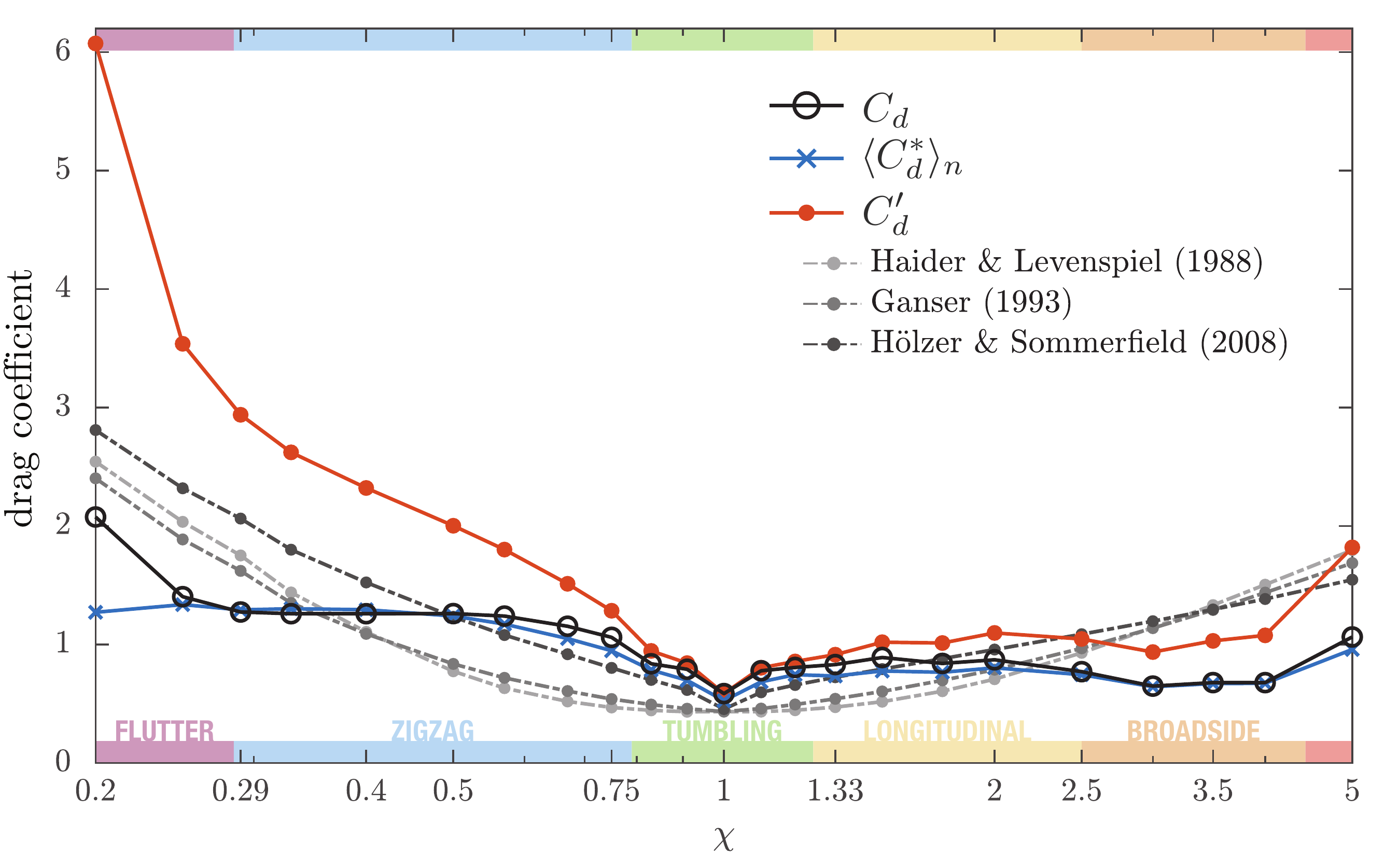}}
	\caption{ Drag coefficient for the current set of experiments using varying definitions. The definition of $C_d'$ (red symbols)  is identical to the ones in the cited studies on heavy particles and can thus be directly compared showing the effects of particle (rotation) inertia through $\Gamma$. }
	\label{fig:App_drag_coefficient}
\end{figure}
The current study focuses on particles with a density ratio significantly smaller than one. Most other studies dealing with anisotropic particles have been performed with particles close to neurally buoyant (at low $Re$) or with heavy particles $\Gamma \gg 1$ (at high $Re$). In this section, we will compare the present results to those reported in the literature for heavy particles in order to highlight differences in the particle behaviour that are related to inertia through $\Gamma$ and $\mathsfbi{I}^*$. To enable this comparison, we revert to a definition of the drag coefficient to the most basic form using the volume equivalent diameter $D$ as the relevant length scale, resulting in:
\begin{equation}
	C_d' = \dfrac{4}{3}\dfrac{g|\Gamma -1|D}{\langle v_z \rangle ^2_n}. \label{eq:CD}
\end{equation}
The results for $C_d'$ are shown as a function of $\chi$ in figure \ref{fig:App_drag_coefficient} as red filled circles. It is immediately obvious that the spread in the data increases drastically when using the definition in (\ref{eq:CD}), with values of $C_d'$ now covering a full order of magnitude ranging from 0.6 to 6.0.

In the literature on heavy particles, there have been numerous attempts to capture the effects of particle geometry in terms of so-called shape factors, such as the  `Corey shape factor' and `sphericity'. We employ three such empirical relations, namely the ones by \cite{Haider:1989}, \cite{Ganser:1993}, and \cite{Holzer:2008}, (based on experimental studies similar to \cite{alger1964} which deserve a lot of credit) for the geometries of our particles and plot the results in figure \ref{fig:App_drag_coefficient}. Where needed, we employ the Reynolds numbers measured here to evaluate the models. It should be noted that these models were not developed for any specific geometry and are based on data sets that contain a large degree of variability in the shape of the particles. Therefore, the predictions based on  these models contain quite a large uncertainty. The comparison with the current data, however, is still interesting and the uncertainty does not affect the  conclusions we draw regarding the fundamental differences between heavy and light anisotropic particles.

Figure \ref{fig:App_drag_coefficient} shows that for all oblate particles in the zigzag (blue) and flutter (purple) regimes the drag of the rising particles is significantly higher compared to their settling counterparts of comparable geometry and $Re$. This difference becomes increasingly pronounced as $\chi$ deviates from 1. Also in the tumbling (green) regime lighter particles experience a larger drag but the difference is less strong. The result for prolate geometries is interesting since we find that for the broadside (orange) regime the light particle drag is in fact lower than that observed for heavy particles at the same $Re$ and with the same shape-parameter. This inversion is very specific for this regime and sets it apart from all the others. Generally, the difference in $C_d'$ between heavy and light particles is smaller for prolate bodies compared to oblate ones.

An important general conclusion that can be drawn from the comparison in this section is that, similar to what has been observed for spheres already, the density ratio plays a very important and complex role in governing the dynamics of anisotropic bodies. This has a large impact on relevant high level parameters such as the settling/rising velocity (drag) and, as we will see in section \S \ref{sec:frequency}, on the oscillation frequency.

\section{Characterisation of the particle trajectory}\label{sec:App_horizontal_motion}
\subsection{Oscillation amplitude}\label{sec:Amplitude}
\begin{figure}
	\centerline{\includegraphics[width=1\textwidth]{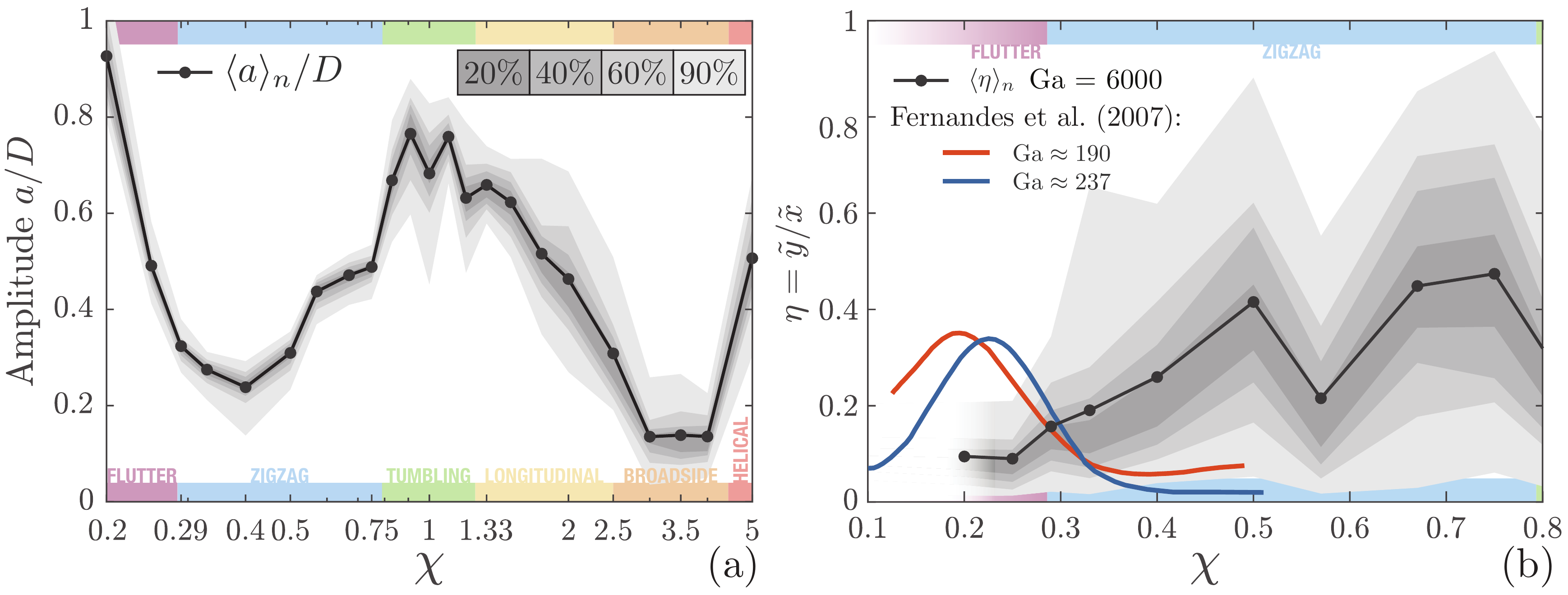}}
	\caption{({\it a\/}) Amplitude of particle path oscillations normalized by the equivalent volumetric diameter as a function of the aspect ratio. The solid symbols show the mean and the grey shaded areas give the spread in the data. ({\it b\/}) Eccentricity of the particle trajectory in the horizontal plane, the markers and the grey areas show the mean and the distribution respectively. Along with the current data we show the results for lower $Ga$ numbers from \cite{fernandes2007}, suggesting a shift towards lower anisotropy for increasing $Ga$.}
	\label{fig:Amp_and_Ecc}
\end{figure}
The mean amplitude of the horizontal periodic motion $\langle a \rangle_n $ and its distribution are shown as a function of $\chi$ in figure \ref{fig:Amp_and_Ecc}\,({\it a\/}). The amplitude in this figure is non-dimensionalised by the volume equivalent sphere diameter $D$. In contrast to the results for $St$, which show a close to constant value, the results for $\langle a \rangle_n/D$ depend strongly on $\chi$. The evolution of $\langle a \rangle_n/D$ with $\chi$ is also non-monotonic towards both the oblate and prolate geometries.

For non-tumbling oblate particles, we observe a decrease in amplitude for increasing anisotropy towards a minimum amplitude in the zigzag (blue) regime around $\chi = 0.4$. Towards the flutter (purple) regime, we see a strong increase of the amplitude. These results are inline with  findings by \cite{fernandes2007}, who observe a similar trend for the amplitude. However, their minimum appears to be shifted to lower values of $\chi \approx 0.17$. One may thus speculate that for increasing $Ga$, the different regimes appear at lower anisotropy (higher $\chi$ for the oblate spheroids). And, indeed, we find  trends consistent with this hypothesis also for the eccentricity (Appendix \ref{sec:ecc}) and the phase-lag of the oscillations (\S\ref{sec:phase}).

The general trend in the oscillation amplitude for prolate particles mirrors the one observed for their oblate counterparts. Initially, in the longitudinal regime, there is a sharp decrease in $\langle a \rangle_n/D$, towards a minimum in the broadside regime, where the amplitude remains constant. It should be noted that the amplitude of the pointing vector oscillation for the broadside regime also decreases significantly, which is not the case for the minimum of the oblate particles (shown in figure \ref{fig:alignment_theta}). In the longitudinal regime, we see also that there is a significant amount of spread in the distribution of the observed amplitudes, as can be seen from the extent of the grey shaded regions. This appears to be caused by the two modes of the oscillations in this regime as was described in \S\ref{sec:prolate_ellipsoids}. At $\chi = 5$, data for the non-helical case is shown for which the amplitude is significantly higher than in the neighbouring broadside regime. In \S\ref{sec:phase}, we will show that the similarity of the trends for $\langle a \rangle_n/D$ between $\chi<1$ and $\chi>1$ is not coincidental but are also associated with a distinct phase delay in the alignments between the particle's velocity and its pointing vector.

When focussing on the tumbling regime, it appears that a small deviations from $\chi =1$ enhance the amplitude of the oscillations slightly compared to a sphere, but this effect drops off quite rapidly for more extreme values of the anisotropy. The value of $\langle a \rangle_n/D$ for $\chi=1$ obtained from our measurements is 0.683, which agrees well with data by \cite{Horowitz:2010}, from which we extract $\langle a \rangle_n/D \approx 0.675$  for our parameters.

\subsection{Eccentricity}\label{sec:ecc}
Another property that characterises the horizontal motions beside their amplitude is the eccentricity of the oscillations. We quantify this by defining the eccentricity 
\begin{equation}
\eta_n \equiv \frac{\tilde{y}_n}{\tilde{x}_n}, \label{eq:eccentricity}
\end{equation}
where $\tilde{y}_n$ and $\tilde{x}_n$ are the axis intercepts in the precession-corrected reference frame (see figure \ref{fig:trajectories}\,({\it c,\,d\/}) and the corresponding  discussion in \S \ref{sec:data_processing}; note also that $\tilde{x}_n = a_n$) similarly $n$ is an index over individual oscillation periods. Due to the complex motion and the presence of the two competing modes, it is not meaningful to compute values of $\eta_n$ for both prolate and tumbling particles. Therefore, the discussion is restricted to oblate non-tumbling particles in this section.

We present our results for the eccentricity in figure \ref{fig:Amp_and_Ecc}\,({\it b\/}), where the mean $\langle \eta \rangle_n$ is plotted and shadings represent the spread in the data. We observe that for increasing anisotropy $\langle \eta \rangle_n$ tends to decrease in general, which implies that the particle paths become more planar. Only the case at $\chi = 0.57$ does not follow this trend.

For comparison, we also show two fits determined by \cite{fernandes2007} for their low $Ga$ datasets in figure \ref{fig:Amp_and_Ecc}\,({\it b\/}). The peak values are similar to our results and the trends are consistent with a shift of the curves `to the right', i.e. towards lower anisotropy with increasing $Ga$. Towards larger values of $\chi$ (closer to $\chi = 1$), the geometry of the disks employed in \cite{fernandes2007} differs significantly from the spheroids used here. This might explain the fact that we observe an increase in $\langle \eta_n \rangle_n$ beyond $\chi = 0.67$ while the data of \cite{fernandes2007} appears to flatten out at larger values of $\chi$.

\bibliographystyle{jfm}
\bibliography{Citations_Ellipsoids.bib}

\end{document}